\documentclass[journal]{new-aiaa}
\usepackage[utf8]{inputenc}
\usepackage{textcomp}
\usepackage{graphicx}
\graphicspath{{../Allfigs/}}
\usepackage{amsmath}

\usepackage{amssymb}
\usepackage{bm}
\usepackage[version=4]{mhchem}
\usepackage{siunitx}
\usepackage{longtable,tabularx}
\usepackage{booktabs}
\usepackage[dvipsnames]{xcolor}
\newcommand{\action}[1]{\textcolor{black}{#1}}

\newcommand{\VGT}{\mathbf{G}_{ij}}
\newcommand{\Strain}{\mathbf{S}_{ij}}
\newcommand{\Vort}{\mathbf{W}_{ij}}
\newcommand{\NS}{\mathbf{G}_{ij}^N}   
\newcommand{\PS}{\mathbf{G}_{ij}^S}   
\newcommand{\RR}{\mathbf{G}_{ij}^R}   
\newcommand{\RS}{\mathbf{G}_{ij}^{RS}}   
\newcommand{\norm}[1]{\left\|#1\right\|}

\title{PhysMiner: An Agentic AI Framework for Automated Flow Component Analysis}

\author{Jiawei Chen\footnote{Postdoctoral Research Associate, Department of Aerospace Engineering.}
  , Han Gao\footnote{Assistant Professor, Department of Aerospace Engineering.}
  and Ping He\footnote{Assistant Professor, Department of Aerospace Engineering. Corresponding Author. Email: phe@iastate.edu}}
\affil{Iowa State University, Ames, IA 50011, USA}

\begin{document}

\maketitle

\begin{abstract}

\action{In-depth flow analysis is essential for understanding flow physics and improving turbulence models. Such analysis, however, still relies heavily on human expertise and effort.
This study presents PhysMiner, an automated agentic AI framework that combines the triple decomposition method (TDM) with large language model (LLM)-driven reasoning for turbulent flow analysis. TDM decomposes the velocity gradient tensor into rigid rotation, pure shearing, and normal straining. This turns raw flow fields into physically interpretable components for LLM analysis.
PhysMiner stores these components in a triple decomposition library. The library supports knowledge reuse and cross-case comparison for later flow analysis.
Existing LLM-based approaches rely heavily on subjective, user-defined prompting. PhysMiner instead grounds its reasoning in the literature or in user-specified research questions. This improves the determinism of the agentic workflow.
PhysMiner is demonstrated on three flow cases: periodic hill, iced airfoil, and propeller wake. For the periodic hill flow, the agent uses the literature database and TDM to modify an existing subgrid-scale model by adding a rotation-suppression correction term. The modified model improves Reynolds-stress predictions over the standard baseline model. For the iced airfoil and propeller cases, PhysMiner uses TDM to separate shear from genuine rotation. It correctly identifies separating shear layers and vortex structures that conventional methods may misidentify.
PhysMiner is open to the public. It provides a foundation for long-term, collaborative progress in automated turbulence-physics discovery.}

\end{abstract}

\section*{Nomenclature}

{\renewcommand\arraystretch{1.0}
\noindent\begin{longtable*}{@{}l @{\quad=\quad} l@{}}
$\VGT$          & velocity gradient tensor \\
$\Strain$       & symmetric strain-rate tensor \\
$\Vort$         & antisymmetric vorticity tensor \\
$\NS$           & normal straining tensor \\
$\PS$           & pure shearing tensor \\
$\RR$           & rigid rotation tensor \\
$\RS$           & rigid rotation - shear tensor \\
$\norm{\cdot}$  & Frobenius norm \\
$\mathrm{{gg}_{ss}}$ & relative contribution of strain-rate tensor, $\mathbf{S}_{ij} \mathbf{S}_{ij} / \mathbf{G}_{ij} \mathbf{G}_{ij}$ \\
$\mathrm{{gg}_{ww}}$ & relative contribution of vorticity tensor, $\mathbf{W}_{ij} \mathbf{W}_{ij} / \mathbf{G}_{ij} \mathbf{G}_{ij}$ \\
$\mathrm{{gg}_{rr}}$ & relative contribution of rigid rotation, $\mathbf{G}_{ij}^R \mathbf{G}_{ij}^R / \mathbf{G}_{ij} \mathbf{G}_{ij}$ \\
$\mathrm{{gg}_{ps}}$ & relative contribution of pure shear, $\mathbf{G}_{ij}^S \mathbf{G}_{ij}^S / \mathbf{G}_{ij} \mathbf{G}_{ij}$ \\
$\mathrm{{gg}_{ns}}$ & relative contribution of normal straining, $\mathbf{G}_{ij}^N \mathbf{G}_{ij}^N / \mathbf{G}_{ij} \mathbf{G}_{ij}$ \\
$\mathrm{{gg}_{rs}}$ & relative contribution of rotation--shear interaction $\mathbf{G}_{ij}^R \mathbf{G}_{ij}^S / \mathbf{G}_{ij} \mathbf{G}_{ij}$ \\
$Re$            & Reynolds number \\
$\nu_t$         & eddy (subgrid-scale) viscosity \\
\multicolumn{2}{@{}l}{Abbreviations}\\
LLM  & large language model \\
LES  & large-eddy simulation \\
SGS  & subgrid-scale \\
VGT  & velocity gradient tensor \\
TDM  & triple decomposition method \\
DIT  & decaying isotropic turbulence \\
BFS  & backward-facing step \\
CFD  & computational fluid dynamics \\
\end{longtable*}}

\section{Introduction}
\lettrine{F}{luids} are ubiquitous in the natural and engineered world, and turbulent flow plays a fundamental role in a wide range of physical phenomena and practical applications. From the foundational decay of isotropic turbulence (DIT)~\cite{comtebellot1971}, which serves as a simplified conceptual benchmark free of wall effects, to wall-bounded flows~\cite{lee2015direct}, the complexity of fluid motion scales rapidly. This complexity is further compounded by geometric intricacies and adverse pressure gradients, as manifested in backward-facing steps ~\cite{vogel1985combined}, periodic hills~\cite{breuer2009flow, rapp2011flow, krank2018direct}, marine propeller wakes ~\cite{posa2023anisotropy}. Understanding the physical mechanisms underlying these diverse flow configurations requires a systematic analysis of the local velocity kinematics, for which a principled decomposition of the velocity gradient tensor is essential.

The velocity gradient tensor contains rich information about local flow kinematics. Its systematic analysis enables a deeper understanding of the underlying flow structures and mechanisms. Traditionally, the Cauchy--Stokes decomposition~\cite{batchelor2000} has been the cornerstone of fluid kinematics, partitioning the velocity gradient into a symmetric strain rate tensor and an anti-symmetric vorticity tensor. The vorticity vector, derived from the vorticity tensor, has been widely applied in vortex identification. Accurate identification of vortical structures is essential, as vortices are ubiquitous in real-world flows. However, Cauchy--Stokes decomposition fails to distinguish between pure shearing and actual rigid rotation, often leading to ``ghost'' vortex signatures in high-shear regions. To address this, the triple decomposition method (TDM) has gained prominence in recent years. Kolář ~\cite{kolavr2007vortex} developed a method to isolate pure shearing effects by identifying a basic reference frame in which the residual vorticity after removing the pure shear contribution represents rigid-body rotation. However, this approach requires solving complex pointwise optimization problems. Liu et al. ~\cite{liu2018rortex} subsequently proposed a computationally efficient alternative based on real Schur decomposition, which identifies a possible rotational axis and enables a fast algorithm for computing the rotation component. Keylock ~\cite{keylock2018schur} used a complex Schur decomposition, but its relationship with basic reference frames in physical space remains unclear ~\cite{kronborg2023triple}. The triple decomposition has provided a clearer understanding of various flow phenomena. Liu et al. ~\cite{gao2019rortex} and Haller et al. ~\cite{haller2021can} improved vortex identification accuracy by extracting the rigid-body rotation component. Das et al. ~\cite{das2020revisiting} investigated forced isotropic turbulence and found that shear dominates the velocity gradient magnitude at all Reynolds numbers, while rigid-body rotation plays a minimal role. This shear dominance becomes more pronounced in regions of intense velocity gradients, often linked to intermittent turbulent structures. Enoki et al. ~\cite{enoki2023statistical} adapted the Biot–Savart law to reconstruct shear and non-shear velocity components and analyzed the statistical properties of these components in isotropic turbulence and turbulent jets. Arun et al. ~\cite{arun2024} have shown that the interplay between shear and rigid-body rotation characterizes the transition and turbulent decay of colliding vortex rings mediated by elliptic instability. Chen et al.~\cite{chen2026numerical} applied triple decomposition to investigate large-scale separation over iced wings, demonstrating its effectiveness in analyzing complex engineering flows. These studies collectively demonstrate the power of triple decomposition in revealing the physical structure of turbulent velocity gradients.  By further isolating the velocity gradient into pure shear, normal straining, and rigid rotation, triple decomposition provides a more granular lens through which to view flow physics. 

Parallel to these developments in fluid analysis methods, large language models (LLMs) and artificial intelligence (AI) agents have recently been introduced into fluid mechanics and CFD workflows. Early studies mainly focused on reducing the operational barrier of CFD by allowing users to describe simulation tasks in natural language. Representative systems, such as OpenFOAMGPT ~\cite{pandey2025openfoamgpt}, OpenFOAMGPT 2.0 ~\cite{feng2025openfoamgpt}, Foam-Agent/Foam-Agent 2.0 ~\cite{yue2025foam} ~\cite{pan2026automating}, FoamGPT ~\cite{yue2025foamgpt}, CFDagent ~\cite{xu2025cfdagent}, ChatCFD ~\cite{fan2025chatcfd}, FlamePilot ~\cite{xiao2026towards}, and natural-language-to-simulation intermediate-representation frameworks ~\cite{shenoy2026natural}, demonstrate that LLM-based agents can automate case setup, mesh generation, solver configuration, error correction, post-processing, and visualization across OpenFOAM-based and other CFD environments. A related line of work extends this capability to aerodynamic and multidisciplinary design optimization, where multi-agent systems such as specialized aerodynamic design agents ~\cite{lee2025aerodynamic}, AirfoilAgent ~\cite{fan2026airfoilagent}, airfoil and wing agents ~\cite{fang2026agentic}, multidisciplinary design optimization agents ~\cite{guo2025multidisciplinary}, vision-language models for engineering design evaluation~\cite{picard2025concept}, AI design agents for aesthetic and aerodynamic car design~\cite{elrefaie2025ai}, and risk-aware set-based engineering design agents ~\cite{kumar2026agentic} couple LLMs with CAD, CFD, optimization, and knowledge-retrieval tools to support geometry generation, design-space exploration, and iterative performance improvement. More importantly for flow-mechanism discovery, recent studies suggest that LLMs can move beyond workflow automation and participate in model formulation and scientific reasoning. AutoTurb uses LLM-guided symbolic regression and evolutionary search to discover algebraic turbulence closure corrections ~\cite{zhang2025autoturb}, while LLM-driven turbulence-model development frameworks treat the LLM as an interactive collaborator that proposes, refines, and evaluates physically interpretable wall-model modifications under complex flow conditions ~\cite{yang2025large}. In parallel, emerging agentic infrastructures based on retrieval-augmented generation, Model Context Protocol tool use, automated tool standardization, and latent foundation models for PDE spaces provide a foundation for closed-loop workflows in which agents can formulate hypotheses, query simulations or surrogate models, analyze flow responses, and verify candidate mechanisms ~\cite{ouyang2025code2mcp}~\cite{di2026toolrosetta}~\cite{yue2026static}~\cite{vishwasrao2026agentic}. These studies indicate that LLMs may serve not only as interfaces to CFD software, but also as reasoning agents that bridge raw numerical data, symbolic model structures, and the discovery of new physical insight. Despite these advances, most existing AI agent frameworks focus primarily on automating CFD workflows, including case setup, mesh generation, and solver configuration. In contrast, autonomous agent-based discovery of physical mechanisms remains largely unexplored. Consequently,  the reasoning capability of agentic AI has not yet been fully exploited in the field of fluid mechanics.

This paper presents an agentic AI framework, PhysMiner, for in-depth analysis of flow components. The scope of this contribution should be clearly distinguished from the physical foundations of the triple decomposition method. \action{TDM reveals several physical characteristics: the dominance of pure shear in turbulent velocity gradients, the vanishing of rigid rotation and normal straining at no-slip walls, and the removal of shear contamination in vortex identification. These characteristics are intrinsic properties of the method, established in prior studies~\cite{kolavr2007vortex,liu2018rortex,arun2024,das2020revisiting}.} PhysMiner does not claim to rediscover these mechanisms. Instead, TDM is used to automatically transform raw flow fields into physically interpretable components. These components provide the basis for LLM-driven analysis of latent physical patterns and the formulation of new directions for turbulence-model development. This study highlights the potential of LLM-based reasoning to accelerate the transition from flow data to physical knowledge and modeling hypotheses. \action{Three aspects of the framework stand out in this study. First, PhysMiner supports both fully automatic, literature-driven reasoning and user-specified research questions. This reduces the framework's dependence on manually crafted prompts. Second, a continuously expanding triple decomposition library enables structured comparison between a target flow and previously analyzed cases. It reveals subtle physical differences even among superficially similar flows. Third, the framework is demonstrated on periodic hill, iced-airfoil, and propeller-wake flows. It yields an improved subgrid-scale model for periodic hill flow and a clearer interpretation of shear-contaminated flow structures for the other two cases.} The source code of PhysMiner and its Triple Decomposition Library is publicly available\footnote{\url{https://github.com/iDesign-Lab/PhysMiner}}. The framework also provides a community contribution mechanism that allows external researchers to submit new flow cases and expand the underlying knowledge base. The remainder of this paper is organized as follows. Sec.~II describes the PhysMiner architecture. Sec.~III presents the results and discussion. Sec.~IV summarizes the main findings and concludes the paper.

\section{Methodology}

\subsection{Overall Architecture of the PhysMiner Framework}

\begin{figure}[hbt!]
\centering
\includegraphics[width=0.9\textwidth]{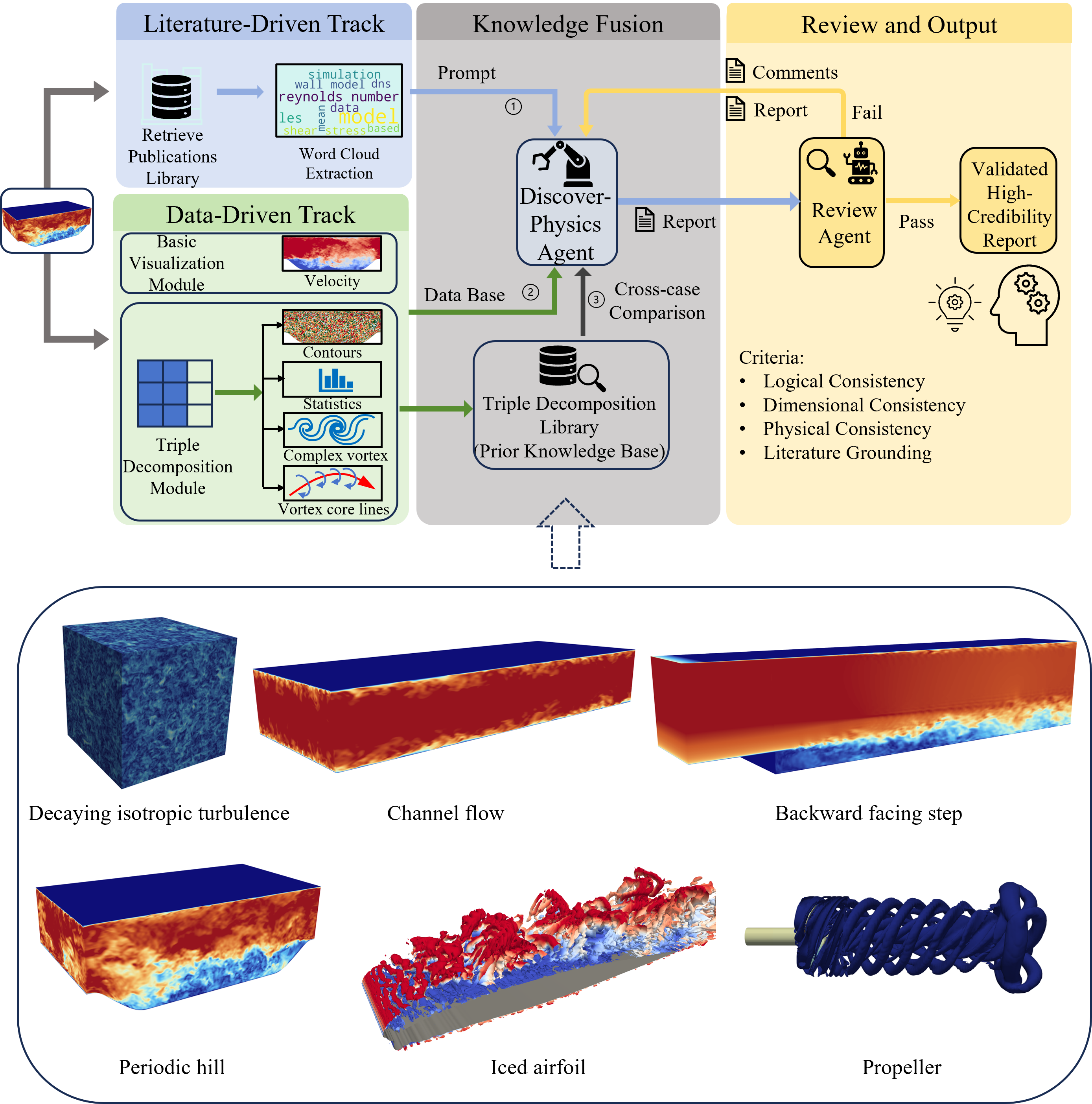}
\caption{Architecture of the proposed multi-agent framework PhysMiner.}
\label{fig:architecture}
\end{figure}

The PhysMiner framework is implemented as an automated, multi-agent pipeline designed for automated flow-component analysis and further in-depth insights, as illustrated in Fig.~\ref{fig:architecture}. The architecture operates through a dual-track concurrent processing strategy followed by a closed-loop refinement stage. The Literature-Driven Track focuses on contextual grounding by retrieving a specialized publications library and performing word cloud extraction to identify high-frequency physical keywords. When a specific research question is provided by the user, the Literature-Driven Track is bypassed in favor of that user-specified prompt. Simultaneously, the Data-Driven Track executes the numerical core, utilizing a basic visualization module and a Triple Decomposition module to transform raw CFD data into physically interpretable components, including contours, statistical distributions, complex vortex topologies, and vortex core lines. In the Knowledge Fusion stage, the Discover-Physics Agent synthesizes these inputs by integrating real-time calculation data with literature-derived keywords and a Triple Decomposition Library experience (Table~\ref{tab:cases} lists the flow cases used in this paper). This synthesis leads to the generation of initial physical discoveries. To ensure the fidelity of these insights, the framework adopts a Closed-Loop Calibration mechanism. A Review Agent, acting as a domain expert, audits the generated discoveries for physical consistency and accuracy. If discrepancies are detected, a feedback loop is triggered, prompting the system to re-assess and rectify physical insights or re-extract information. Once validated, the pipeline culminates in the generation of a high-credibility, structured report. This iterative design ensures that the framework’s analytical capability evolves progressively, adapting its heuristics to the specific flow topologies encountered.

\begin{table}[hbt!]
\caption{Summary of the six flow configurations used in this study.}
\label{tab:cases}
\centering
\begin{tabular}{l p{0.26\textwidth} l >{\raggedright\arraybackslash}p{0.26\textwidth} c c}
\toprule
Case & Geometry \& Domain & Data & Reynolds number & Separation & Ref.\\
\midrule
DIT &
  Free-shear turbulence; $256^3$ cubic grid; snapshot at $t=0.285$ &
  LES & $Re_M=34{,}000$, based on grid width $M$ & No & \cite{comtebellot1971} \\[4pt]
Channel flow &
  Internal, flat walls; fully developed inflow &
  LES & $Re_{\tau}=543.5$, based on channel half-height $h$ & No & \cite{lee2015direct} \\[4pt]
BFS &
  Internal, abrupt step expansion; channel expansion ratio ER$\,{=}\,5/4$ &
  LES & $Re_H=28{,}000$, based on step height $H$ & Yes & \cite{vogel1985combined} \\[4pt]
Periodic hill &
  Internal, curved hills; domain $9h\times3.035h\times4.5h$ &
  LES & $Re_h=10{,}595$, based on hill height $h$ & Yes & \cite{breuer2009flow, rapp2011flow, krank2018direct} \\[4pt]
Ice airfoil &
  External, curved surface; domain $20c$ &
  IDDES & $Re_c=3.5\times10^6$, based on airfoil chord $c$ & Yes & \cite{broeren2006flowfield} \\[4pt]
Propeller wake &
  Internal rotating wake; four-bladed marine propeller &
  URANS & $Re_D=1{,}000{,}000$, based on propeller diameter $D$ & No & \cite{jasak2007openfoam} \\
\bottomrule
\end{tabular}
\end{table}

\subsection{Literature-Driven Track}

The framework initiates with a dual-track concurrent processing strategy to ensure a comprehensive understanding of the target flow. One track is the Literature-Driven Track, while the other is the Data-Driven Track. The Literature-Driven Track analyzes published studies to identify key challenges and unresolved issues in specific flow configurations, thereby providing directions for scientific discovery. In parallel, the Data-Driven Track decomposes the flow field into physically interpretable components to achieve a deeper understanding of the underlying flow structures and mechanisms, thereby establishing the foundation for subsequent discoveries.

The Literature-Driven Track retrieves relevant studies and utilizes word cloud extraction to distill high-frequency keywords. To construct the literature dataset, a target keyword (e.g., periodic hill'') is searched in the Web of Science using two ranking strategies: "Citations: highest first" and "Date: newest first". The retrieved papers are then downloaded through the university library access. A word cloud is then generated using Python-based tools. Specifically, the PyMuPDF library (\texttt{fitz}) is employed to extract selectable text from each PDF document on a page-by-page basis. The raw text is subsequently cleaned using Python's built-in \texttt{re} module to remove numerical characters, punctuation, and domain-generic stopwords (e.g., ``figure,'' ``equation,'' ``reference''), so that only content-bearing terms are retained. The cleaned corpus is then passed to the \texttt{wordcloud} library, which computes word frequencies across all documents and renders each term at a size proportional to its occurrence count; bigram collocation detection is enabled to preserve meaningful two-word phrases. The resulting image is composited and exported via \texttt{matplotlib}. This pipeline provides an efficient means of visualizing keyword frequency across the literature. 

In addition to fully automatic operation, the Literature-Driven Track supports an optional user-specified mode. When a user provides a specific research question, the literature retrieval and word-cloud extraction steps are frozen, and the discover-physics agent instead follows the user-provided prompt directly during the Knowledge Fusion stage. This prompt only needs to state the question of interest. It does not need to specify implementation details such as which decomposition method to apply, so the burden on the user remains minimal. This dual-mode design allows PhysMiner to operate either autonomously, guided by literature-derived priorities, or interactively, guided by a user-specified question, as demonstrated for the iced-airfoil and propeller-wake cases in Sec.~III.E and III.F.

\subsection{Data-Driven Track}

Simultaneously, the Data-Driven Track serves as the computational engine of the framework. Raw CFD snapshots are processed through two modules: a basic visualization module and a Triple Decomposition module. The basic visualization module automatically generates velocity and vortex visualizations, while the Triple Decomposition module separates the velocity gradient tensor into physically distinct components. This track generates high-fidelity visual and quantitative inputs, including contours, statistical distributions, complex vortex topologies, and vortex core lines, providing a multi-dimensional representation of the flow field. The three-component decomposition provides deeper insight into flow physics by separating the flow into physically interpretable constituents. Systematic analysis of each component facilitates the identification of previously unrecognized mechanisms and underlying physical processes. Such an enhanced understanding of the flow not only advances fundamental knowledge but also offers valuable guidance for improving and developing turbulence modeling approaches.

Traditionally, the velocity gradient tensor $\VGT$ is partitioned into a symmetric strain-rate tensor $\Strain$ and an antisymmetric vorticity tensor $\Vort$:
\begin{equation}
\label{eq:CS}
\VGT = \Strain + \Vort
\end{equation}

where $\Strain = \tfrac{1}{2}(\VGT + \VGT^T)$ and $\Vort = \tfrac{1}{2}(\VGT - \VGT^T)$.

Correspondingly, the strength of the velocity gradients can be expressed as
\begin{equation}
\label{eq:CS_strength}
\VGT \VGT = \Strain \Strain + \Vort \Vort
\end{equation}

The two terms represent the strengths of the constituents in Eq.~\eqref{eq:CS}. Furthermore, the velocity gradient partitioning is defined in terms of the relative contributions of these constituents to $\mathbf{G}_{ij} \mathbf{G}_{ij}$. The relative contribution of the strain-rate tensor is $\mathrm{{gg}_{ss}} = \mathbf{S}_{ij} \mathbf{S}_{ij} / \mathbf{G}_{ij} \mathbf{G}_{ij}$, and the relative contribution of the vorticity tensor is $\mathrm{{gg}_{ww}} = \mathbf{W}_{ij} \mathbf{W}_{ij} / \mathbf{G}_{ij} \mathbf{G}_{ij}$.

A more refined framework, known as triple decomposition, partitions the VGT into three physically distinct components: normal straining, pure shearing, and rigid-body rotation~\cite{arun2024}. The VGT can be expressed in its principal reference frame as
\begin{equation}
\label{eq:TD}
\VGT = \NS + \RR + \PS
\end{equation}

where $\NS$, $\RR$, and $\PS$ denote the normal straining, rigid rotation, and pure shearing tensors, respectively. The rigid rotation tensor $\RR$ is directly related to a vortex identification method $\mathbf{R}$ proposed by Liu et al.~\cite{liu2018rortex}, which has also been employed to develop a subgrid-scale model ~\cite{ding2022liutex} ~\cite{hossen2025liutex} ~\cite{chen2025near} for large eddy simulation. These triple decomposition tensors can be determined and transformed to the original coordinate system using the ordered real Schur decomposition of $\VGT$~\cite{arun2024}.

Correspondingly, the strength of the velocity gradients can be expressed as
\begin{equation}
\label{eq:TD_strength}
\VGT \VGT = \PS \PS + \NS \NS + \RR \RR + \RR \PS
\end{equation}

The first three terms represent the strengths of the constituents in Eq.~\eqref{eq:TD} and the last term represents the interaction between shearing and rigid rotation. Furthermore, the velocity gradient partitioning is defined in terms of the relative contributions of these constituents to $\VGT \VGT$. The relative contribution of normal straining is $\mathrm{gg}_{ns} = \NS \NS / \VGT \VGT$, the relative contribution of rigid rotation is $\mathrm{gg}_{rr} = \RR \RR / \VGT \VGT$, the relative contribution of pure shearing is $\mathrm{gg}_{ps} = \PS \PS / \VGT \VGT$, and the relative contribution of the rotation--shear interaction is $\mathrm{gg}_{rs} = \RR \PS / \VGT \VGT$.

The rigid rotation measure $\mathrm{gg}_{rr}$ provides a threshold-insensitive method for vortex identification, capable of capturing both strong and weak vortical structures ~\cite{chen2025relative}. Building upon such flow field analysis, a large language model (LLM) can further leverage the $\mathrm{gg}_{rr}$-based results to facilitate scientific discovery in two aspects. First, discrepancies between the vortex structures identified by $\mathrm{gg}_{rr}$ and those reported in prior studies or obtained using conventional methods may indicate potential new physical insights. Second, the physical interpretation of $\mathrm{gg}_{rr}$ also provides new insights for turbulence modeling~\cite{chen2025dynamic}. Large language models (LLMs) can further assist in identifying potential replacements for existing model terms, thereby facilitating the development of improved turbulence modeling approaches.

The pure shear measure $\mathrm{gg}_{ps}$ characterizes the dissipative effects in turbulent flows. Accurate identification of pure shear is essential for revealing the intrinsic flow structures and underlying physical mechanisms. In conventional vortex identification methods, the presence of pure shear often contaminates the results, leading to structures that do not faithfully represent true vortical motion. By analyzing the distribution of $\mathrm{gg}_{ps}$, the shear contribution can be identified and separated, yielding a clearer depiction of the intrinsic vortex structure. Although removing shear may eliminate certain apparent features, this process in fact produces a more physically accurate representation of vortices. Building upon this analysis, an LLM can further exploit $\mathrm{gg}_{ps}$-based results to uncover new scientific insights. First, differences between shear-filtered vortex structures and those obtained using conventional methods may highlight previously unrecognized mechanisms or misinterpretations associated with shear contamination. Second, since pure shear is closely related to vortex attenuation and dissipation, the LLM can use $\mathrm{gg}_{ps}$ distributions to infer and model the processes governing vortex weakening and decay, thereby supporting the development of improved physical models with enhanced interpretability.

The normal straining measure $\mathrm{gg}_{ns}$ characterizes the extensional and compressional deformation in a flow field, capturing the effects of normal strain that are fundamentally distinct from both rigid rotation and pure shear. By analyzing the spatial distribution of $\mathrm{gg}_{ns}$, regions dominated by stretching and compression can be clearly identified, which are often associated with flow separation, interface deformation, and energy transfer across scales. Unlike vortex-focused diagnostics, $\mathrm{gg}_{ns}$ highlights the non-rotational mechanisms that govern the evolution of flow structures, thereby providing a complementary perspective for interpreting complex flow dynamics. In particular, the interaction between normal strain and vortical motion can be elucidated by jointly examining $\mathrm{gg}_{ns}$ with other measures, offering a more complete picture of the local kinematics. Building upon such flow field analysis, an LLM can further leverage $\mathrm{gg}_{ns}$-based results to enable scientific discovery. \action{First, normal strain may show anomalous or previously overlooked distributions, especially in regions where conventional analyses focus mainly on vortices. Such distributions may reveal new mechanisms related to flow instability, structural deformation, or energy redistribution.} Second, by correlating $\mathrm{gg}_{ns}$ with vortex evolution and dissipation processes, the LLM can help identify and model the coupling between strain-dominated and rotation-dominated dynamics, thereby facilitating the development of more comprehensive and physically grounded models for complex flows.


As illustrated in Fig.~\ref{fig:architecture}, the data-driven track of the PhysMiner framework consists of two modules: a basic visualization module and a triple decomposition module. This section focuses on the automated triple decomposition module, which plays the central role in extracting flow features and enabling scientific discovery. This module operates fully automatically. Once a raw CFD case is provided, all analyses are performed without additional user configuration or manual intervention.  To use the framework, one simply needs to place the ``PhysMiner" folder in the CFD case directory and run the provided Python script. PhysMiner then automatically generates four outputs: (1) contours of the flow field, (2) statistics of the components, (3) complex vortex system identification, and (4) vortex core lines. Some selected functions are demonstrated using a decaying isotropic turbulence case in this section, while additional functions are presented in Sec.~III Results and discussion.

\subsubsection{Contours of the Flow Field}

\begin{figure}[hbt!]
\centering
\begin{minipage}[b]{0.6\textwidth}
  \centering
  \includegraphics[width=\textwidth]{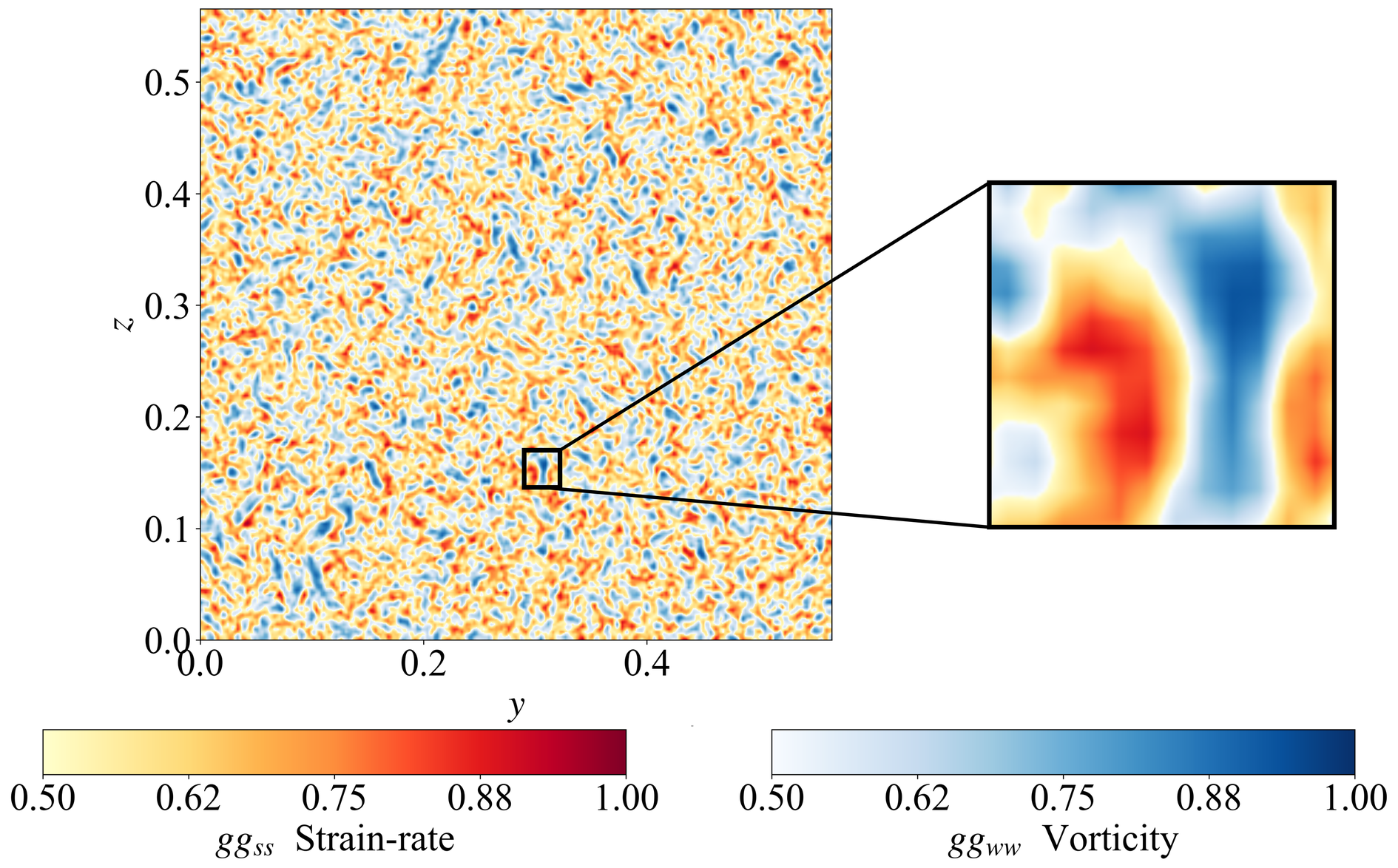}\\[2pt]
  a) Cauchy--Stokes decomposition
\end{minipage}
\hfill
\begin{minipage}[b]{0.6\textwidth}
  \centering
  \includegraphics[width=\textwidth]{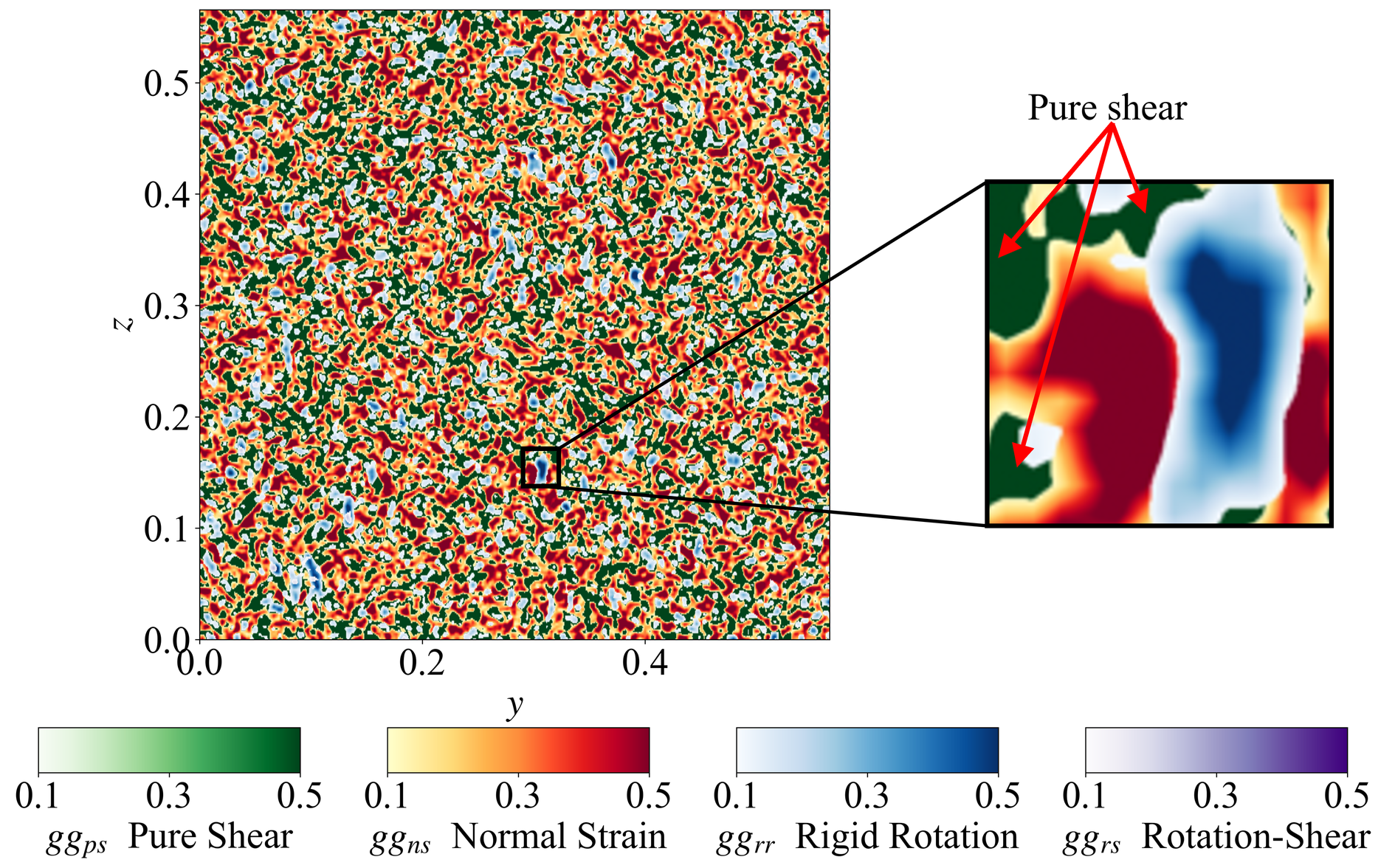}\\[2pt]
  b) Triple decomposition
\end{minipage}
\caption{Comparison of the Cauchy--Stokes decomposition and the triple decomposition in decaying isotropic turbulence.}
\label{fig:DIT_decomp}
\end{figure}

The automated triple decomposition module enables the visualization of contours for each decomposed component. To facilitate a clearer understanding of the triple decomposition, the Cauchy--Stokes decomposition is also presented for comparison. Fig.~\ref{fig:DIT_decomp} compares the Cauchy--Stokes decomposition and the triple decomposition in decaying isotropic turbulence. As shown in Fig.~\ref{fig:DIT_decomp}(a), the vorticity component obtained from the Cauchy--Stokes decomposition is contaminated by pure shear, which obscures the accurate identification of vortex structures. In contrast, Fig.~\ref{fig:DIT_decomp}(b) shows that the triple decomposition effectively separates rigid rotation from pure shear, thereby providing a clearer and more physically meaningful representation of the flow structures.

\subsubsection{Statistics of the components}

\begin{figure}[hbt!]
\centering
\begin{minipage}[b]{0.48\textwidth}
  \centering
  \includegraphics[width=\textwidth]{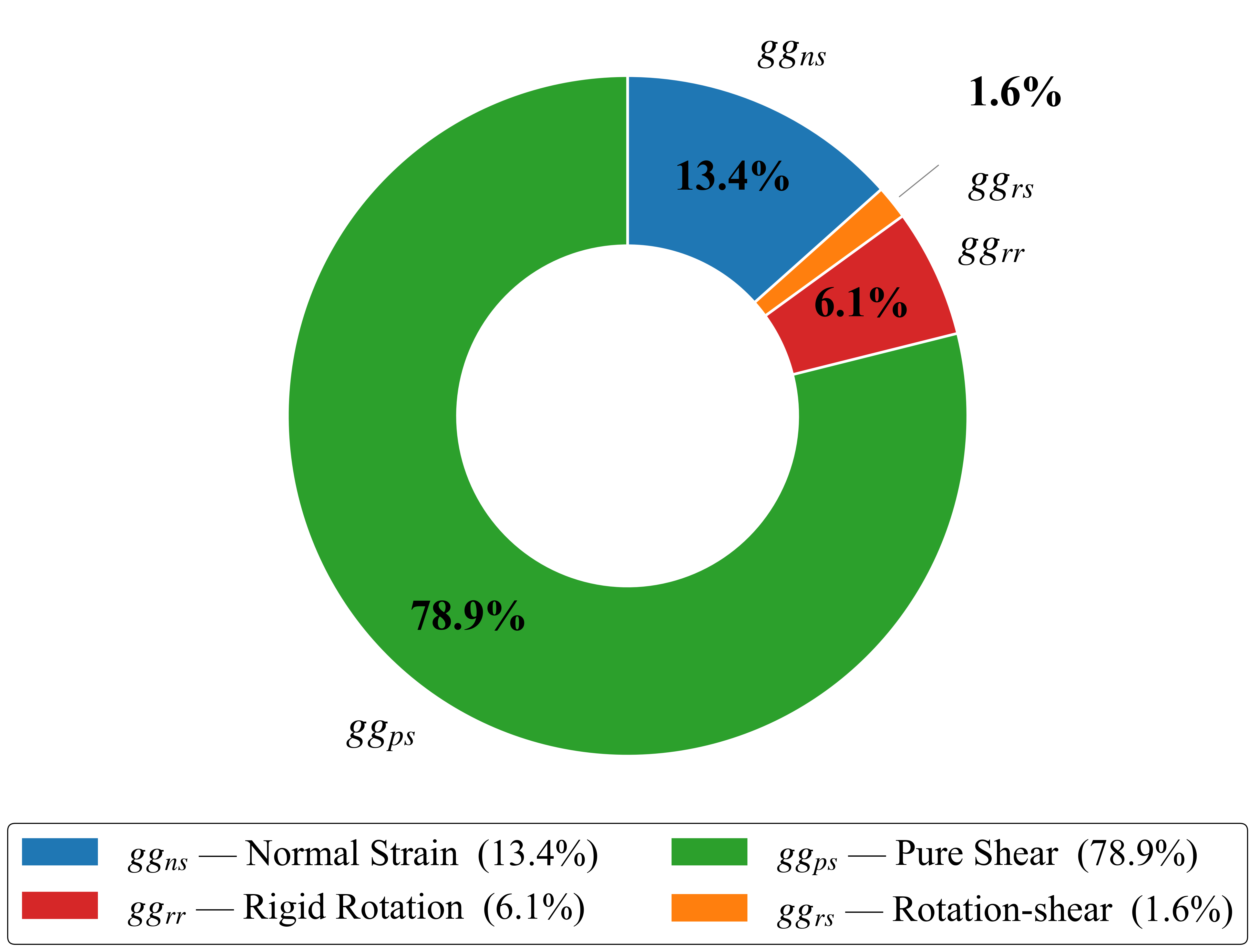}\\[2pt]
\end{minipage}
\caption{Statistical analysis of the components obtained from the triple decomposition for decaying isotropic turbulence.}
\label{fig:DIT_stats}
\end{figure}

The automated triple-decomposition module supports statistical analysis over both the entire computational domain and selected streamwise locations. Global statistics characterize the overall flow behavior, while local statistics reveal the spatial evolution of the flow. Figure~\ref{fig:DIT_stats} presents the statistics of the triple-decomposition components for decaying isotropic turbulence. In Figure~\ref{fig:DIT_stats}, the pure shear component dominates the overall flow in this case. In addition to statistical analysis, the framework provides spatial distributions along prescribed directions and at selected locations. These distributions enable a more detailed examination of the spatial characteristics and evolution of each component. Representative results are presented in the following Sec.~III Results and Discussion.

\subsubsection{Complex Vortex System Analysis}

Unlike conventional methods that depend on user-defined thresholds, often leading to a trade-off between isolating strong vortices and preserving weaker ones, PhysMiner can use $\mathrm{gg}_{rr}$ to enable a more complete and physically consistent representation of the vortex system. By analyzing the spatial distribution of $\mathrm{gg}_{rr}$ and comparing it with established results in the literature, one can assess whether the identified vortex structures align with the commonly accepted flow organization or reveal previously unresolved features. This capability is demonstrated in Sec.~III.D using the propeller-wake flow.

\subsubsection{Vortex Core Lines}

The automated triple-decomposition module enables the extraction of vortex core lines. Since the vortex identification method $\mathbf{R}$ proposed by Liu et al.~\cite{liu2018rortex} is directly associated with the rigid-rotation component in the triple decomposition, the extraction of vortex core lines based on the $\mathbf{R}$ vector is also incorporated into the automated triple-decomposition module. The explicit formula of $\mathbf{R}$ vector~\cite{wang2019explicit} is given by 

\begin{equation}
\label{eq:RortexVector}
\bm{R} = R\bm{r} = \left[ \bm{\omega} \cdot \bm{r} - \sqrt{(\bm{\omega} \cdot \bm{r})^2 - 4\lambda_{ci}^2} \right] \bm{r}
\end{equation}

Here $\bm{\omega}$ is the vorticity vector, $\lambda_{ci}$ is the imaginary part of the complex conjugate eigenvalue of the velocity gradient tensor, and $\bm{r}$ is the eigenvector corresponding to the real eigenvalue.

\begin{figure}[hbt!]
\centering
\includegraphics[width=0.5\textwidth]{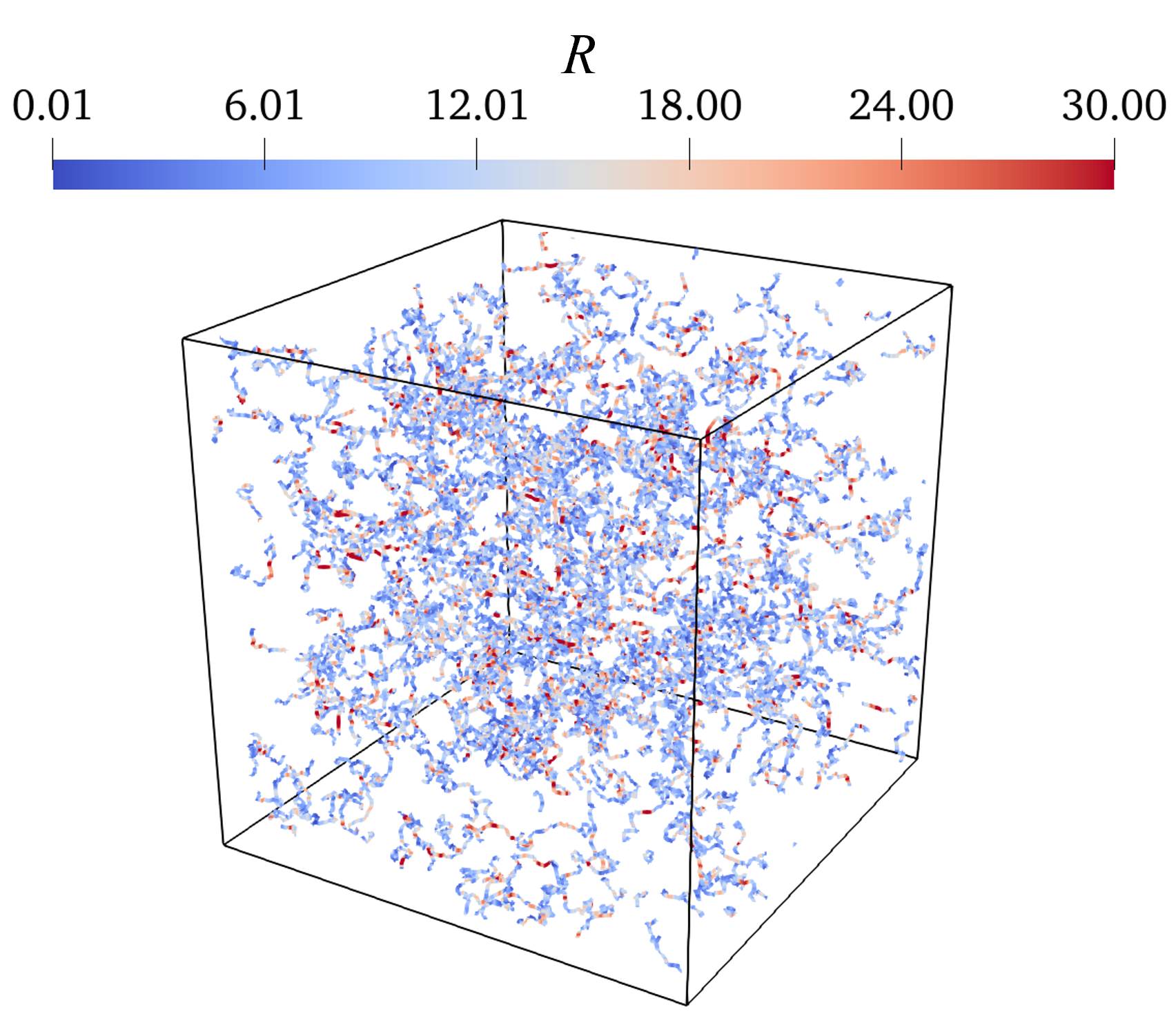}
\caption{Extracted vortex core lines in decaying isotropic turbulence (DIT).}
\label{fig:vortex_core}
\end{figure}

Fig.~\ref{fig:vortex_core} presents the extracted vortex core lines in decaying isotropic turbulence. The vortical structures are distributed throughout the computational domain, exhibiting highly complex and irregular spatial patterns. The orientations of the vortices are randomly distributed without any preferred direction, which is consistent with the fundamental characteristic of isotropic turbulence. The automated extraction of vortex core lines also captures the tangled topology of vortex structures, consistent with previous studies~\cite{xiong2019identifying}. Although decaying isotropic turbulence is employed here for demonstration, the automated vortex core line extraction approach can be readily extended to complex engineering flows.

\subsection{Knowledge Fusion}

At the heart of the pipeline is the Knowledge Fusion stage, where the Discover-Physics Agent acts as a central integrator. This agent synergizes three disparate streams of information: (1) the semantic insights (keywords) from the literature-driven track, (2) the quantitative triple decomposition data from the data-driven track, and (3) the cross-case comparison by the triple decomposition library. In particular, the literature-driven track directs the discover-physics agent toward the most critical and challenging physical issues associated with a specific flow.

The triple decomposition library facilitates a deeper understanding of new flow cases through comparisons with historical flow cases stored in the database. Identifying subtle differences among similar flow structures helps reveal potentially critical underlying flow physics. Moreover, as PhysMiner continues to be applied to new flow cases, new flow instances are continuously incorporated into the library, thereby expanding the knowledge base and further enhancing its comparative and analytical capabilities.

\subsection{Review and Output}

To ensure the reliability of new findings, a review stage is incorporated into the framework. A Review Agent, acting as a virtual domain expert, evaluates each generated report against four predefined criteria. The objective is to identify and eliminate hallucinations, logical inconsistencies, and physically invalid conclusions that may arise during reasoning and discovery processes. The Review Agent generates a structured comments file. The first line of the file contains the overall review result, either Pass or Fail. A Pass indicates that the report satisfies all review criteria. A Fail indicates that at least one criterion has been violated. In this case, the comments file lists each failed criterion, provides an explanation for the failure, and identifies the corresponding location in the report. When a review process fails, both the report and the comments file are returned to the previous physics discovery stage. The discovery and review cycle continues iteratively until the first line of the comments file becomes Pass. Only the report that successfully passes all review criteria is accepted as final output. The review criteria are embedded within the Review Agent, which systematically evaluates the report against each criterion. The four criteria embedded in the review agent are listed as follows:

(1) Internal Logical Consistency: All findings must remain logically consistent throughout the report. Terminology, definitions, and quantitative descriptions should be applied uniformly. For example, a quantity identified as the maximum value in one section must not be referred to as the minimum value in another section.

(2) Dimensional consistency: Any newly proposed equation or model must satisfy dimensional consistency, meaning that both sides of the equation must have identical physical dimensions. If dimensional inconsistency is detected, the proposed formulation is considered invalid and should be either revised or discarded.

(3) Physical consistency: All statements in the report should satisfy physical constraints. Specifically, the values of $\mathrm{gg}_{ps}$, $\mathrm{gg}_{rr}$, and $\mathrm{gg}_{ns}$ must remain within the physically admissible range of 0 to 1.  

(4) Literature grounding: Any newly proposed finding or expression should be compared with existing methods and supported by appropriate references. If the study addresses a completely new problem for which no direct comparison is available, the significance and novelty of the proposed finding should be clearly justified.

\subsection{Evolving Triple Decomposition Library}

\begin{figure}[hbt!]
\centering
\includegraphics[width=0.9\textwidth]{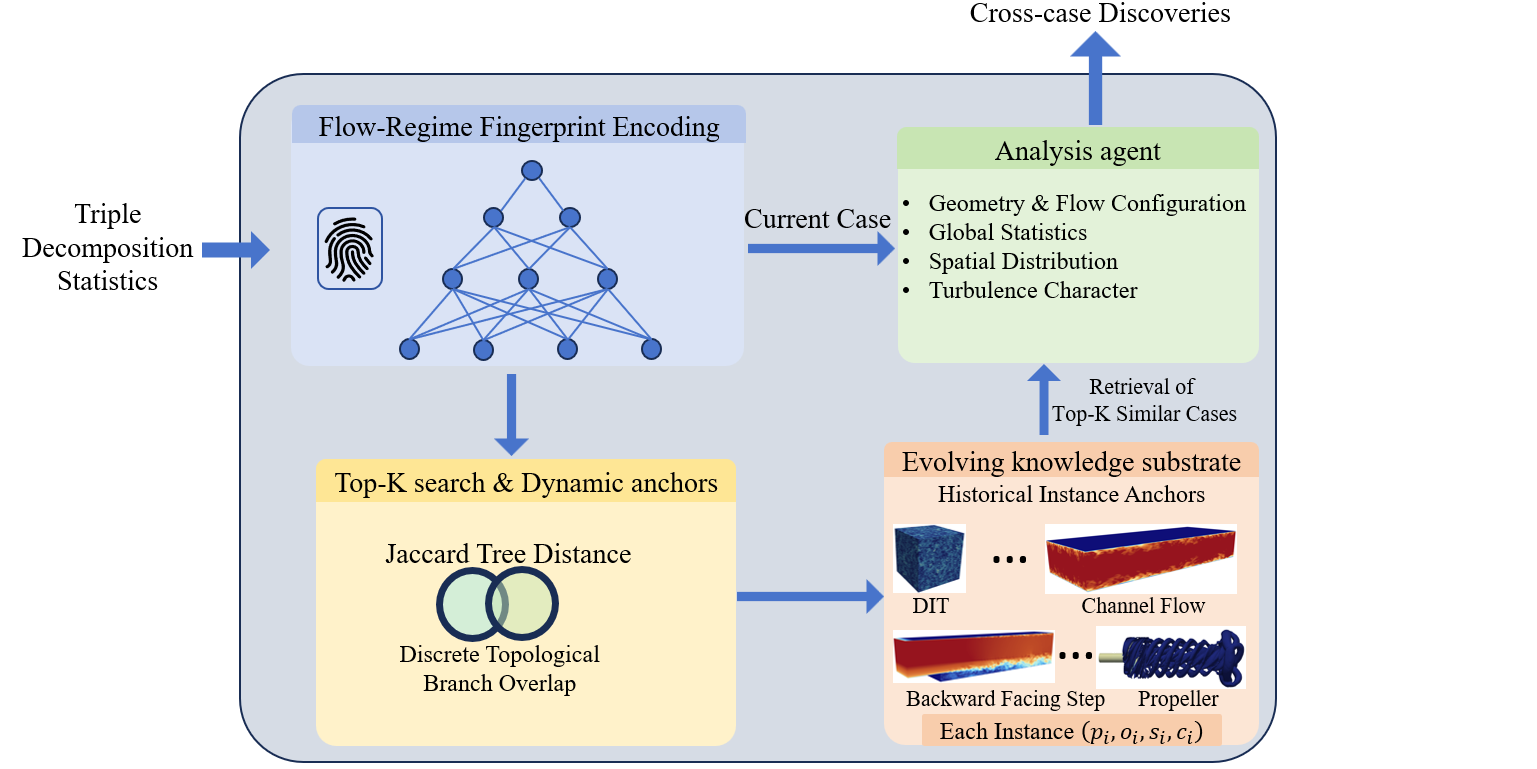}
\caption{Architecture of the proposed Triple Decomposition library.}
\label{fig:TDL_architecture}
\end{figure}

The Triple Decomposition Library is designed as a non-parametric, continuously expanding knowledge paradigm. Given the currently limited number of cases (five), the term `continuously expanding'' is used instead of `self-evolving'' to avoid overstating inductive generalization, which will be investigated as the library grows. Fig.~\ref{fig:TDL_architecture} illustrates the overall architecture of the proposed library. The library takes triple decomposition results as inputs and performs comparative physical analyses by transferring knowledge from historical cases.

The workflow begins with the mapping process and the closest historical problem is identified for a new target problem. The target problem is first mapped into the problem tree, where a unique fingerprint is assigned to characterize its flow attributes. Each fingerprint consists of a series of problem descriptors, as listed in Table 2. For example, a backward-facing step flow can be represented by the fingerprint
$[1,3,5,7,10,12,13,18,19,23]$, corresponding to Single-phase flow, Internal flow, Viscous flow, Newtonian flow, Unsteady flow, Separated flow, Isothermal flow, Three-dimensional flow, Stationary boundary, and Incompressible flow, respectively, according to Table 2. After the fingerprint generation, the library scans the problem tree to identify the closest historically and successfully solved case. The similarity between the target problem and the historical cases is quantified using the Jaccard tree distance. The Jaccard Tree Distance ($D_{\text{tree}}$) measures the similarity between the unsolved target case and an historical case by evaluating the overlap of their traversal paths in the hierarchical problem tree. Let $\mathcal{P}_{\text{target}}$ and $\mathcal{P}_{\text{hist}}$ denote the sets of activated tree nodes from the root to the leaf for the target case and a historical anchor case, respectively. The distance is defined as

\begin{equation}
\label{eq:JaccardTree}
D_{\text{tree}}(\mathcal{P}_{\text{target}}, \mathcal{P}_{\text{hist}}) 
= 1 - 
\frac{
|\mathcal{P}_{\text{target}} \cap \mathcal{P}_{\text{hist}}|
}{
|\mathcal{P}_{\text{target}} \cup \mathcal{P}_{\text{hist}}|
}
\end{equation}

After the closest historical case is identified, cross-case comparison is performed by the Analysis Agent. Historical experience, including the full-domain statistics, streamwise profiles and the spatial distribution of the triple decomposition components, is used for comparative analysis. The comparison results are then passed to the Discover-Physics Agent for subsequent physical interpretation and knowledge discovery. Meanwhile, the current case is stored in the knowledge library to support future knowledge transfer and inference. Each case is represented as a structured tuple $(p_i, o_i, s_i, c_i)$, where $p_i$ denotes the unique problem fingerprint, $o_i$ represents the full-domain statistical features, $s_i$ corresponds to the streamwise profiles, and $c_i$ represents the spatial distribution of the triple decomposition components. This non-parametric and continuously expanding database architecture follows the same conceptual principle as recent framework GRAFT-ATHENA~\cite{toscano2026graft} that employ self-improving agentic teams for autonomous scientific discovery and evolutionary algorithm development.

\begin{table}[hbt!]
\caption{Summary of the classification and different types of fluid flows.}
\label{tab:flow_cases_revised_v2}
\centering
\begin{tabularx}{\textwidth}{l X r}
\toprule
\textbf{Classification Dimension} & \textbf{Typical Options / Categories} & \textbf{No.} \\
\midrule

Phase Condition
& Single-phase flow;
  Multiphase flow
& 1, 2 \\[4pt]

Flow Configuration
& Internal flow;
  External flow
& 3, 4 \\[4pt]

Viscous Effect
& Viscous flow;
  Inviscid flow ($Re \to \infty$)
& 5, 6 \\[4pt]

Rheological Behavior
& Newtonian;
  Non-Newtonian
& 7, 8 \\[4pt]

Temporal Behavior
& Steady flow ($\partial/\partial t = 0$);
  Unsteady flow
& 9, 10 \\[4pt]

Flow Separation
& Attached flow (no separation);
  Separated flow
& 11, 12 \\[4pt]

Thermal Coupling
& Isothermal flow;
  Non-isothermal flow;
  Conjugate heat transfer
& 13, 14, 15 \\[4pt]

Spatial Dimensionality
& 1D;
  2D;
  3D
& 16, 17, 18 \\[4pt]

Boundary Motion
& Stationary boundary;
  Moving boundary (translating/oscillating);
  Rotating boundary (rotating machinery, MRF);
  Deforming boundary (fluid--structure interaction)
& 19, 20, 21, 22 \\[4pt]

Compressibility
& Incompressible flow ($Ma < 0.3$, $\Delta\rho/\rho < 5\%$);
  Compressible flow ($Ma \ge 0.3$);
  Transonic ($0.8 \le Ma \le 1.2$);
  Supersonic ($Ma > 1$);
  Hypersonic ($Ma > 5$)
& 23, 24, 25, 26, 27 \\

\bottomrule
\end{tabularx}
\end{table}

\subsection{Implementation Details}
\label{sec:implementation}

To support reproducibility, this subsection specifies the concrete implementation underlying the agentic pipeline described above. The exact model identifiers used to generate the reported results are Claude Sonnet 5 for the flow-fingerprint, flow-comparison, and discover-physics agents, and Claude Haiku 4.5 for the review agent, consistent with the token and cost breakdown reported in Sec.~III.B. \action{Each pipeline stage shown in Fig.~\ref{fig:architecture} is implemented as an independently invocable ``skill'' (for example, a discover-physics skill for the Knowledge Fusion stage and a review-report skill for the closed-loop Review Agent). Each skill is launched non-interactively (\texttt{claude -p /\textless skill-name\textgreater}) with permission prompts disabled for unattended execution.} The closed-loop Review--Discovery cycle is bounded by a maximum of 5 re-generation attempts. If this limit is reached without a Pass verdict, the run is flagged for manual inspection rather than being silently accepted as a final output. Across the five reproducibility runs reported in Sec.~III.B, this limit was never reached, as every run achieved a Pass verdict (Table~\ref{tab:reproducibility}). Inter-agent communication uses structured text/JSON files written to a shared case directory, so that each stage's output (contours, statistics, fingerprints, comments file) is both machine-readable by the downstream stage and human-inspectable for auditing. To further demonstrate the reproducibility and ease of use of PhysMiner, the complete end-to-end workflow is provided in a demonstration video at \url{https://github.com/iDesign-Lab/PhysMiner}.

\section{Results and Discussion}

This section demonstrates the main capabilities of the proposed PhysMiner framework. Sec.~III.A first illustrates how the triple decomposition library enables cross-case comparison. Secs.~III.B--III.D then demonstrate the complete PhysMiner pipeline on three representative flow cases. Periodic hill flow (Sec.~III.B) is a literature-guided discovery case, in which the framework identifies new physical insights and leads to an improved subgrid-scale model, followed by posterior validation. Iced-airfoil flow (Sec.~III.C) and propeller wake (Sec.~III.D) instead use user-specified prompts to guide the discovery process. The iced-airfoil case employs a more specific and detailed prompt, whereas the propeller-wake case uses a simpler prompt. The successful discovery from the latter demonstrates that PhysMiner can remain effective even with a relatively simple user-specified prompt.

\subsection{Automated cross-case analysis using triple decomposition library}

This section introduces the triple decomposition library and its role within the PhysMiner framework. After the automated triple decomposition module completes the decomposition process, the statistical results, including domain-averaged statistics and wall-normal statistical profiles, are first transferred to the triple decomposition library. The library is designed as a continuously expanding database that is continuously enriched through the incorporation of validated flow cases. Subsequently, the processed statistical results are delivered to the Discover-Physics agent to support the identification of new physical insights for the target case.  In this section, the workflow of the triple decomposition library is demonstrated using the periodic hill case as an input case to the triple decomposition library. It should be noted that the PhysMiner framework has previously been applied to several canonical and engineering flow cases, including decaying isotropic turbulence (DIT), channel flow and backward-facing step flow. These cases have been reviewed according to the library criteria and stored in the database as historical knowledge. 

\subsubsection{Inputs to the triple decomposition library}

This section describes the inputs to the triple decomposition library, which correspond to the outputs of the data-driven track. The analysis focuses on the differences between the conventional Cauchy--Stokes decomposition and the triple decomposition for flow-component analysis. The two decompositions are compared in terms of their spatial distributions, wall-normal profiles, and component values at the no-slip wall in the periodic hill flow.

\begin{figure}[hbt!]
\centering
\begin{minipage}[b]{0.48\textwidth}
  \centering
  \includegraphics[width=\textwidth]{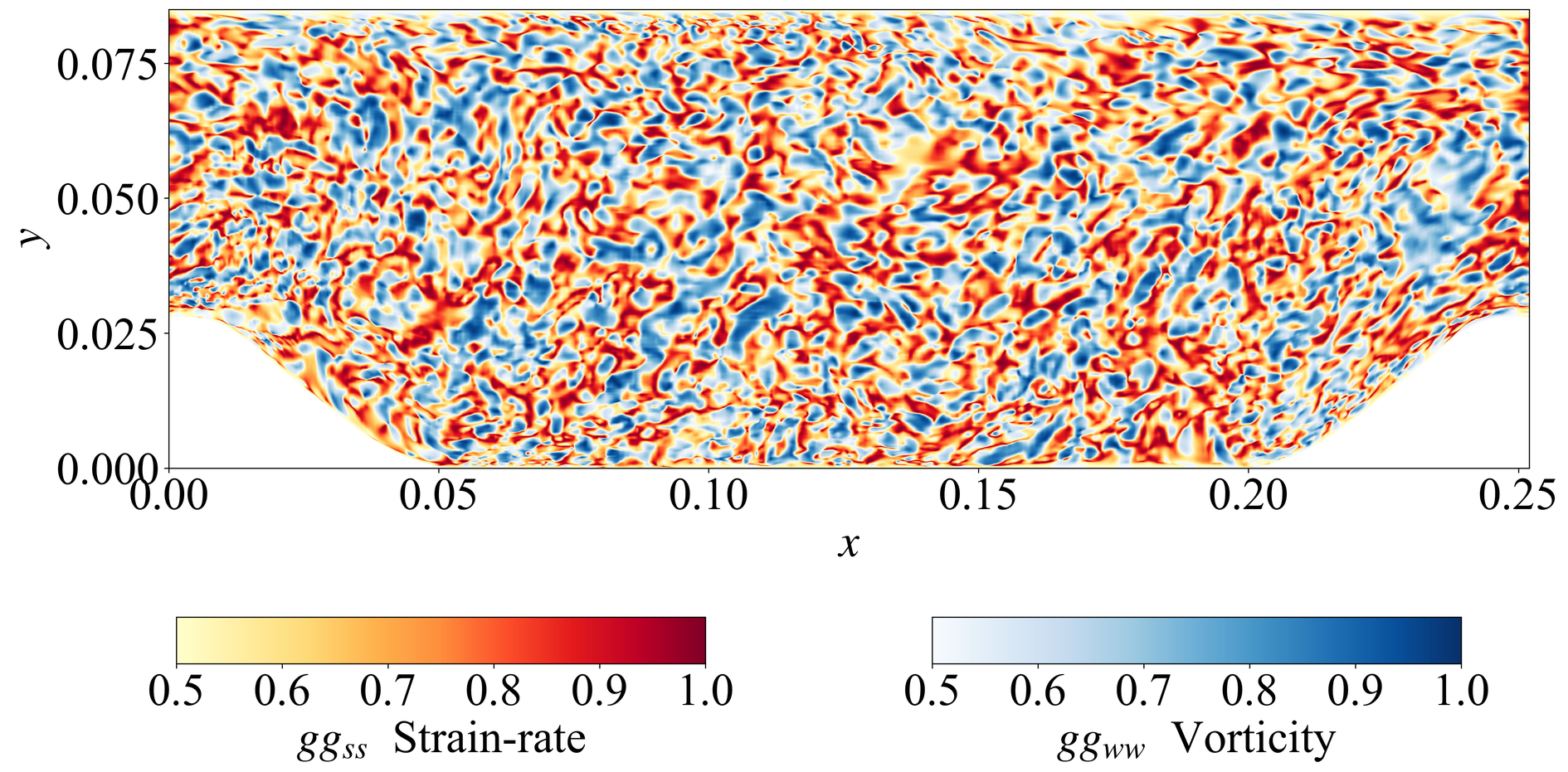}\\[2pt]
  a) Cauchy--Stokes decomposition
\end{minipage}
\hfill
\begin{minipage}[b]{0.48\textwidth}
  \centering
  \includegraphics[width=\textwidth]{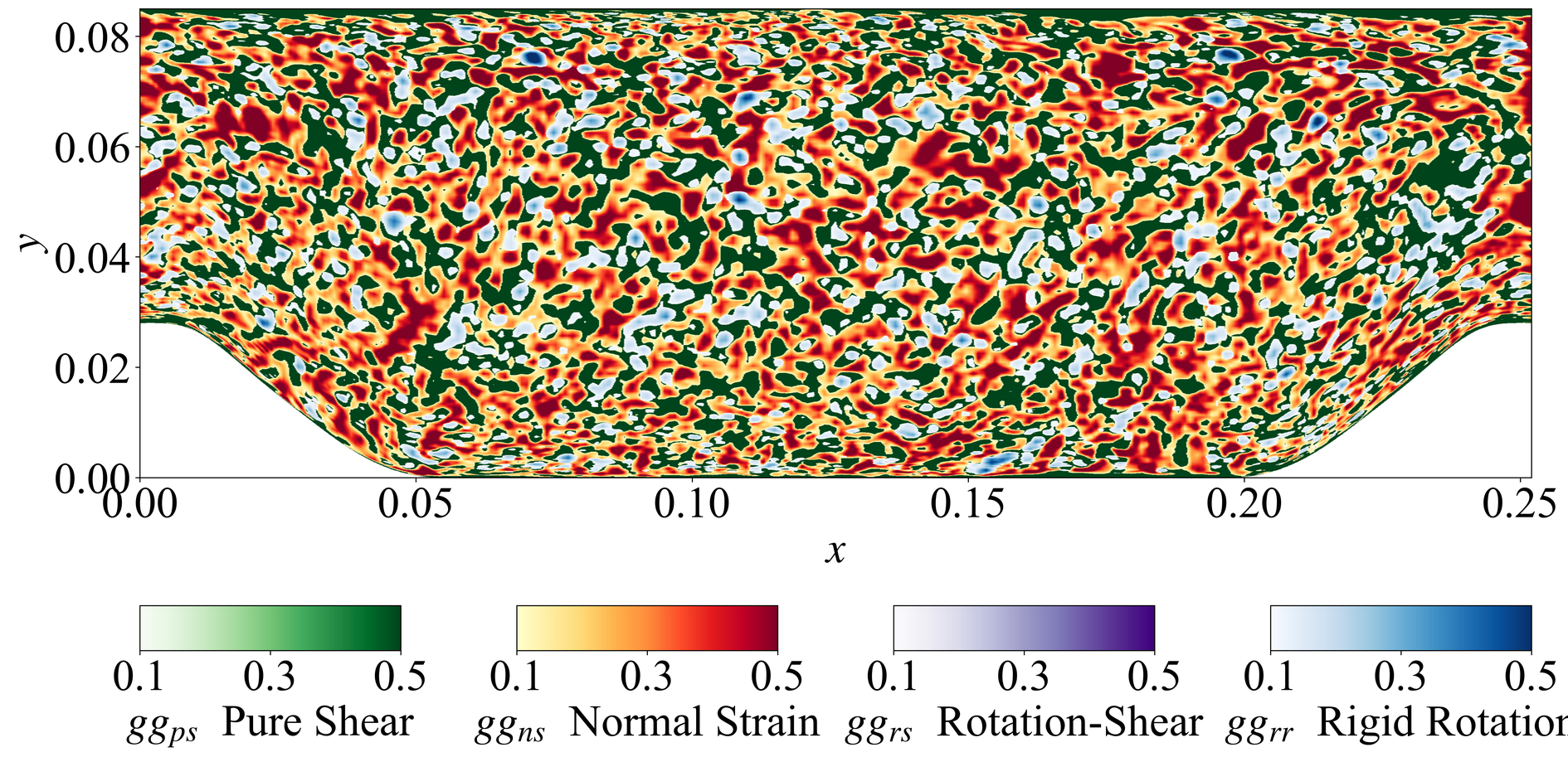}\\[2pt]
  b) Triple decomposition
\end{minipage}
\caption{Comparison between the Cauchy--Stokes decomposition and the triple decomposition at the mid-span section of the periodic hill.}
\label{fig:periodic_hill_CS_vs_TDM}
\end{figure}

Figure~\ref{fig:periodic_hill_CS_vs_TDM} compares the Cauchy--Stokes decomposition with the triple decomposition in the periodic hill flow. Compared with the Cauchy--Stokes decomposition, the triple decomposition separates the flow into more distinct and physically interpretable components.

\begin{figure}[hbt!]
\centering
\begin{minipage}[b]{0.6\textwidth}
  \centering
  \includegraphics[width=\textwidth]{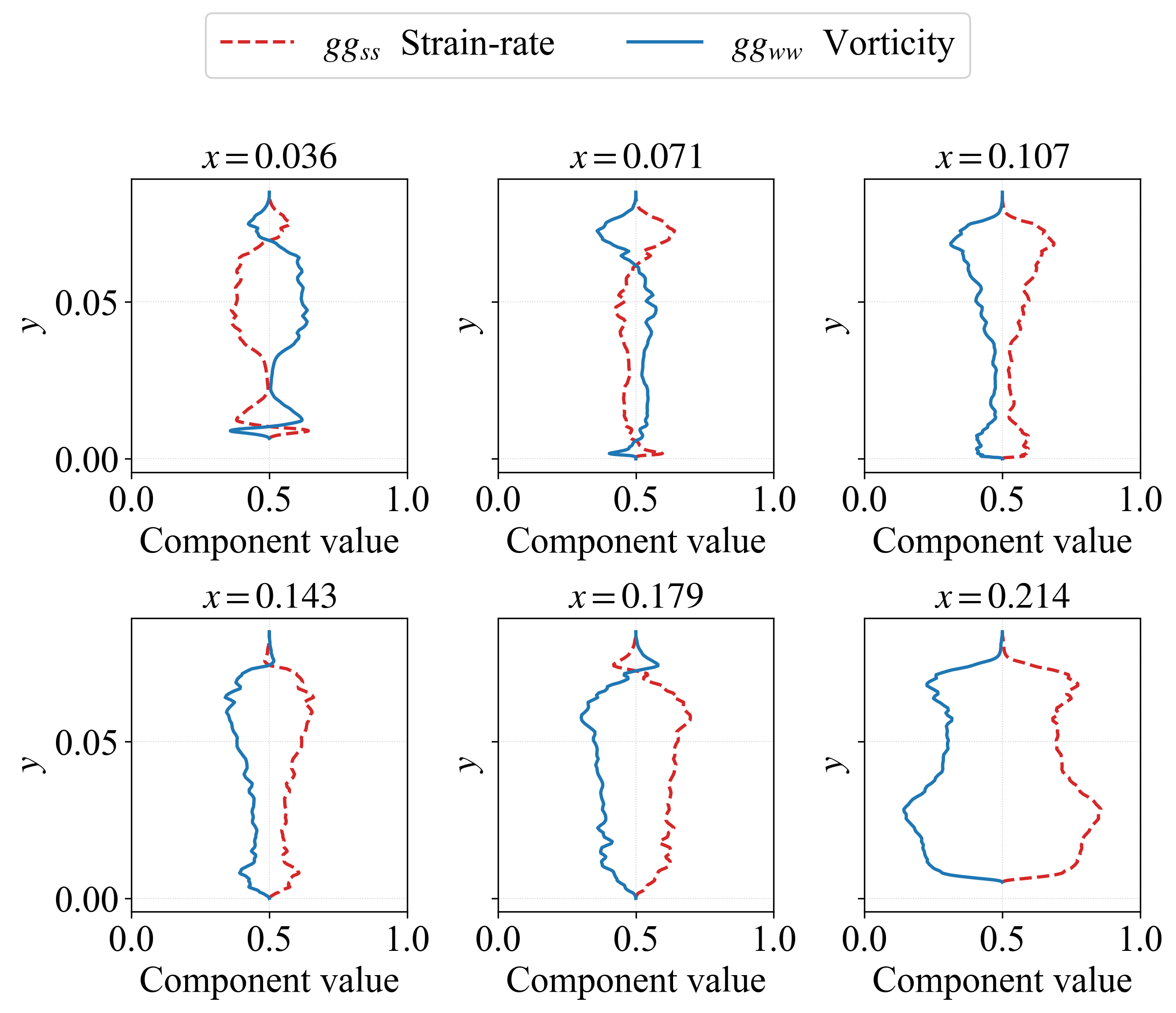}\\[2pt]
  a) Cauchy--Stokes decomposition
\end{minipage}
\hfill
\begin{minipage}[b]{0.6\textwidth}
  \centering
  \includegraphics[width=\textwidth]{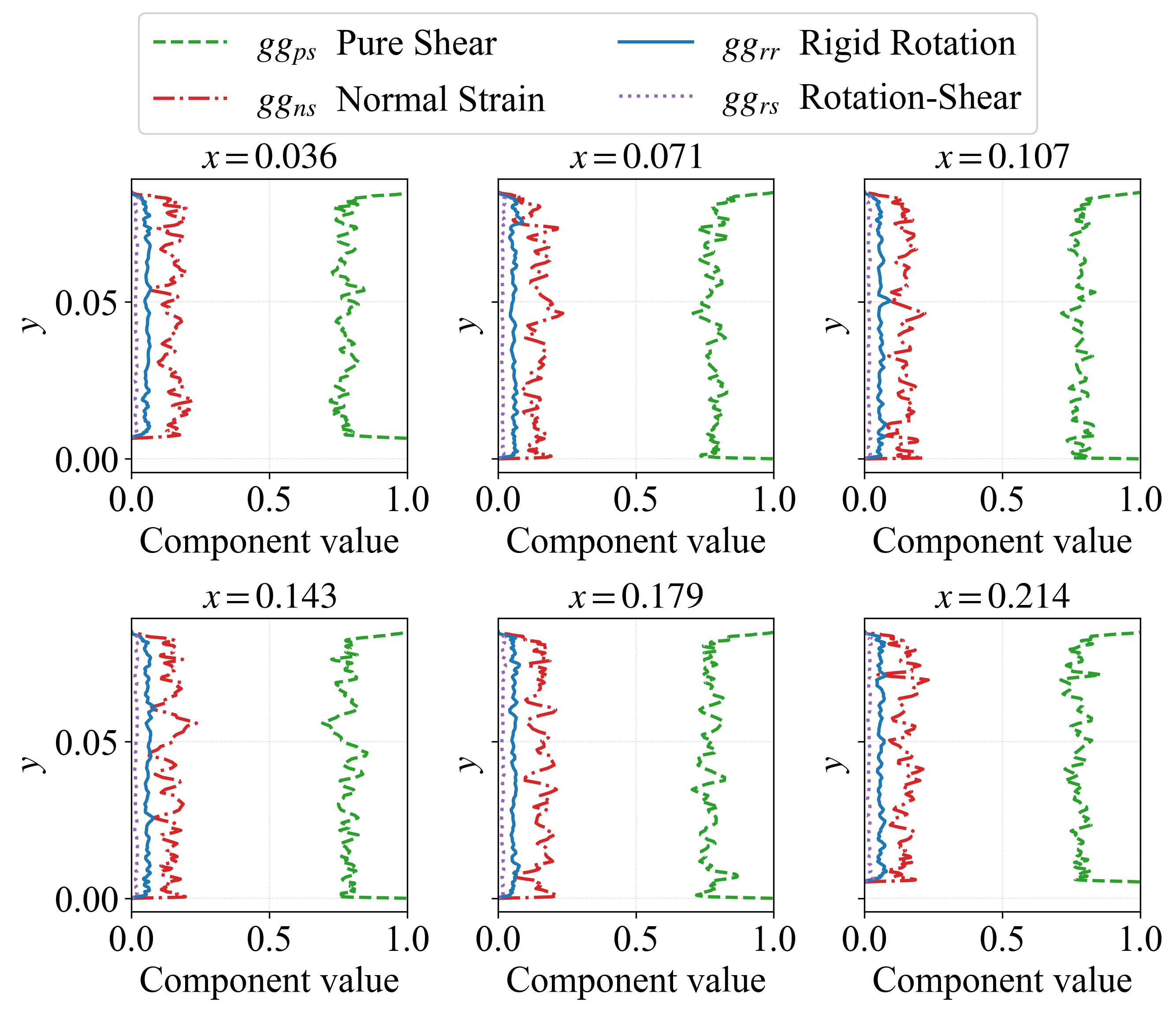}\\[2pt]
  b) Triple decomposition
\end{minipage}
\caption{Wall-normal distributions of the components obtained from the Cauchy--Stokes decomposition and the triple decomposition in periodic hill.}
\label{fig:periodic_hill_CS_vs_TDM_curves}
\end{figure}

Figure~\ref{fig:periodic_hill_CS_vs_TDM_curves} shows the wall-normal distributions of the components obtained from the Cauchy--Stokes decomposition and the
triple decomposition in periodic hill. In Figure~\ref{fig:periodic_hill_CS_vs_TDM_curves}(a), the strain-rate component and the vorticity component exhibit symmetric distributions throughout the computational domain, and both are equal to 0.5 at the no-slip wall. In Figure~\ref{fig:periodic_hill_CS_vs_TDM_curves}(b), the components exhibit distinctly different behaviors. The pure shear component dominates the other components throughout the computational domain. The normal strain, rigid rotation, and rotation-shear components decrease to zero at the wall, whereas the pure shear component approaches 1.0 in the near-wall region.

\begin{figure}[hbt!]
\centering
\includegraphics[width=1.0\textwidth]{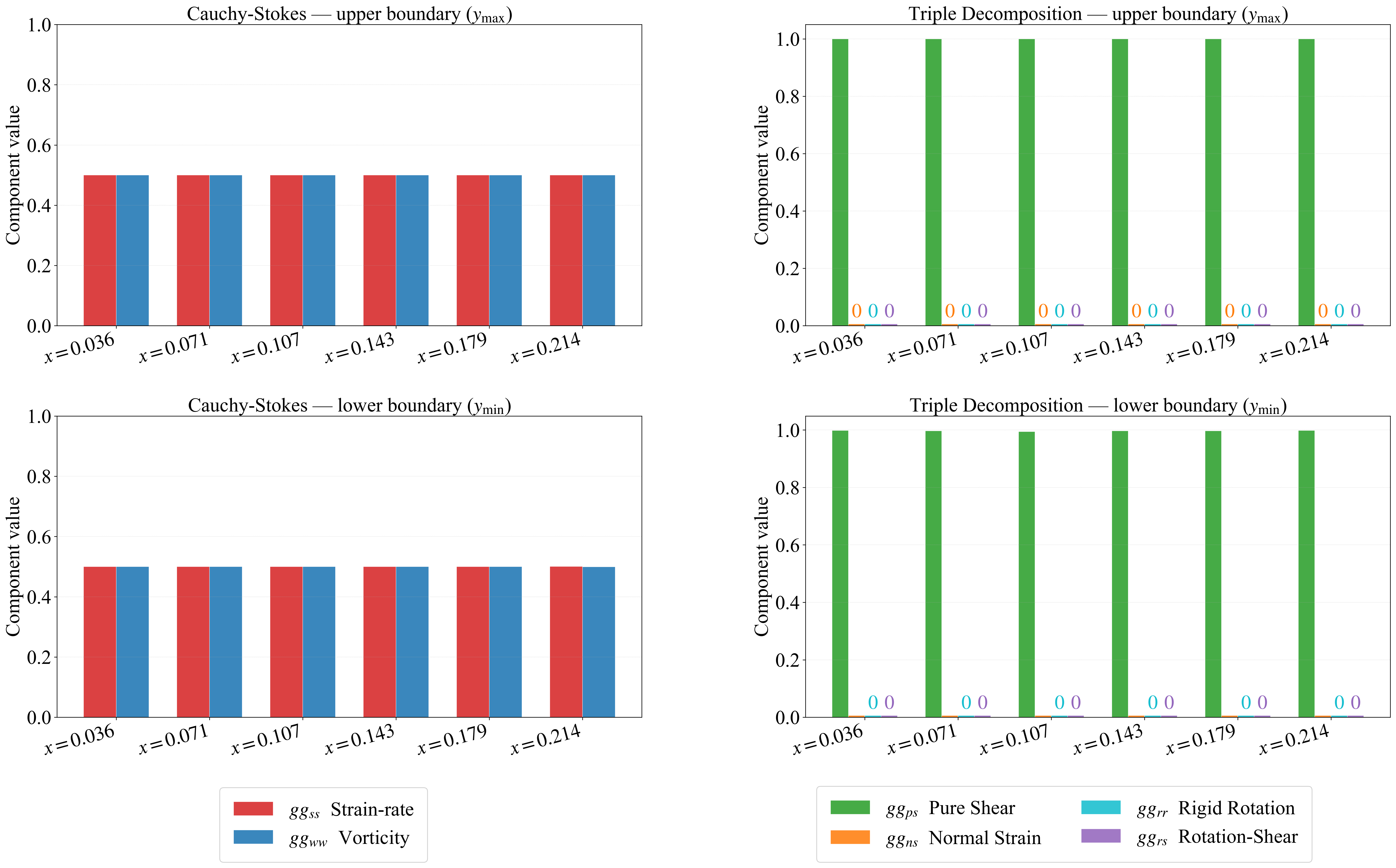}
\caption{Decomposition component values at no-slip walls in periodic hill flow: (a) Cauchy--Stokes decomposition and (b) triple decomposition.}
\label{fig:periodic_hill_wall_values}
\end{figure}

Figure~\ref{fig:periodic_hill_wall_values} further compares the distributions obtained from the Cauchy--Stokes decomposition and the triple decomposition on no-slip walls. In the periodic hill flow, both the upper and lower boundaries are set as no-slip walls. As shown in Figure~\ref{fig:periodic_hill_wall_values}(a), the strain-rate component and the vorticity component decomposed by the Cauchy--Stokes decomposition are both equal to 0.5. In Figure~\ref{fig:periodic_hill_wall_values}(b), however, only the pure shear component exists and its value is 1.0, while the other three components, namely the normal strain, rigid rotation, and rotation-shear components, are all zero. These results indicate that the Cauchy--Stokes decomposition cannot clearly separate normal strain and rotation from pure shear, leading to incorrect values at the wall. In contrast, the triple decomposition correctly isolates the pure shear component and therefore provides physically consistent wall values.

\subsubsection{Flow-fingerprint agent}

Upon receiving the new periodic hill case, the flow-fingerprint agent searches the triple-decomposition library for the most similar historical case. The triple-decomposition library evaluates the Jaccard tree distance to identify the closest matching case in the database, which is found to correspond to the backward-facing step flow. The corresponding scanning results are summarized as follows. According to Table 2, the target case, namely the periodic hill flow, is characterized by the unique fingerprint $[1,3,5,7,10,12,13,18,19,23]$. The historical cases have the following fingerprints: decaying isotropic turbulence $[1,4,5,7,10,11,13,18,19,23]$, channel flow $[1,3,5,7,10,11,13,18,19,23]$, and backward-facing step flow $[1,3,5,7,10,12,13,18,19,23]$. Based on the Jaccard tree distance defined in Equation~(5), the distance between the periodic hill flow and the backward-facing step flow is 

\begin{equation}
D_{\text{tree}}(\mathcal{P}_{\text{target}}, \mathcal{P}_{\text{hist}})
=
1-
\frac{
|\mathcal{P}_{\text{target}}
\cap
\mathcal{P}_{\text{hist}}|
}{
|\mathcal{P}_{\text{target}}
\cup
\mathcal{P}_{\text{hist}}|
}
=
1-\frac{10}{10}
=
0.00.
\end{equation}

This result indicates an identical fingerprint structure between the two cases. In contrast, the distances between the periodic hill flow and the decaying isotropic turbulence, and channel flow are calculated as 0.33, and 0.18, respectively. These results demonstrate that the backward-facing step flow exhibits the highest structural similarity to the periodic hill flow within the current library.

\subsubsection{Flow-comparison agent}

After the closest flow case is identified, a systematic comparison is conducted between the two flows, namely the periodic hill and backward-facing step. The comparison includes the spatial distributions, overall statistical characteristics, and wall-normal distributions of the triple-decomposition quantities. The flow-comparison agent automatically analyzes the differences in triple-decomposition characteristics between the two flows and generates a comparative summary.

\begin{figure}[hbt!]
\centering
\begin{minipage}[b]{0.48\textwidth}
  \centering
  \includegraphics[width=\textwidth]{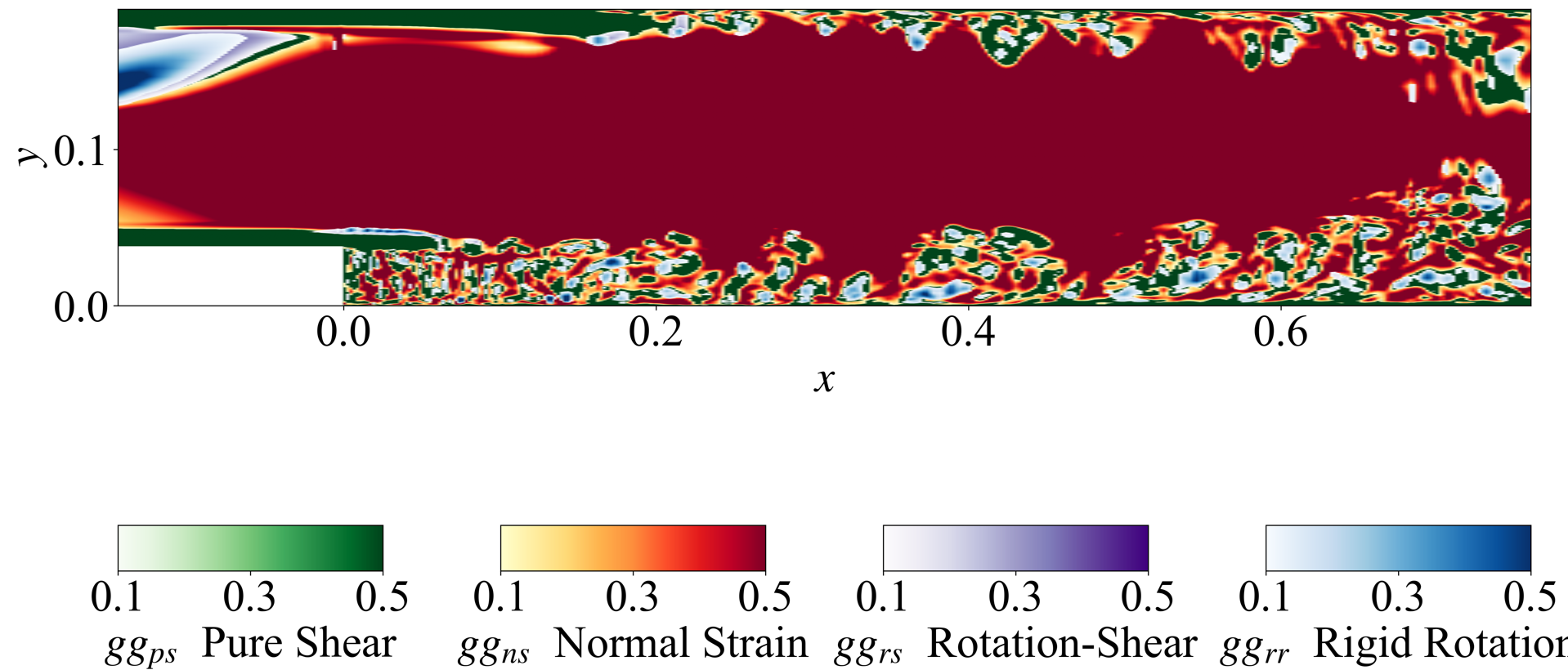}\\[2pt]
  a) Backward-facing step
\end{minipage}
\hfill
\begin{minipage}[b]{0.48\textwidth}
  \centering
  \includegraphics[width=\textwidth]{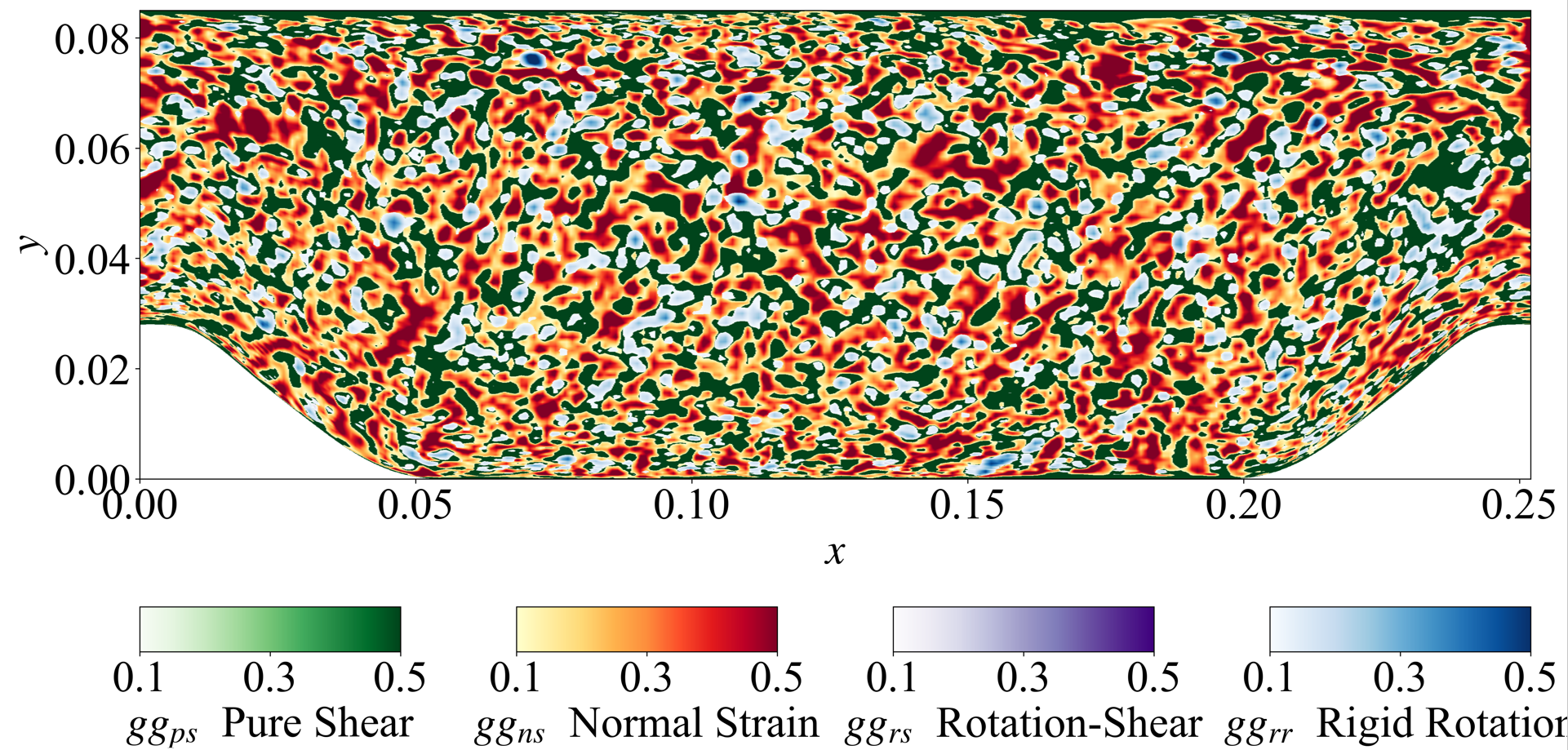}\\[2pt]
  b) Periodic hill
\end{minipage}
\caption{Comparison of triple-decomposition components between the backward-facing step and the periodic hill.}
\label{fig:bfs_vs_periodic_hill}
\end{figure}

Figure~\ref{fig:bfs_vs_periodic_hill} shows a comparison of triple-decomposition components between the backward-facing step and the periodic hill. In the backward-facing step case, the inflow boundary layer develops within the computational domain, causing the freestream region to be predominantly characterized by the normal straining component. In contrast, the periodic hill case has a fully developed turbulent inflow, resulting in a more uniform distribution of the different decomposition components throughout the freestream region.  

\begin{figure}[hbt!]
\centering
\begin{minipage}[b]{0.48\textwidth}
  \centering
  \includegraphics[width=\textwidth]{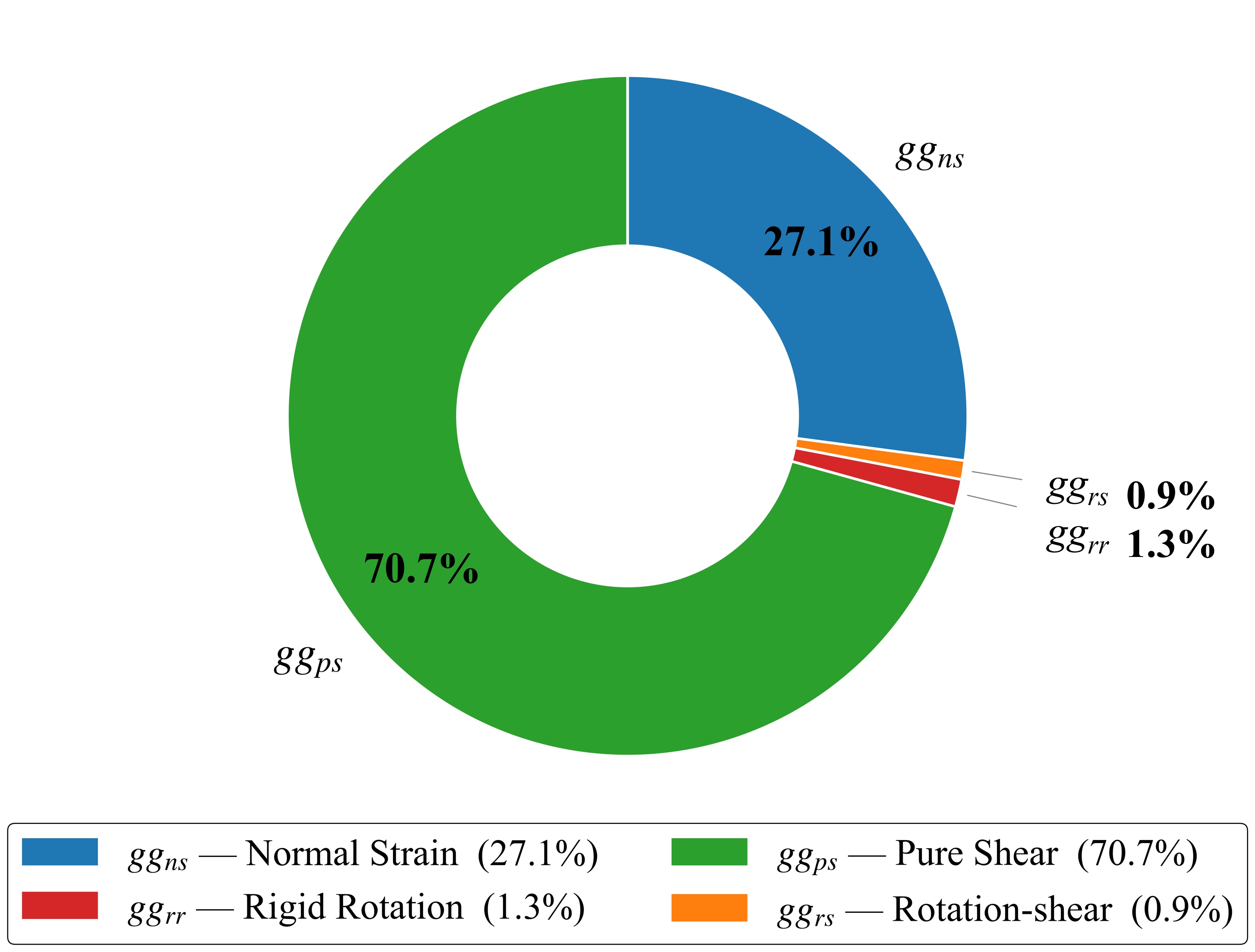}\\[2pt]
  a) Backward-facing step
\end{minipage}
\hfill
\begin{minipage}[b]{0.48\textwidth}
  \centering
  \includegraphics[width=\textwidth]{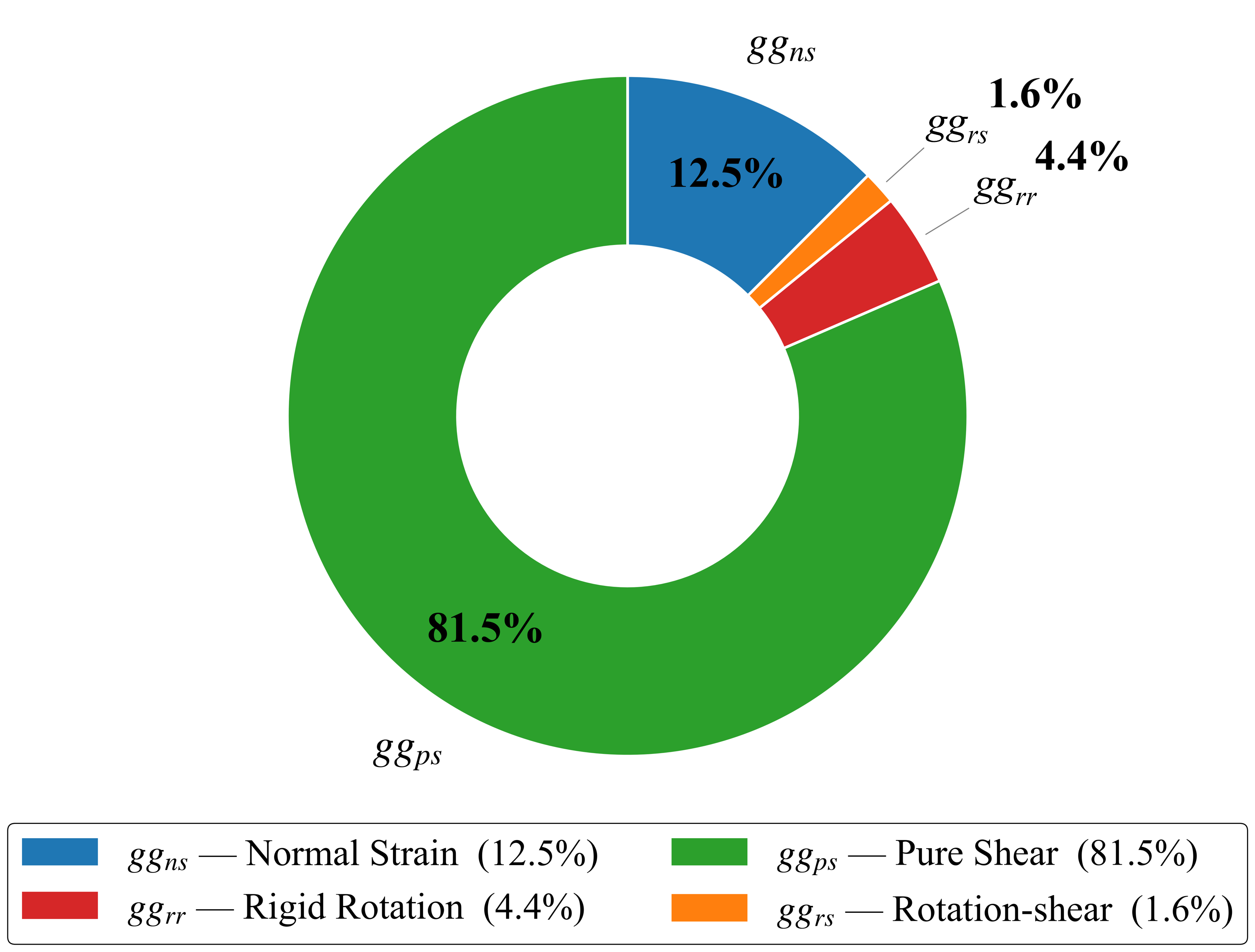}\\[2pt]
  b) Periodic hill
\end{minipage}
\caption{Comparison of the statistical triple decomposition between the backward-facing step and the periodic hill.}
\label{fig:bfs_vs_periodic_hill_statistics}
\end{figure}

Figure~\ref{fig:bfs_vs_periodic_hill_statistics} shows the comparison of the statistical triple decomposition between the backward-facing step and the periodic hill. The backward-facing step yields $\mathrm{gg}_{ps}= 70.7\%$, $\mathrm{gg}_{ns}= 27.1\%$, $\mathrm{gg}_{rr}= 1.3\%$ , and $\mathrm{gg}_{rs}= 0.9\%$ , reflecting a mixed shear-and-strain character. The substantial $\mathrm{gg}_{ns}$ contribution arises from the developing-boundary-layer freestream, which inherently carries a normal-strain background across the entire domain. The periodic hill, by contrast, shows a markedly higher $\mathrm{gg}_{ps}= 81.5\%$ and a substantially reduced $\mathrm{gg}_{ns}= 12.5\%$, consistent with its fully developed inflow that eliminates any freestream normal-strain background.

\begin{figure}[hbt!]
\centering
\begin{minipage}[b]{0.6\textwidth}
  \centering
  \includegraphics[width=\textwidth]{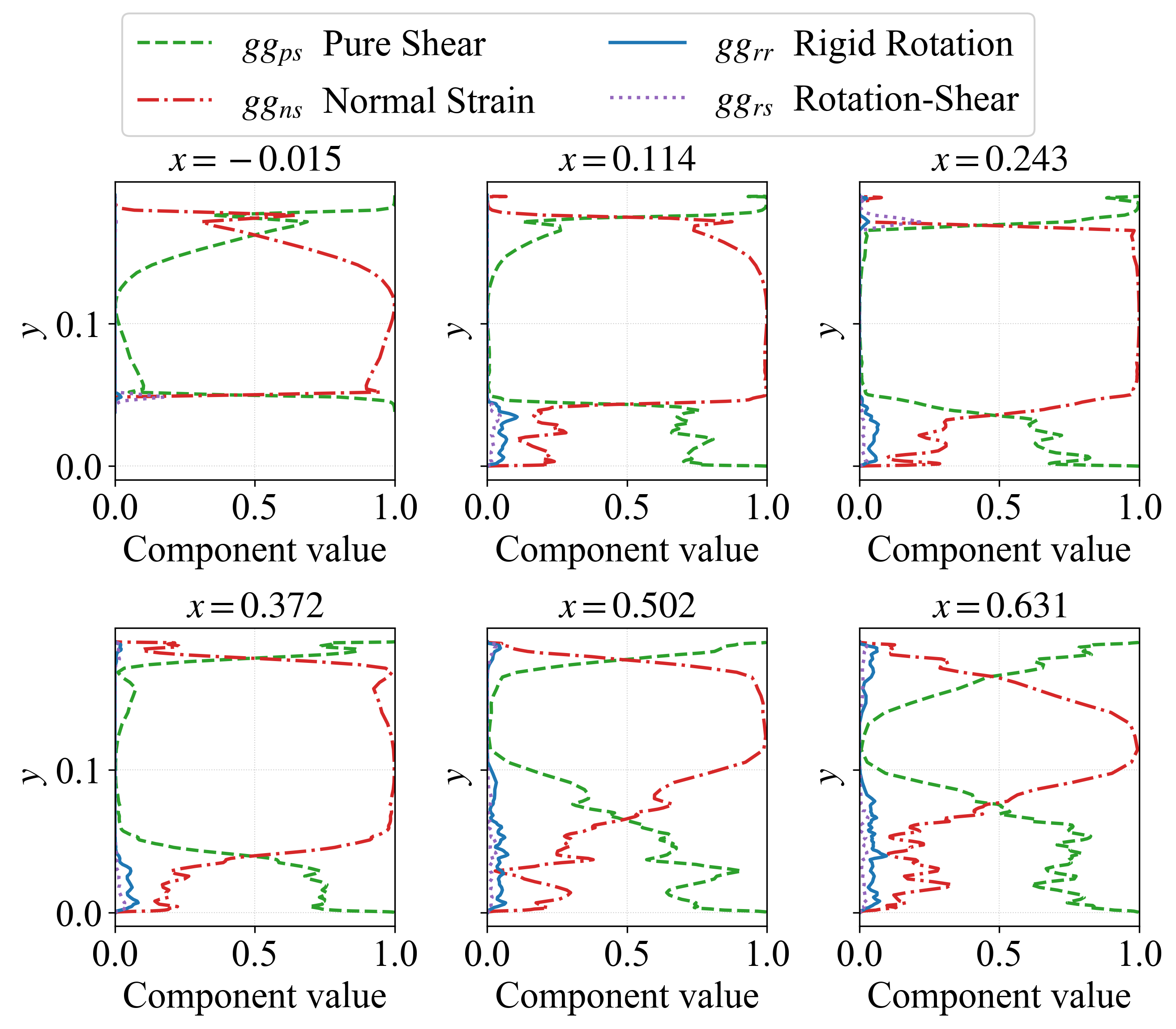}\\[2pt]
  a) Backward-facing step
\end{minipage}
\hfill
\begin{minipage}[b]{0.6\textwidth}
  \centering
  \includegraphics[width=\textwidth]{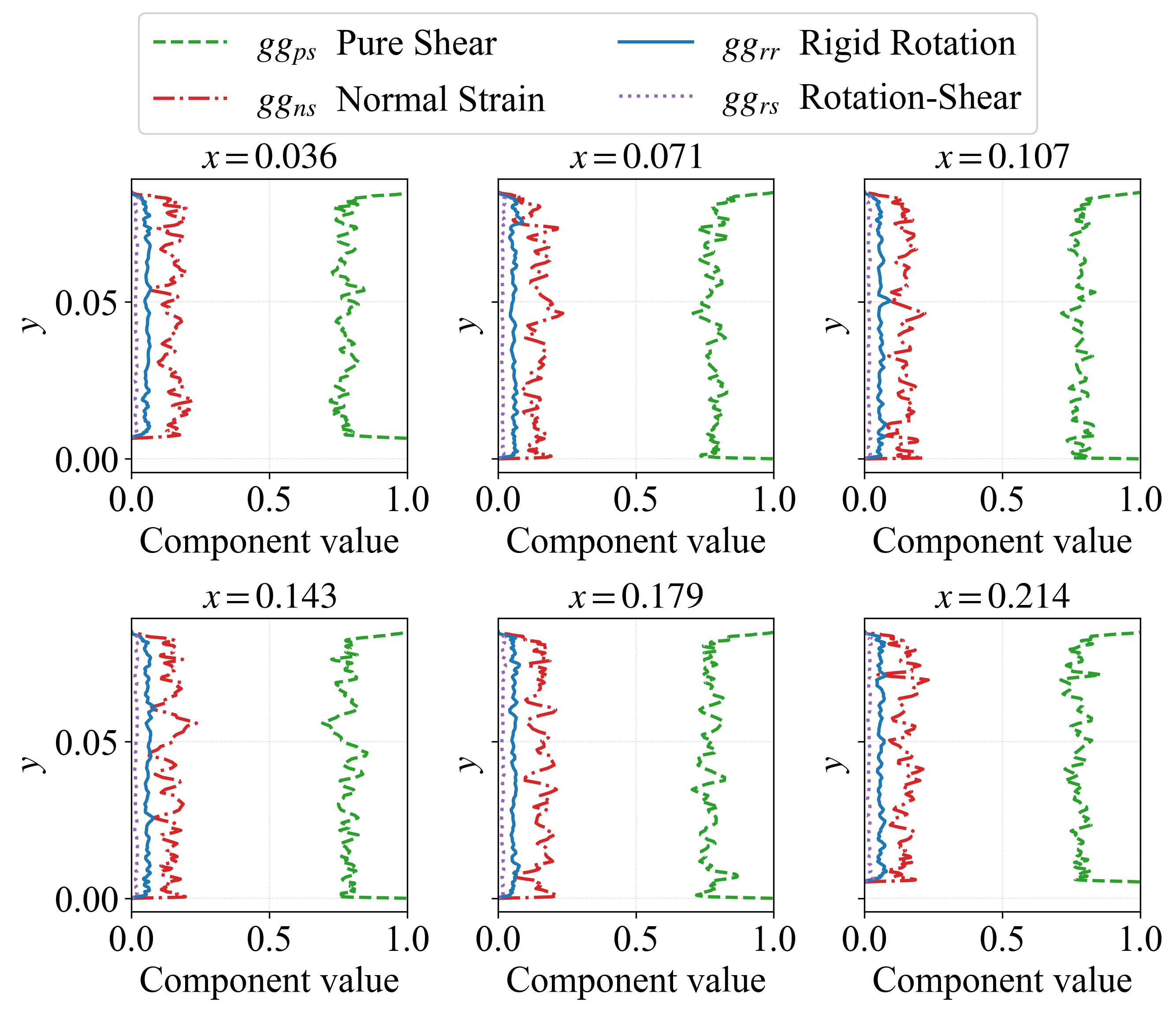}\\[2pt]
  b) Periodic hill
\end{minipage}
\caption{Wall-normal distributions of the triple-decomposition components in the backward-facing step and the periodic hill.}
\label{fig:bfs_vs_periodic_hill_wallnormal}
\end{figure}

Figure~\ref{fig:bfs_vs_periodic_hill_wallnormal} compares the wall-normal distributions of the triple-decomposition 
components in the backward-facing step flow and the periodic hill flow. In the 
backward-facing step, the profiles vary dramatically between stations. This strong streamwise 
non-equilibrium evolution is driven by the superposition of two mechanisms: 
the abrupt geometric expansion causing a sudden state jump and the 
developing-boundary-layer freestream continuously supplying a $\mathrm{gg}_{ns}$ 
background across all stations. In the periodic hill, by 
contrast, the profiles remain remarkably stable across all streamwise stations 
($x = 0.036$--$0.214$), with $\mathrm{gg}_{ps}$ consistently dominating the full 
wall-normal height at every location. This spatial uniformity reflects three 
mutually reinforcing mechanisms: the fully developed inflow eliminating any 
freestream $\mathrm{gg}_{ns}$ background, the smooth hill curvature distributing 
separation and reattachment disturbances over a long streamwise extent, and 
the periodic boundary conditions constraining all stations toward the same 
quasi-equilibrium state.

Subsequently, the triple decomposition library performs cross-comparison against the backward-facing step case by integrating historical knowledge with the reasoning capability of the LLM. Table~\ref{tab:triple_decomp_comparison}, automatically generated by the PhysMiner AI agent, presents the cross-comparison from four perspectives, including geometry and flow configuration, triple decomposition global statistics, spatial distributions, and turbulence characteristics. The following subsection provides a detailed interpretation of the comparisons presented in Table~\ref{tab:triple_decomp_comparison}, together with related results from both flow cases to illustrate the physical similarities and differences identified by the framework.

\begin{table}[hbt!]
\caption{Automatically generated comparison of triple-decomposition characteristics between the  periodic hill and the backward-facing step}
\label{tab:triple_decomp_comparison}
\centering
\begin{tabular}{lll}
\toprule
Dimension & Backward-Facing Step & Periodic Hill \\
\midrule

\multicolumn{3}{l}{\textit{Geometry \& flow configuration}} \\[4pt]

Geometry &
\begin{minipage}[t]{0.35\textwidth}\raggedright
Sudden expansion at a sharp step edge; single geometric discontinuity
\end{minipage} &
\begin{minipage}[t]{0.35\textwidth}\raggedright
Continuous curved wall with smoothly varying curvature; periodically repeating hills
\end{minipage} \\[14pt]

Separation type &
\begin{minipage}[t]{0.35\textwidth}\raggedright
Abrupt, geometry-forced separation at step corner
\end{minipage} &
\begin{minipage}[t]{0.35\textwidth}\raggedright
Smooth, curvature-driven separation on hill leeward side
\end{minipage} \\[14pt]

\midrule
\multicolumn{3}{l}{\textit{\action{Triple decomposition: global statistics}}} \\[4pt]

$\mathrm{gg}_{ps}$ (pure shear) &
$70.7\%$ &
$\mathbf{81.5\%}$ (higher) \\[6pt]

$\mathrm{gg}_{ns}$ (normal strain) &
$\mathbf{27.1\%}$ (higher) &
$12.5\%$ (lower) \\[6pt]

$\mathrm{gg}_{rr}$ (rigid rotation) &
$1.3\%$ &
$\mathbf{4.4\%}$ (higher) \\[6pt]

$\mathrm{gg}_{rs}$ (rotation--shear) &
$0.9\%$ &
$1.6\%$ (slightly higher) \\[6pt]

\midrule
\multicolumn{3}{l}{\textit{Spatial distribution}} \\[4pt]

Wall-normal profiles &
\begin{minipage}[t]{0.35\textwidth}\raggedright
Strong cross-sectional variation between stations; highly non-equilibrium evolution downstream
\end{minipage} &
\begin{minipage}[t]{0.35\textwidth}\raggedright
$\mathrm{gg}_{ps}$ dominates across the full wall-normal height at every streamwise station; profiles are nearly identical between stations and can be characterised by a single representative profile
\end{minipage} \\[18pt]

Spatial homogeneity &
\begin{minipage}[t]{0.35\textwidth}\raggedright
\action{Lower: large spatial gradients, strongly influenced by the separation point}
\end{minipage} &
\begin{minipage}[t]{0.35\textwidth}\raggedright
\action{Higher: periodic geometry promotes a more spatially uniform decomposition structure}
\end{minipage} \\[14pt]

Cause of streamwise evolution &
\begin{minipage}[t]{0.35\textwidth}\raggedright
Two mechanisms superpose: (1) abrupt geometric expansion causes a sudden jump from attached to separated state; (2) the developing-boundary-layer freestream continuously supplies a $\mathrm{gg}_{ns}$ background across all stations
\end{minipage} &
\begin{minipage}[t]{0.35\textwidth}\raggedright
Three mechanisms reinforce each other: (1) fully developed inflow eliminates any freestream $\mathrm{gg}_{ns}$ background; (2) the smooth hill curvature distributes separation and reattachment disturbances over a long streamwise extent; (3) periodic boundary conditions constrain all stations toward the same quasi-equilibrium state
\end{minipage} \\[26pt]

\midrule
\multicolumn{3}{l}{\textit{Turbulence character}} \\[4pt]

Flow state &
\begin{minipage}[t]{0.35\textwidth}\raggedright
Strongly non-equilibrium; dominated by single, intense shear layer downstream of step
\end{minipage} &
\begin{minipage}[t]{0.35\textwidth}\raggedright
Quasi-periodic equilibrium; turbulence is sustained and self-repeating over each hill cycle
\end{minipage} \\[18pt]

Overall character &
Shear + strain mixed &
Pure shear dominant \\[6pt]

\bottomrule
\end{tabular}
\end{table}

\subsection{Application of PhysMiner to case 1: periodic hill}

This section employs the periodic hill case to demonstrate the overall workflow of PhysMiner, including the literature-driven track, data-driven track, knowledge fusion, and closed-loop calibration framework. Before this case, the PhysMiner framework has been applied to several flow cases, including decaying isotropic turbulence, channel flow, backward-facing step flow, and propeller flow.

\subsubsection{Human-intervention-free and objective prompt generation from literature}

Unlike conventional LLM-based approaches that rely on user-defined prompts, the proposed framework automatically generates prompts based on the user-provided reference. This strategy reduces the difficulty of prompt selection and minimizes subjective bias in prompt design. PhysMiner can automatically identify the key physical problems embedded in a given flow by capturing the research focus through word cloud analysis via the literature-driven track. This process is objective and effectively avoids biases introduced by subjective judgment. The word clouds analysis is divided into three subsets: (a) highly cited literature, which reflects long-standing core themes; (b) recent literature (2026), indicating emerging research trends; and (c) an overall synthesis of the entire dataset. 

\begin{figure}[hbt!]
\centering
\begin{minipage}[b]{0.31\textwidth}
  \centering
  \includegraphics[width=\textwidth]{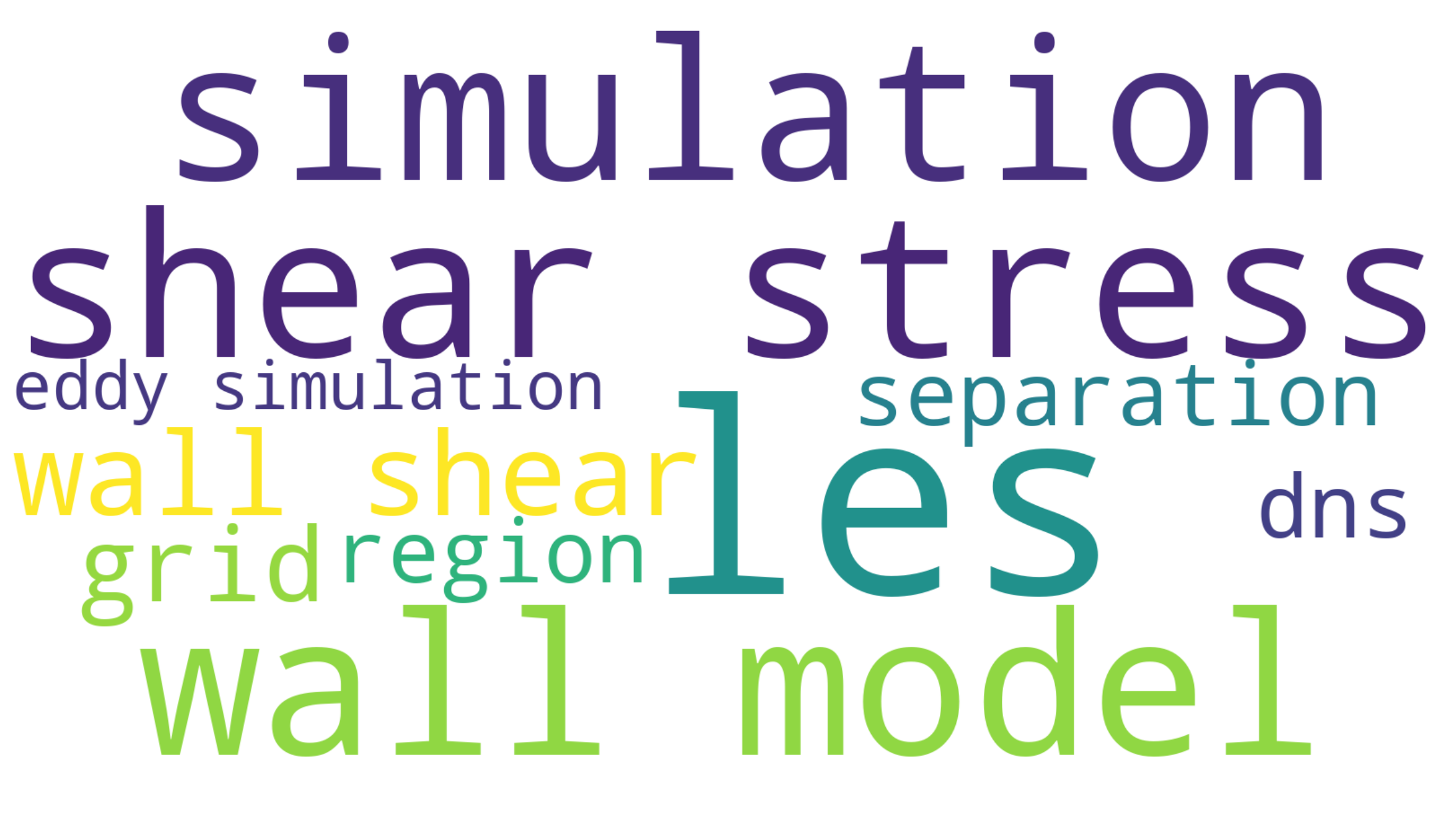}\\[2pt]
  a) Highly cited literature
\end{minipage}
\hfill
\begin{minipage}[b]{0.31\textwidth}
  \centering
  \includegraphics[width=\textwidth]{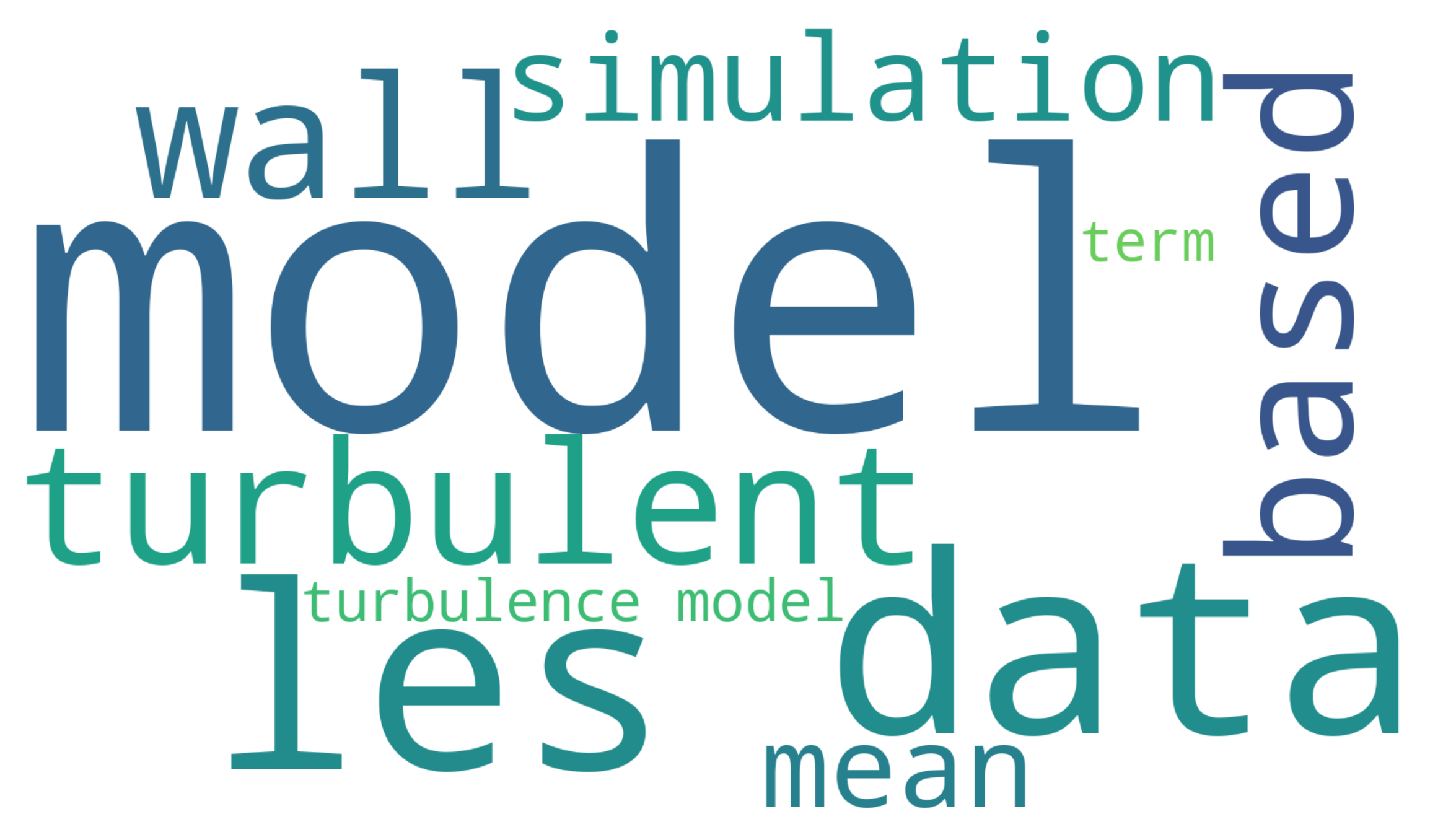}\\[2pt]
  b) Recent literature
\end{minipage}
\hfill
\begin{minipage}[b]{0.31\textwidth}
  \centering
  \includegraphics[width=\textwidth]{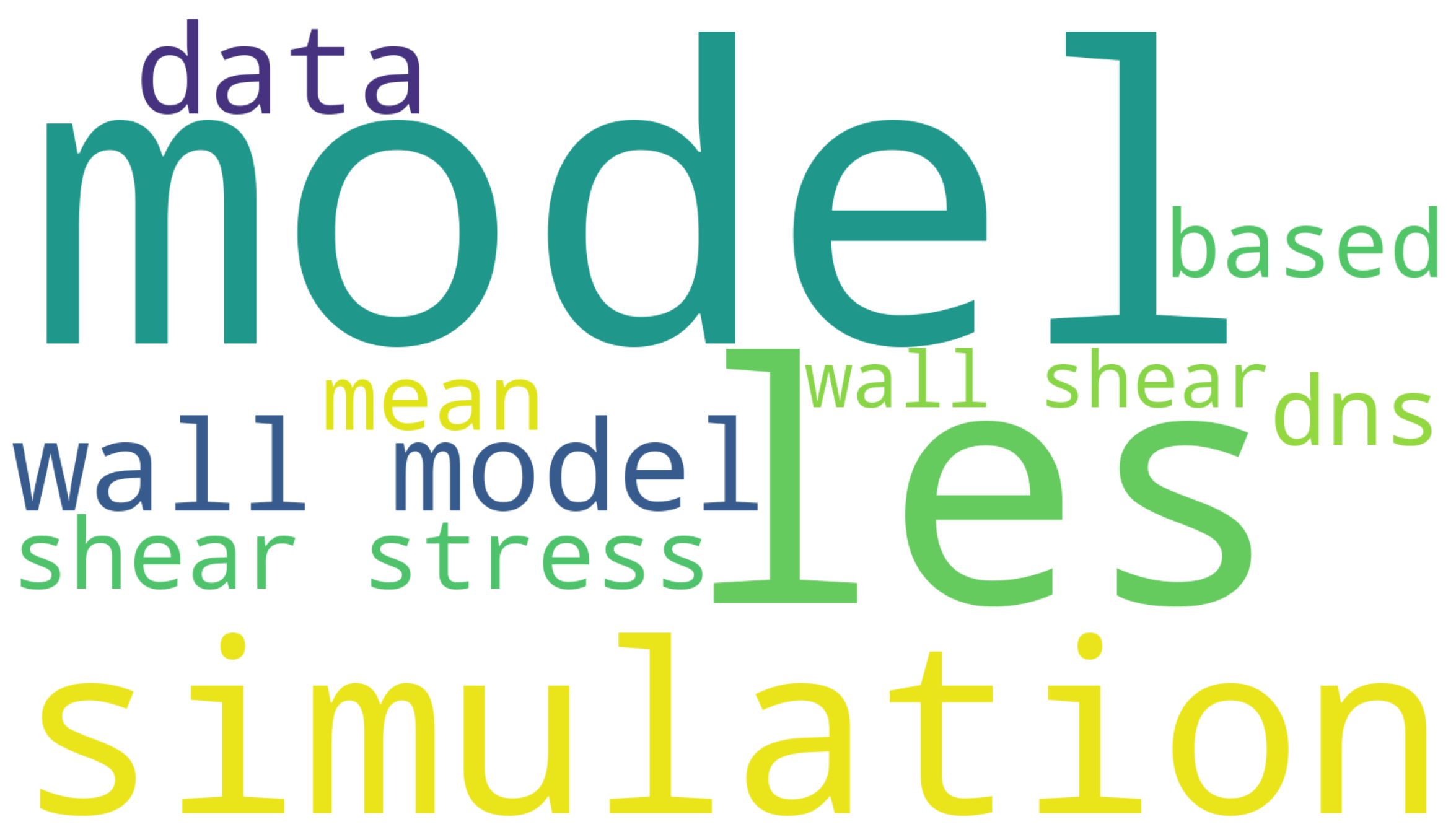}\\[2pt]
  c) Overall synthesis
\end{minipage}
\caption{Word cloud for the periodic hill literature: (a) Highly cited literature, (b) Recent literature (2026), and (c) Overall synthesis.}
\label{fig:wordcloud_PH}
\end{figure}

\begin{figure}[hbt!]
\centering
\includegraphics[width=1.0\textwidth]{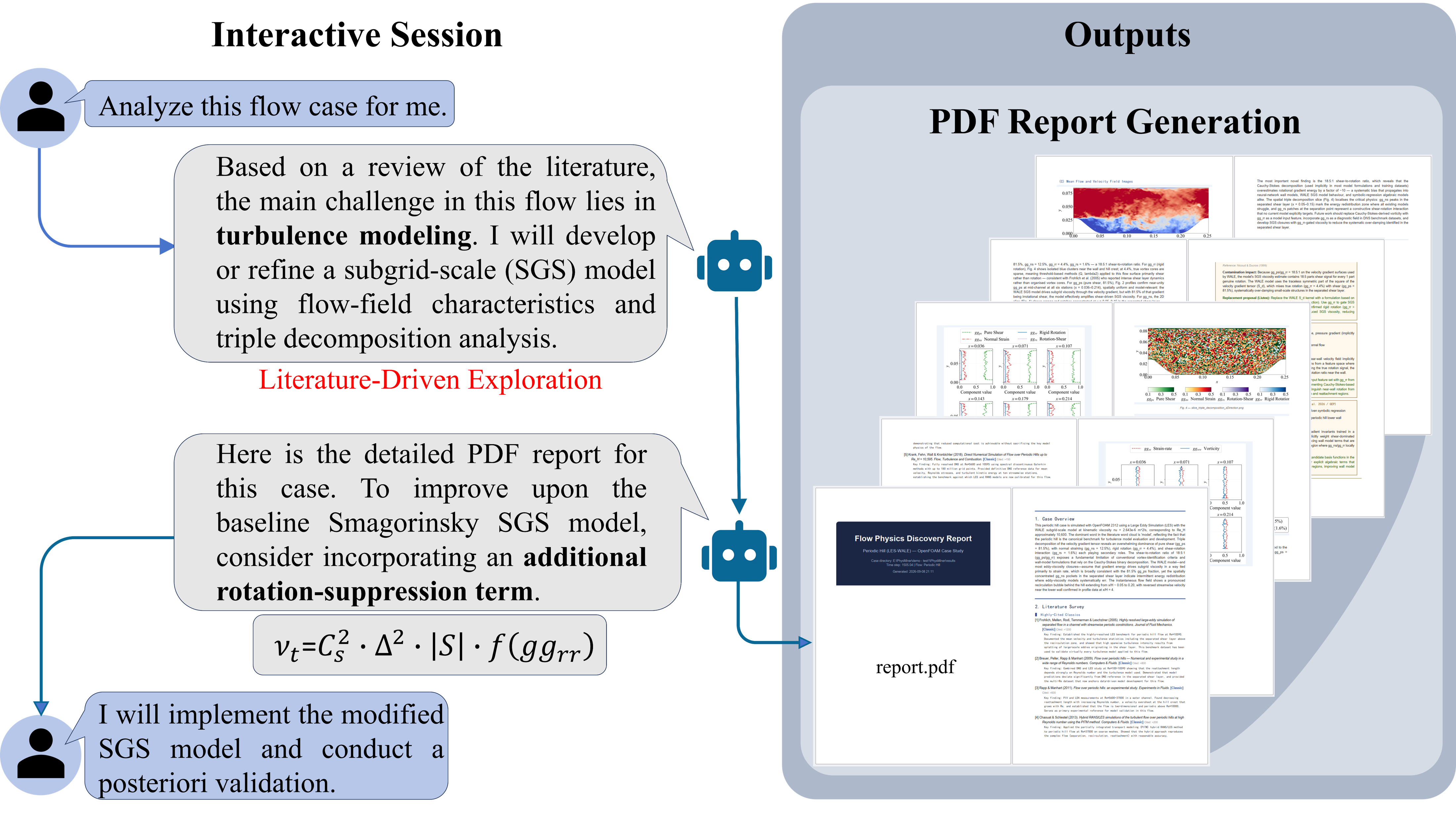}
\caption{Literature-driven exploration and the final PDF report generated by PhysMiner.}
\label{fig:prompt}
\end{figure}

 To construct the literature dataset, "periodic hill” is searched in the Web of Science
using two ranking strategies: "Citations: highest first" and "Date: newest first". The retrieved papers are then downloaded
through the university library access. Figure~\ref{fig:wordcloud_PH} presents the resulting word clouds generated from the periodic hill literature, categorized by citation count and publication recency. In subset (a), the term “LES” appears prominently, whereas in subset (b) and (c), the term “model” exhibits the highest frequency. This phenomenon mainly arises because periodic hill flow involves large-scale flow separation, which remains challenging for Reynolds-averaged Navier–Stokes (RANS) methods. Consequently, large eddy simulation (LES) has become the most widely adopted approach for this flow configuration. The dominance of the term “model” in recent literature indicates that turbulence modeling continues to be a central research focus. Figure~\ref{fig:prompt} presents the literature-driven exploration and the final PDF report generated by PhysMiner. The literature guides the physics-discovery agent to explore an improved SGS model for more accurate prediction of large-scale separated flows over a periodic hill. The final PDF report summarizes the analysis of each automatically generated figure and proposes an improved SGS model. The modified SGS model is presented and subsequently validated in the following sections.

\subsubsection{PhysMiner turbulence-model proposal}

Knowledge from the literature-driven and data-driven tracks is integrated within the physics-discovery agent for knowledge fusion. The resulting hypotheses are then evaluated through a closed-loop calibration process to identify physically consistent modeling strategies. For the periodic hill case, turbulence modeling is the primary challenge. The highest-ranked SGS-model proposal is therefore selected for a posteriori implementation and validation. The proposed SGS model is based on the Smagorinsky model~\cite{smagorinsky1963general} with an additional rotation-suppression term. The modified model is formulated as

\begin{equation}
\nu_t = C_s^2 \cdot \Delta^2 \cdot |S| \cdot f(\mathrm{gg}_{rr}),
\label{eq:sgs_proposal1}
\end{equation}

where $C_s=0.1$ is the model coefficient, consistent with the value proposed by Deardorff~\cite{deardorff1970numerical} for wall-bounded turbulent shear flows. $\Delta$ is the maximum grid-cell width among the three coordinate directions, and $|S| = \sqrt{2 \bar{S}_{ij} \bar{S}_{ij}}$ is the magnitude of the resolved strain-rate tensor, with $\bar{S}_{ij}$ denoting the resolved strain-rate tensor. The suppression function $f(\mathrm{gg}_{rr})$ is defined as

\begin{equation}
f(\mathrm{gg}_{rr}) = 1 - \mathrm{gg}_{rr}.
\label{eq:f_ggrr}
\end{equation}

Here, $\mathrm{gg}_{rr}$ serves as a rotation-suppression factor. Local peaks of $\mathrm{gg}_{rr}$ occur within vortex cores, where conventional SGS models can produce excessive eddy viscosity. The proposed suppression function therefore reduces $\nu_t$ in strongly rotating regions while preserving the standard Smagorinsky form elsewhere. Suppressing eddy viscosity within vortex cores is a physically intuitive modeling strategy that could also be proposed independently by an experienced turbulence expert. The novelty claimed here is therefore not the suppression concept itself. Instead, the novelty of PhysMiner lies in its ability to identify the dominant modeling deficiency for a specific flow and derive a targeted correction from the triple-decomposition components. This approach provides a systematic and physically interpretable route from flow physics to turbulence-model modification.

\subsubsection{Posteriori implementation and validation of the proposed model modification}

\begin{figure}[hbt!]
\centering
\includegraphics[width=0.48\textwidth]{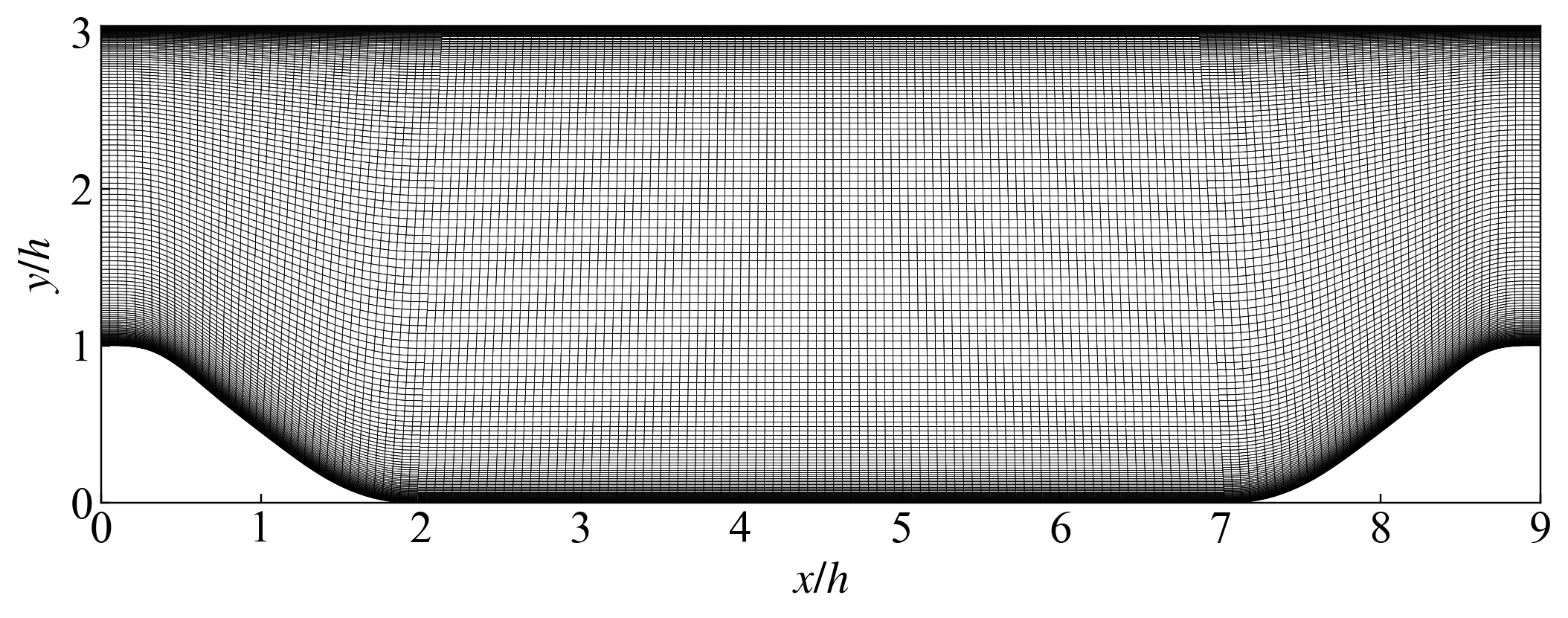}
\caption{Computational domain and medium grid for the periodic hill.}
\label{fig:periodic_hill_mesh}
\end{figure}

This section implements and evaluates the proposed modification to the standard Smagorinsky model~\cite{smagorinsky1963general}. The standard Smagorinsky model with a constant model coefficient of $C_s=0.1$ and the modified model, referred to as the Smagorinsky LLM model, are implemented in OpenFOAM 2312. The computational domain contains two identical hills with a height of $h=0.028 m$, separated by a streamwise distance of $9h$. The domain height is $3.035h$, and the spanwise extent is $4.5h$. The upper boundary is a flat wall, while the lower boundary is a curved channel wall. Periodic boundary conditions are imposed in the streamwise and spanwise directions. No-slip conditions are applied at both the upper and lower walls. The flow is driven from left to right by a uniform body force. The Reynolds number based on the hill height $h$ and the bulk velocity $U_b$ above the hill crest is $Re_h=10,595$. Three structured grid sets are employed to assess grid convergence, with their details summarized in Table~\ref{tab:grid_sets}. The time step is selected to maintain a maximum local CFL number of 0.3. Each simulation is first advanced for 15 flow-through periods, denoted by $15\Delta T$, where one flow-through period is defined as $\Delta T=9h/U_b$. Mean flow quantities are then sampled over an additional $10\Delta T$ and averaged in the spanwise direction. Figure~\ref{fig:periodic_hill_mesh} shows the computational domain and medium grid for the periodic hill case. 

\begin{table}[hbt!]
\caption{Three grid set with different grid resolutions.}
\label{tab:grid_sets}
\centering
\begin{tabular}{l c r r} 
\toprule
\textbf{Grid set} & \textbf{Grid dimension} & \textbf{Total number} & \textbf{$y^+$} \\
\midrule
Coarse grid & $119\times 100\times107$ & 1.3 M & 1.12 \\[4pt]
Medium grid & $178\times150\times160$ & 4.3 M & 0.78 \\[4pt]
Fine grid & $267\times225\times240$ & 14.4 M & 0.41 \\
\bottomrule
\end{tabular}
\end{table}

\begin{figure}[hbt!]
\centering
\begin{minipage}[b]{0.48\textwidth}
  \centering
  \includegraphics[width=\textwidth]{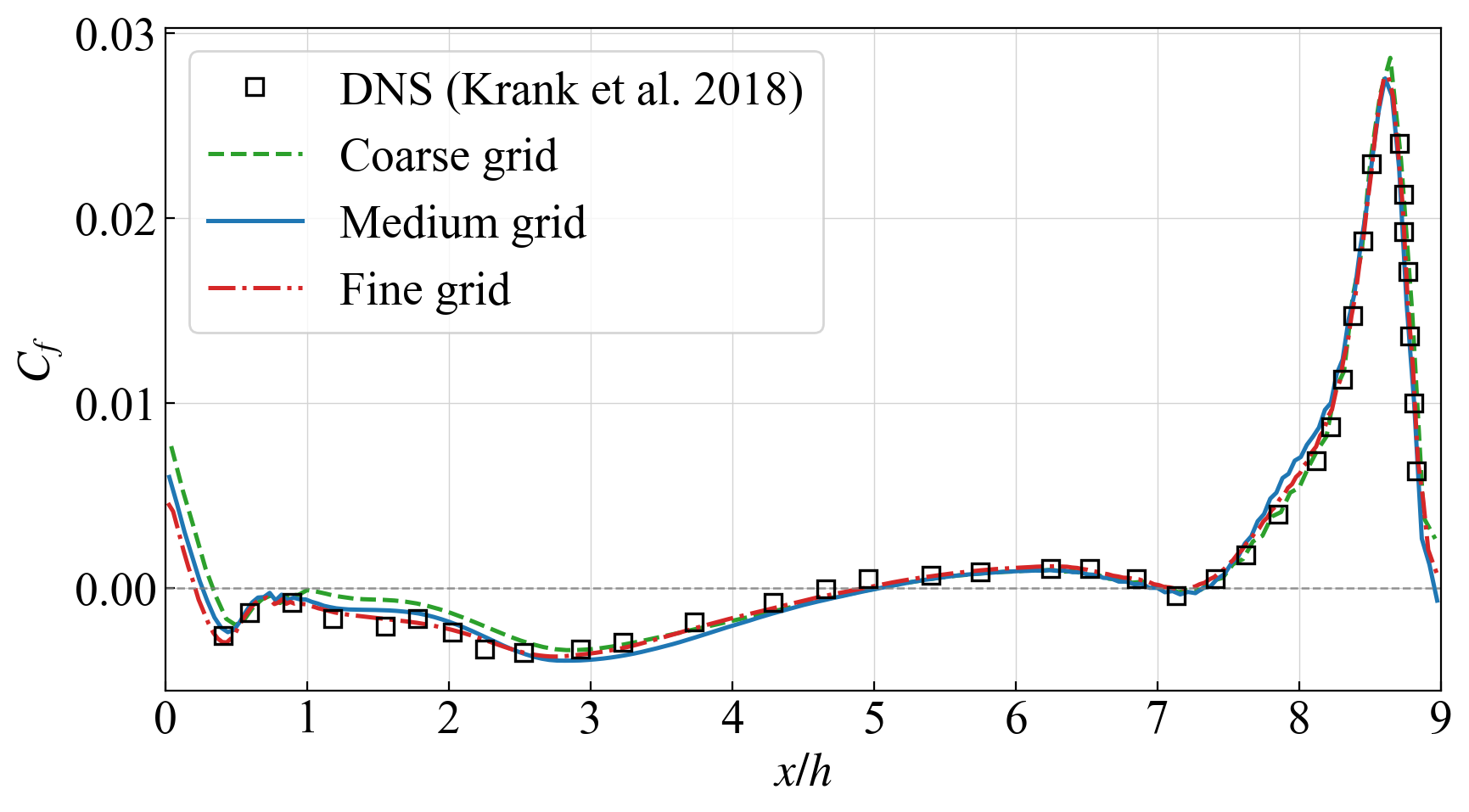}\\[2pt]
  a) Grid convergence using the Smagorinsky-LLM model
\end{minipage}
\hfill
\begin{minipage}[b]{0.48\textwidth}
  \centering
  \includegraphics[width=\textwidth]{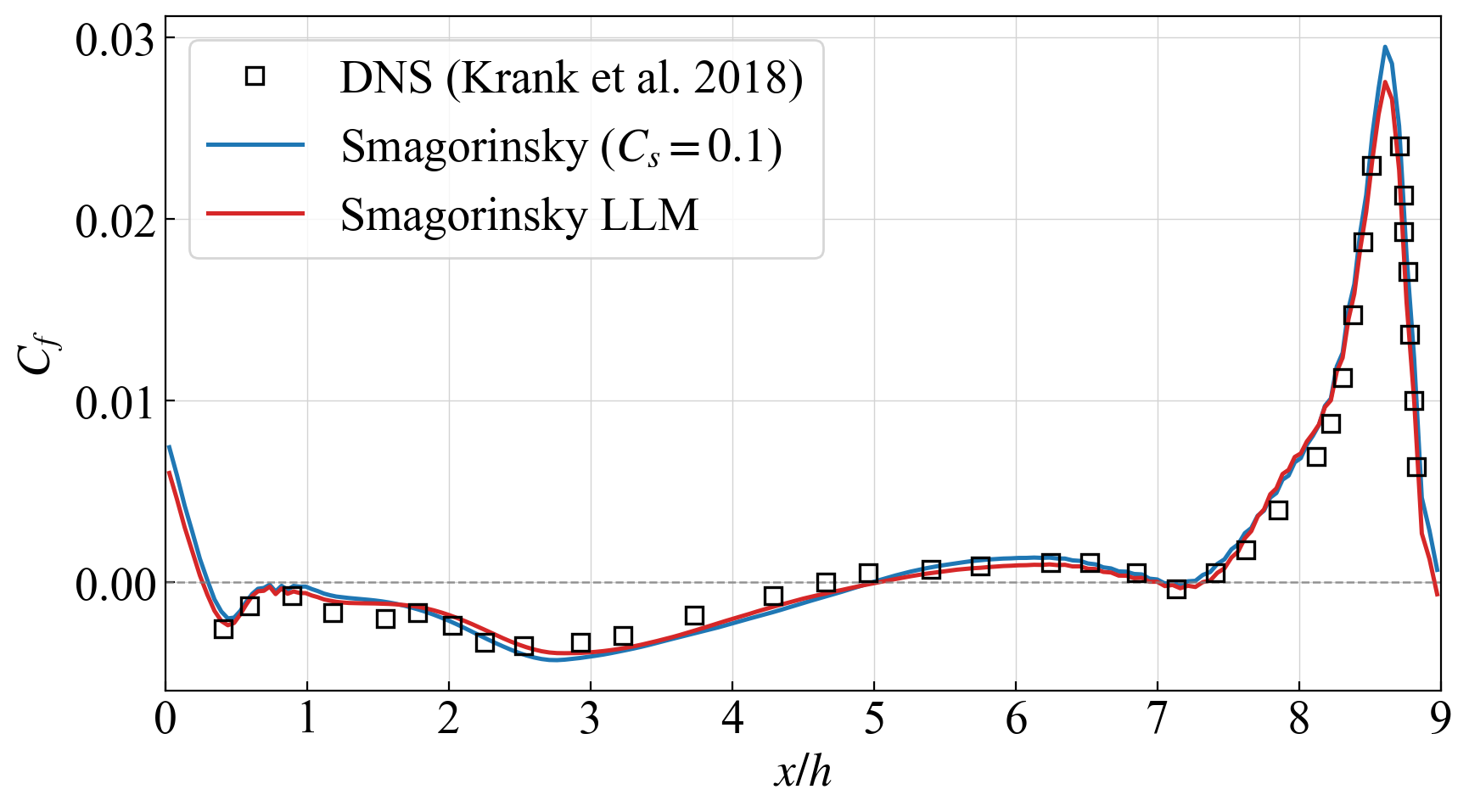}\\[2pt]
  b) Smagorinsky model vs. Smagorinsky-LLM model
\end{minipage}
\caption{Skin-friction coefficient $C_f$ distribution for the periodic-hill case at $Re_h=10,595$}
\label{fig:periodic_hill_Cf}
\end{figure}

 Figure~\ref{fig:periodic_hill_Cf} presents the skin-friction coefficient $C_f$ distribution for the periodic-hill case. Figure~\ref{fig:periodic_hill_Cf}(a) examines grid convergence using the standard Smagorinsky model. As the grid is refined, the predicted $C_f$ distribution, particularly within the separation region, becomes progressively more accurate. The results show clear grid convergence. Figure~\ref{fig:periodic_hill_Cf}(b) compares the standard Smagorinsky model with the Smagorinsky LLM model on the same medium grid. Both models accurately reproduce the $C_f$ distribution and show good agreement with the DNS results reported by Krank et al.~\cite{krank2018direct}. The medium grid is used for all subsequent comparison.

\begin{figure}[hbt!]
\centering
\begin{minipage}[b]{0.48\textwidth}
  \centering
  \includegraphics[width=\textwidth]{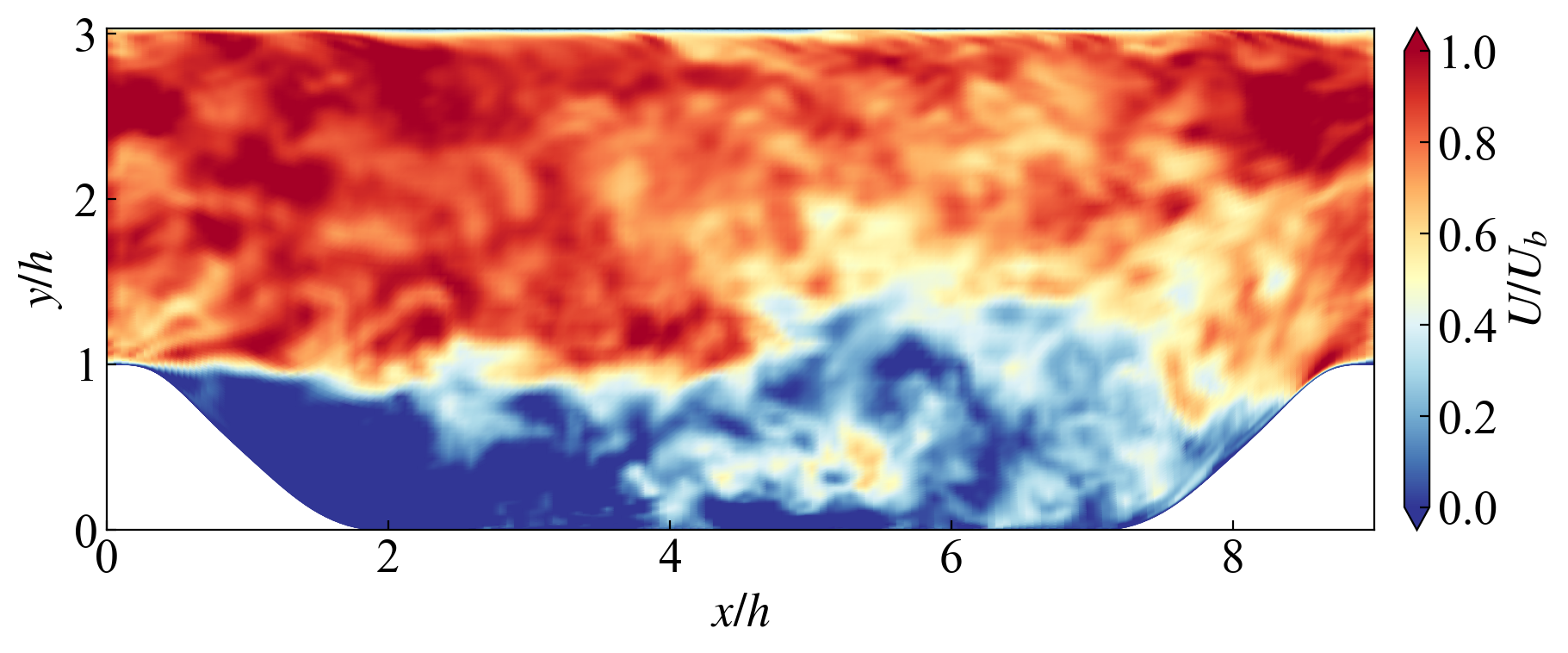}\\[2pt]
  a) Smagorinsky model ($C_s=0.10$)
\end{minipage}
\hfill
\begin{minipage}[b]{0.48\textwidth}
  \centering
  \includegraphics[width=\textwidth]{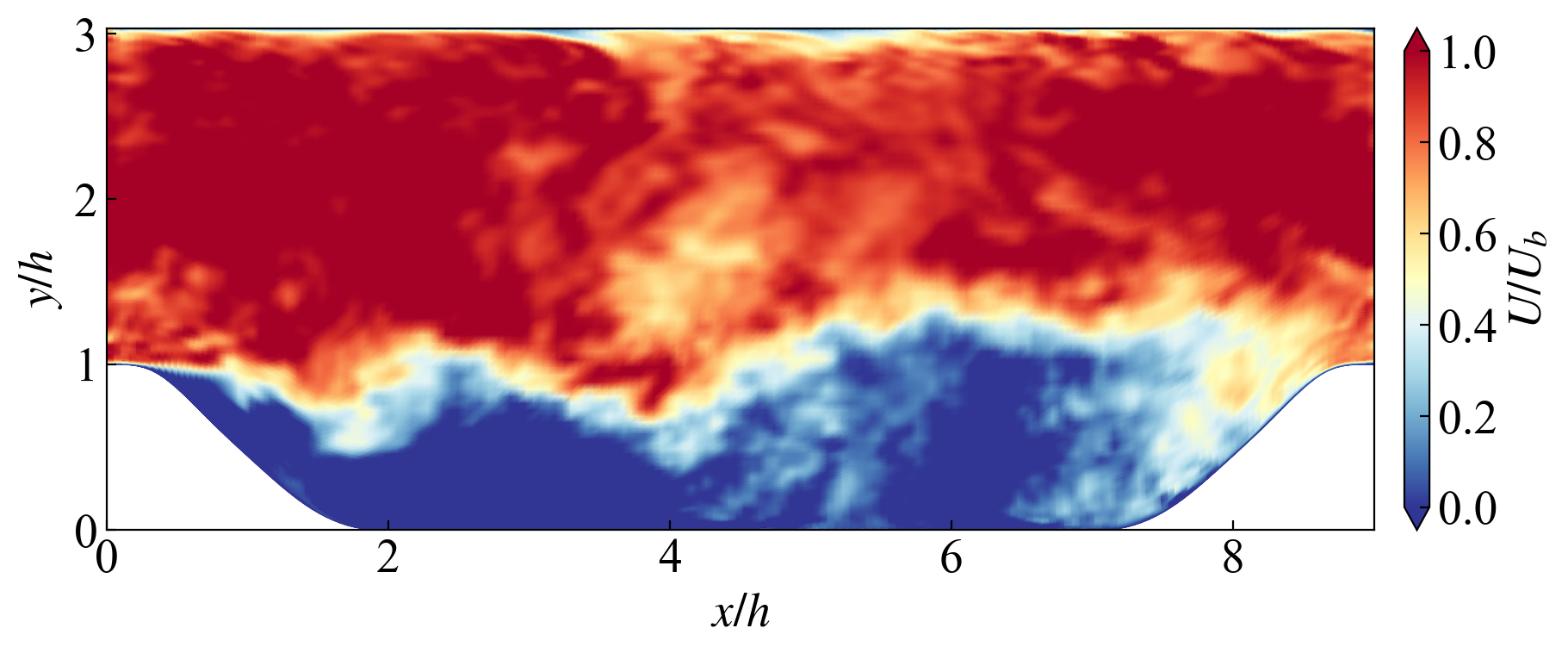}\\[2pt]
  b) Smagorinsky-LLM model
\end{minipage}
\caption{Instantaneous streamwise velocity contours for the periodic hill at $Re_h=10,595$.}
\label{fig:periodic_hill_instantaneous}
\end{figure}

\begin{figure}[hbt!]
\centering
\begin{minipage}[b]{0.48\textwidth}
  \centering
  \includegraphics[width=\textwidth]{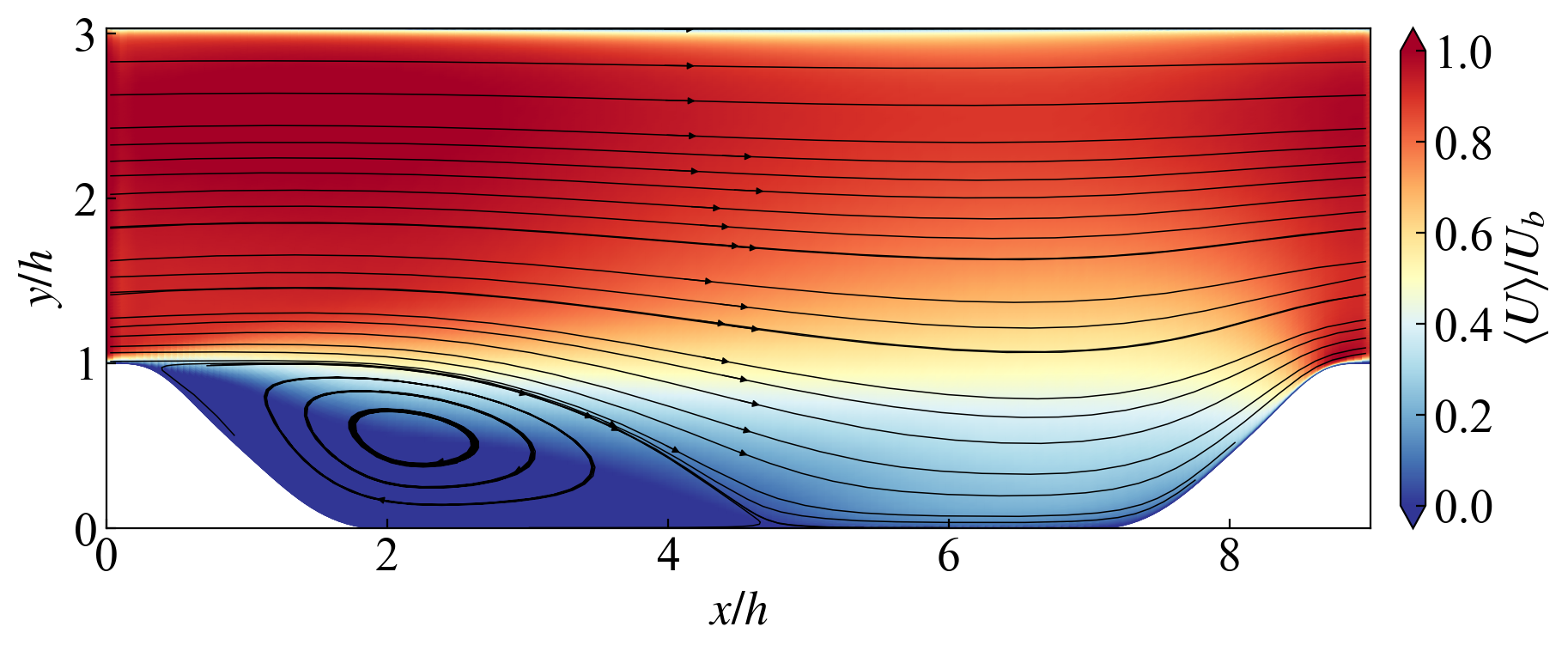}\\[2pt]
  a) Smagorinsky model ($C_s=0.10$)
\end{minipage}
\hfill
\begin{minipage}[b]{0.48\textwidth}
  \centering
  \includegraphics[width=\textwidth]{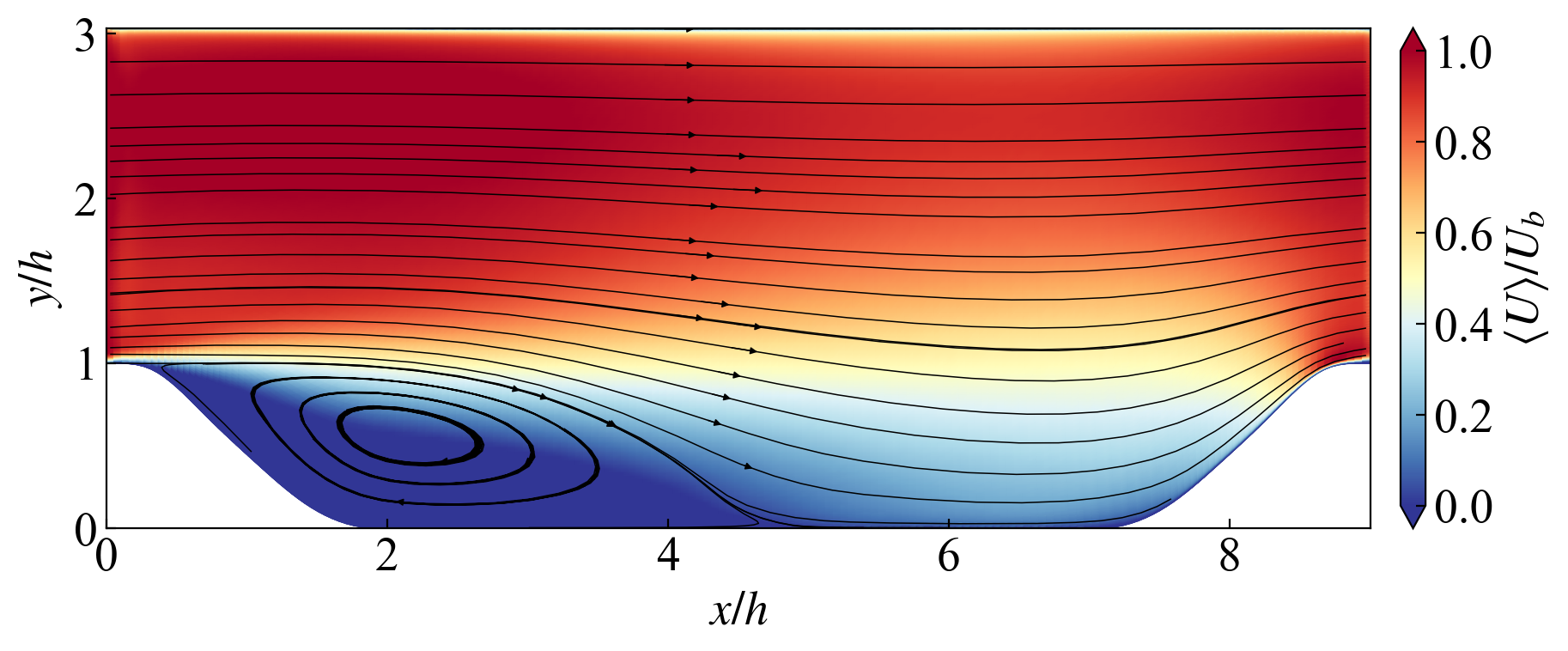}\\[2pt]
  b) Smagorinsky-LLM model
\end{minipage}
\caption{Time-averaged velocity field and streamlines for the periodic hill obtained using the standard Smagorinsky model and the Smagorinsky-LLM model at $Re_h=10,595$.}
\label{fig:periodic_hill_streamlines}
\end{figure}

Figure~\ref{fig:periodic_hill_instantaneous} presents the instantaneous streamwise velocity contours for the periodic hill. A flow separation region is observed downstream of the left hill. The separated flow is significant and extends toward the right hill. This large-scale separation is induced by the adverse pressure gradient on the curved surface and poses a substantial challenge for traditional turbulence modeling methods. Figure~\ref{fig:periodic_hill_streamlines} presents the time-averaged velocity field and streamlines for the periodic hill case obtained using the standard Smagorinsky model and the Smagorinsky-LLM model. According to the experimental results reported by Ch. Rapp et al. \cite{rapp2011flow}, the measured reattachment point is located at $x/h = 4.21$. Both models predict reattachment at $x/h = 4.9$, yielding comparable accuracy relative to the experimental measurement.

\begin{figure}[htbp] 
\centering
\begin{minipage}[b]{1.0\textwidth} 
  \centering
  \includegraphics[width=\textwidth]{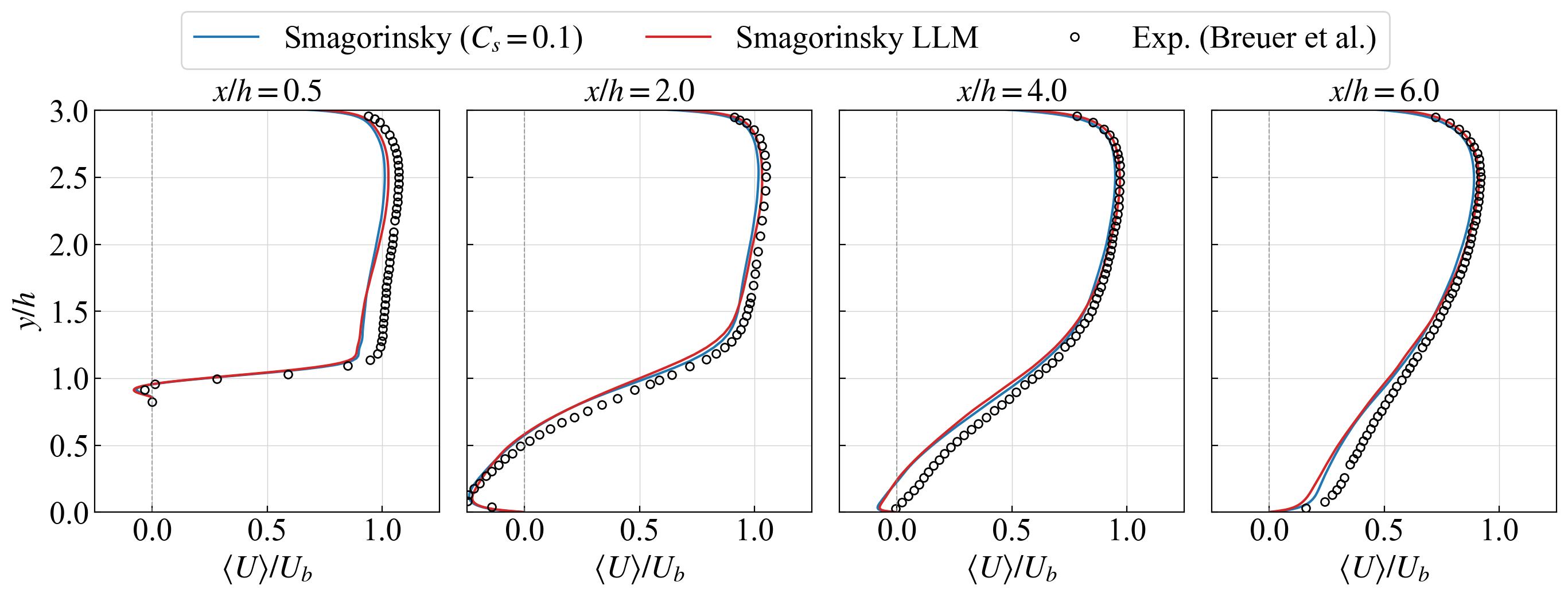}
  \\ a) $\langle U \rangle / U_b$
\end{minipage} \\

\begin{minipage}[b]{1.0\textwidth}
  \centering
  \includegraphics[width=\textwidth]{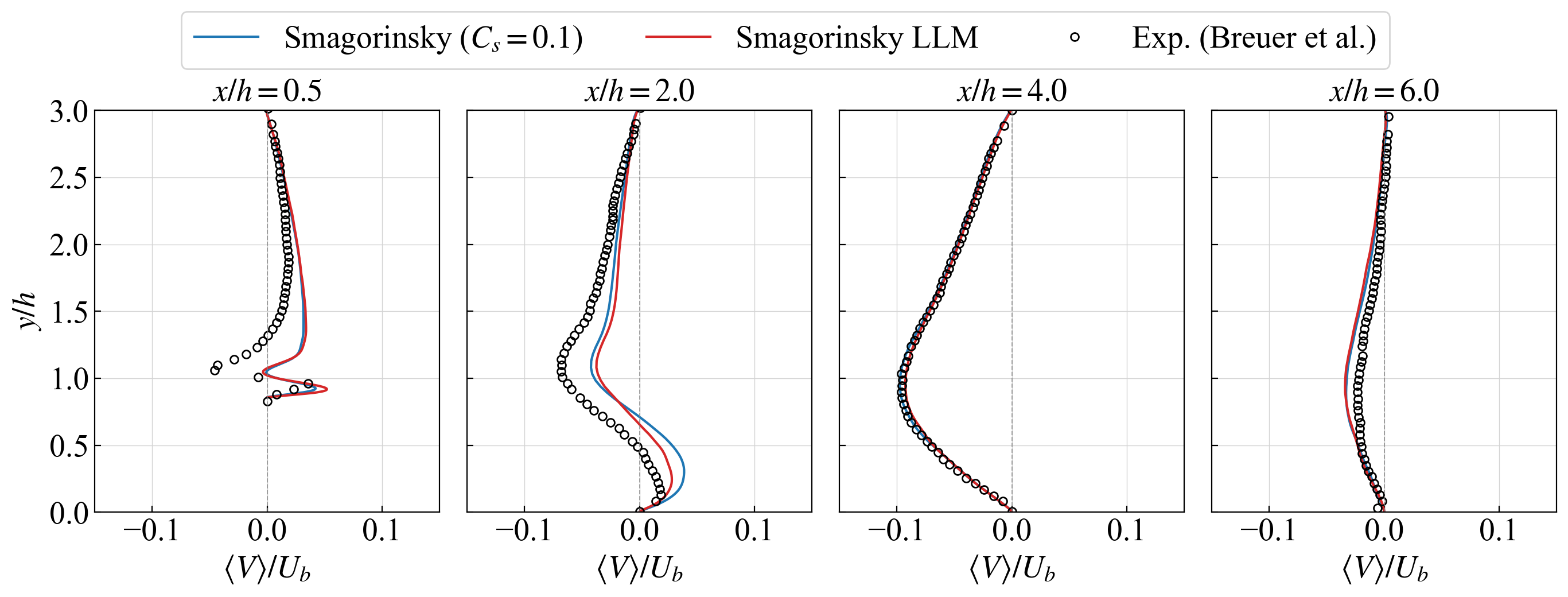}
  \\ c) $\langle V \rangle / U_b$
\end{minipage}
\caption{Time-averaged streamwise and wall-normal velocities, $\langle U \rangle/U_b$ and $\langle V \rangle/U_b$, for the periodic hill flow. Lines represent different SGS models, and symbols denote the experimental data of Breuer et al.~\cite{breuer2009flow}.}
\label{fig:periodic_hill_profiles} 
\end{figure}

Figure~\ref{fig:periodic_hill_profiles} presents the time-averaged streamwise and wall-normal velocities, $\langle U \rangle/U_b$ and $\langle V \rangle/U_b$, for the periodic hill flow. The two models produce very similar mean velocity profiles, with noticeable differences only around $x/h=2.0$. The Smagorinsky LLM model provides better predictions for $\langle V \rangle/U_b$. Overall, the Smagorinsky LLM model shows slightly better agreement with the reference time-averaged flow field. A more pronounced difference is observed in the following prediction of higher-order statistical quantities. The Smagorinsky LLM model shows a clear advantage over the standard Smagorinsky model in the predictions for Reynolds stresses. 

\begin{figure}[htbp] 
\centering
\begin{minipage}[b]{0.8\textwidth} 
  \centering
  \includegraphics[width=\textwidth]{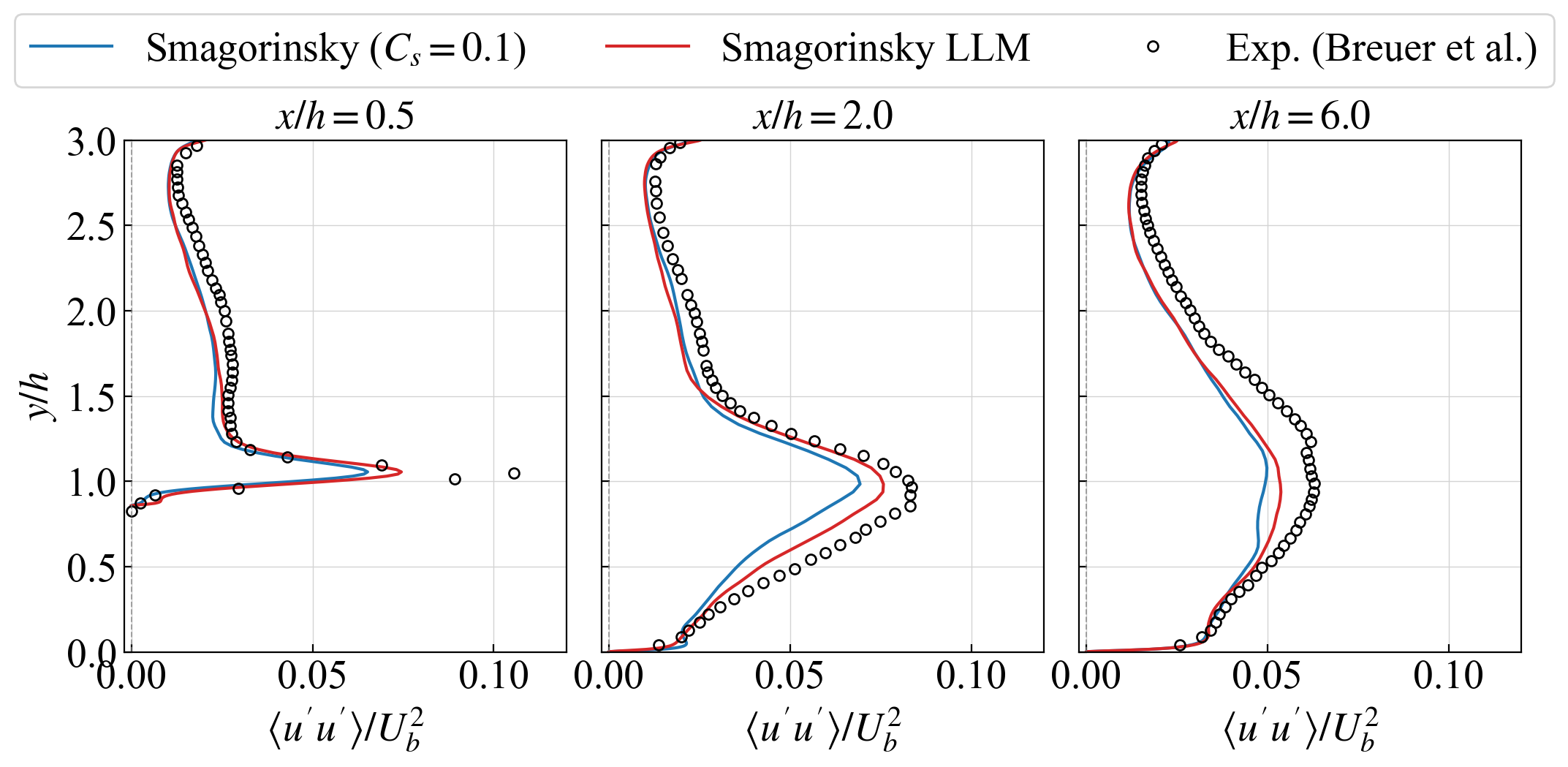}
  \\ a) $\langle u'u' \rangle / U_b^2$
\end{minipage} \\

\begin{minipage}[b]{0.8\textwidth}
  \centering
  \includegraphics[width=\textwidth]{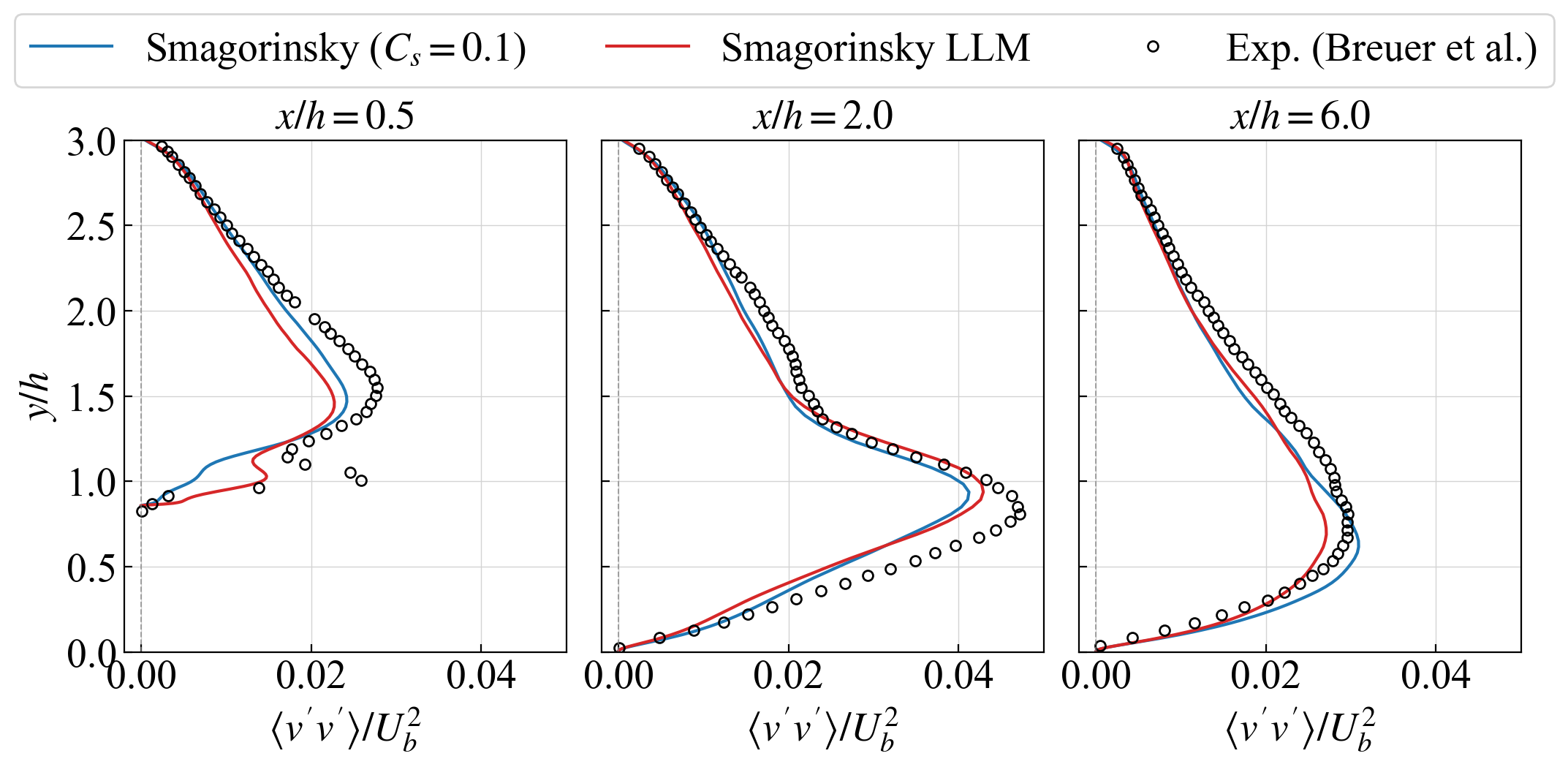}
  \\ b) $\langle v'v' \rangle / U_b^2$
\end{minipage} \\

\begin{minipage}[b]{0.8\textwidth}
  \centering
  \includegraphics[width=\textwidth]{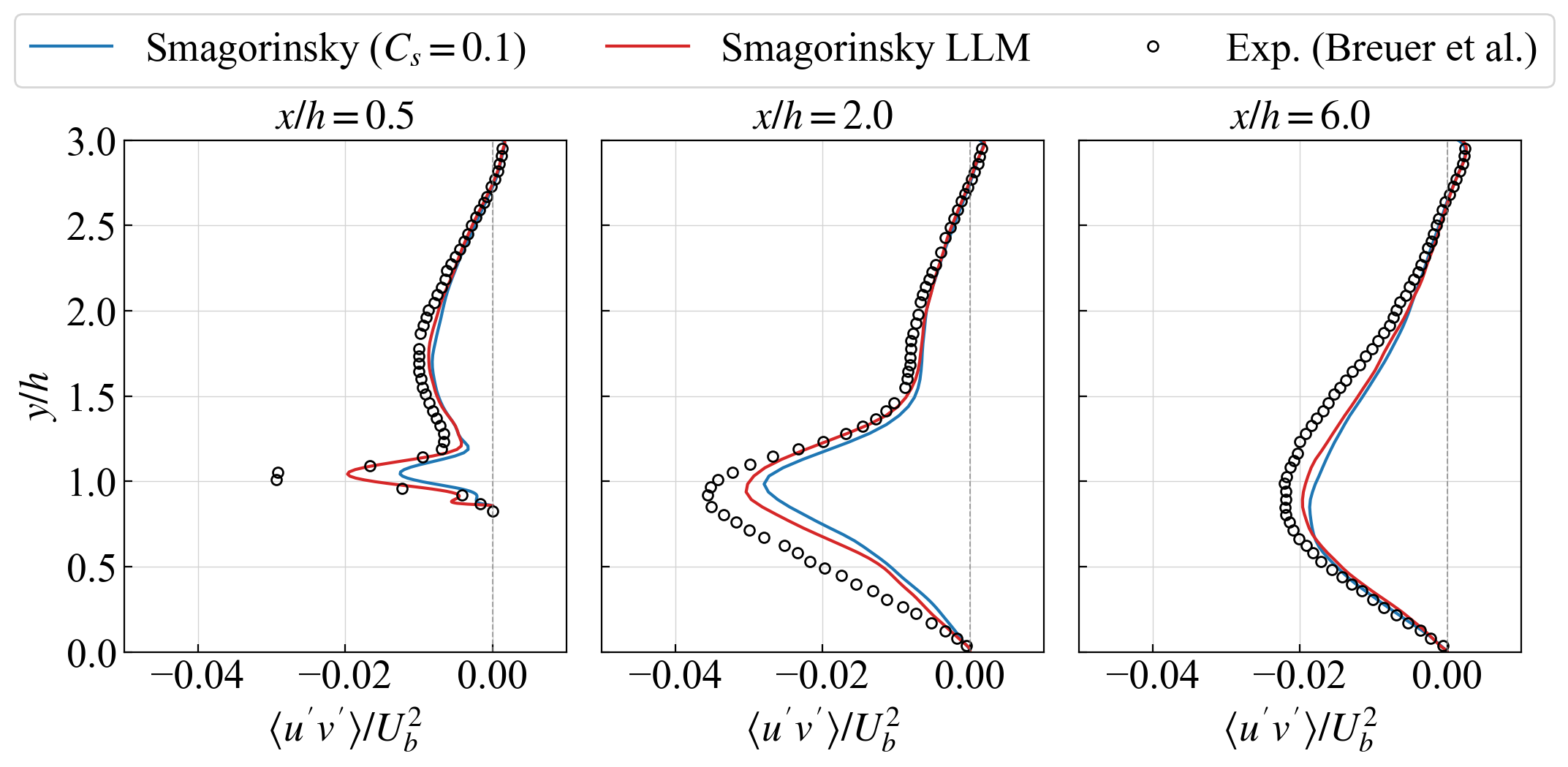}
  \\ c) $\langle u'v' \rangle / U_b^2$
\end{minipage}
\caption{Profiles of Reynolds stresses ($\langle u'u'\rangle$, $\langle v'v'\rangle$ and $\langle u'v'\rangle$) at multiple streamwise locations for the periodic hills case. Lines represent different SGS models, and symbols denote the experimental data of Breuer et al.~\cite{breuer2009flow}.}
\label{fig:periodic_hill_Reynolds_stress} 
\end{figure}

Figure~\ref{fig:periodic_hill_Reynolds_stress} presents profiles of Reynolds stresses ($\langle u'u'\rangle$, $\langle v'v'\rangle$ and $\langle u'v'\rangle$) at multiple streamwise locations for the periodic hills. The standard Smagorinsky model ($C_s=0.10$)~\cite{deardorff1970numerical} and the Smagorinsky LLM are compared. The Smagorinsky LLM model consistently provides better predictions across the investigated streamwise locations. The improvement is particularly pronounced at $x/h=0.5$ and $2$, where a separating shear layer is present. The standard Smagorinsky model exhibits excessive dissipation in this region, leading to substantial discrepancies in the predicted Reynolds stresses. At $x/h=0.5$, for example, the predicted $\langle v'v'\rangle$ profile from the standard Smagorinsky model even deviates qualitatively from the reference profile, with an incorrect shape.
This result demonstrates the effectiveness of the rotation-suppression correction term and its contribution to more accurate Reynolds stress prediction.

\subsection{Token Consumption and Reproducibility Analysis}

Quantifying the token and cost consumption of a complete PhysMiner execution is important for assessing its computational efficiency. The PhysMiner framework consists of four agents: \texttt{flow-fingerprint Agent}, \texttt{flow-comparison Agent}, \texttt{discover-physics Agent}, and \texttt{review agent}. Table~\ref{tab:agent_usage_Haiku_matched} summarizes the token and cost consumption of a complete PhysMiner execution. The \texttt{Discover Agent} accounts for the largest share of the computational cost, consuming 75.00\% of the total tokens and 83.95\% of the total cost. This dominant usage is consistent with its role as the primary agent for physics reasoning and discovery. The \texttt{flow-fingerprint agent}, \texttt{flow-comparison agent}, and \texttt{discover-physics agent} use Sonnet 5, which provides the reasoning capability required for their respective tasks. In contrast, the \texttt{Review Agent} performs a relatively simple review task. Specifically, it reads the generated report and evaluates it according to four predefined criteria. This task does not require the same level of reasoning capability. Therefore, Haiku 4.5 is employed for the \texttt{Review Agent}. Replacing Sonnet 5 with Haiku 4.5 in the \texttt{Review Agent} reduces token consumption by approximately 29.2\%, total cost by 83.2\%, and execution time by 78.0\%, while maintaining the predefined verification criteria.

\begin{table}[hbt!]
\caption{Token and cost consumption of
a complete PhysMiner execution.}
\label{tab:agent_usage_Haiku_matched}
\centering
\begin{tabular}{l l p{0.32\textwidth} r r}
\toprule
Agent & Model & Function & Tokens (share, \%) & Cost, USD (share, \%) \\
\midrule

\texttt{flow-fingerprint} & Sonnet~5 & Classifies the current flow into a 10-dimensional fingerprint vector & 0.11M (3.62) & 0.17 (4.47) \\

\texttt{flow-comparison} & Sonnet~5 & Structured comparison of the current case against historical flow cases & 0.26M (8.55) & 0.31 (8.16) \\

\texttt{discover-physics} & Sonnet~5 & In-depth physical analysis of result images, surveyed literature, and flow-field data; generates \texttt{report.pdf} & 2.28M (75.00) & 3.19 (83.95) \\

\texttt{review-report} & Haiku~4.5 & Reviews \texttt{report.pdf} against four validity criteria and issues a pass/fail verdict & 0.39M (12.83) & 0.13 (3.42) \\

\midrule
\textbf{Total} & \action{N/A} & \action{N/A} & \textbf{3.04M (100.00)} & \textbf{3.80 (100.00)} \\
\bottomrule
\end{tabular}
\end{table}

The reproducibility and end-to-end consumption of PhysMiner is further assessed through five independent runs. The mean and standard deviation of token consumption, cost, and execution time are calculated across the five runs. Table~\ref{tab:reproducibility} summarizes the reproducibility and end-to-end resource consumption of PhysMiner across five independent runs. The average token consumption is 3.24M, with a standard deviation of 0.49M (15.10\% of the mean). The average cost is \$3.89, with a standard deviation of \$0.56 (14.29\% of the mean). The average execution time is 52.97 minutes, with a standard deviation of 10.31 minutes (19.47\% of the mean). The relatively small variations in token consumption, cost and time indicate stable resource requirements and execution across repeated runs. \action{Each run produces a PDF report for direct inspection of the generated content. Although the reports vary in detail, all five runs propose SGS modeling strategies. A run is counted as a 100\% success if the top-ranked recommendation is to introduce an additional rotation-suppression term and the closed loop finishes within the maximum of 5 iterations (the default limit)}.

\begin{table}[hbt!]
\caption{End-to-end resource consumption per full PhysMiner run.}
\label{tab:reproducibility}
\centering
\begin{tabular}{l r r r r}
\toprule
Run & Success & Tokens & Cost(\$) & Time(min) \\
\midrule
Run 1 & 100\% & 3.04M & 3.80 & 50.95 \\
Run 2 & 100\% & 3.22M & 4.18 & 49.89 \\
Run 3 & 100\% & 2.55M & 3.02 & 39.02 \\
Run 4 & 100\% & 3.58M & 3.93 & 66.73 \\
Run 5 & 100\% & 3.81M & 4.51 & 58.25 \\
\midrule
\textbf{Mean}                      & \action{N/A} & \textbf{3.24M} & \textbf{3.89} & \textbf{52.97} \\
\textbf{Std.\ dev.}                & \action{N/A} & 0.49M & 0.56 & 10.31 \\
\textbf{Std.\ dev.\ (\% of mean)}  & \action{N/A} & 15.10\% & 14.29\% & 19.47\% \\
\bottomrule
\end{tabular}
\end{table}

\subsection{Application of PhysMiner to case 2: iced airfoil}

This section evaluates PhysMiner on a more complex engineering flow and demonstrates its capability for user-directed physics discovery. User-specified prompts can guide the discovery process toward a targeted physical question or modeling objective. This mode is enabled by freezing the literature-driven track, allowing the discover-physics agent to follow the specified prompt rather than being automatically guided by the literature (see Sec.~II.B for the corresponding architectural description).

\subsubsection{User-specified prompts}

PhysMiner also supports user-specified research questions through prompts. When a user-defined prompt is provided, PhysMiner identifies the specified research direction and bypasses the literature-driven track. An iced-airfoil case is used to demonstrate this capability. The selected configuration consists of a GLC305 airfoil with the 944 ice shape. The glaze ice forms ice horns on both the upper and lower surfaces, with a relative ice height of $h/c = 3.68\%$. The flow conditions are $\mathit{Ma} = 0.12$, $\mathit{Re} = 3.5 \times 10^6$, and $AoA = 6^\circ$. Relevant experimental results are reported in Ref.~\cite{broeren2006flowfield}. The flow field is simulated using the IDDES method with approximately 10 million cells.

\begin{figure}[hbt!]
\centering
\includegraphics[width=1.0\textwidth]{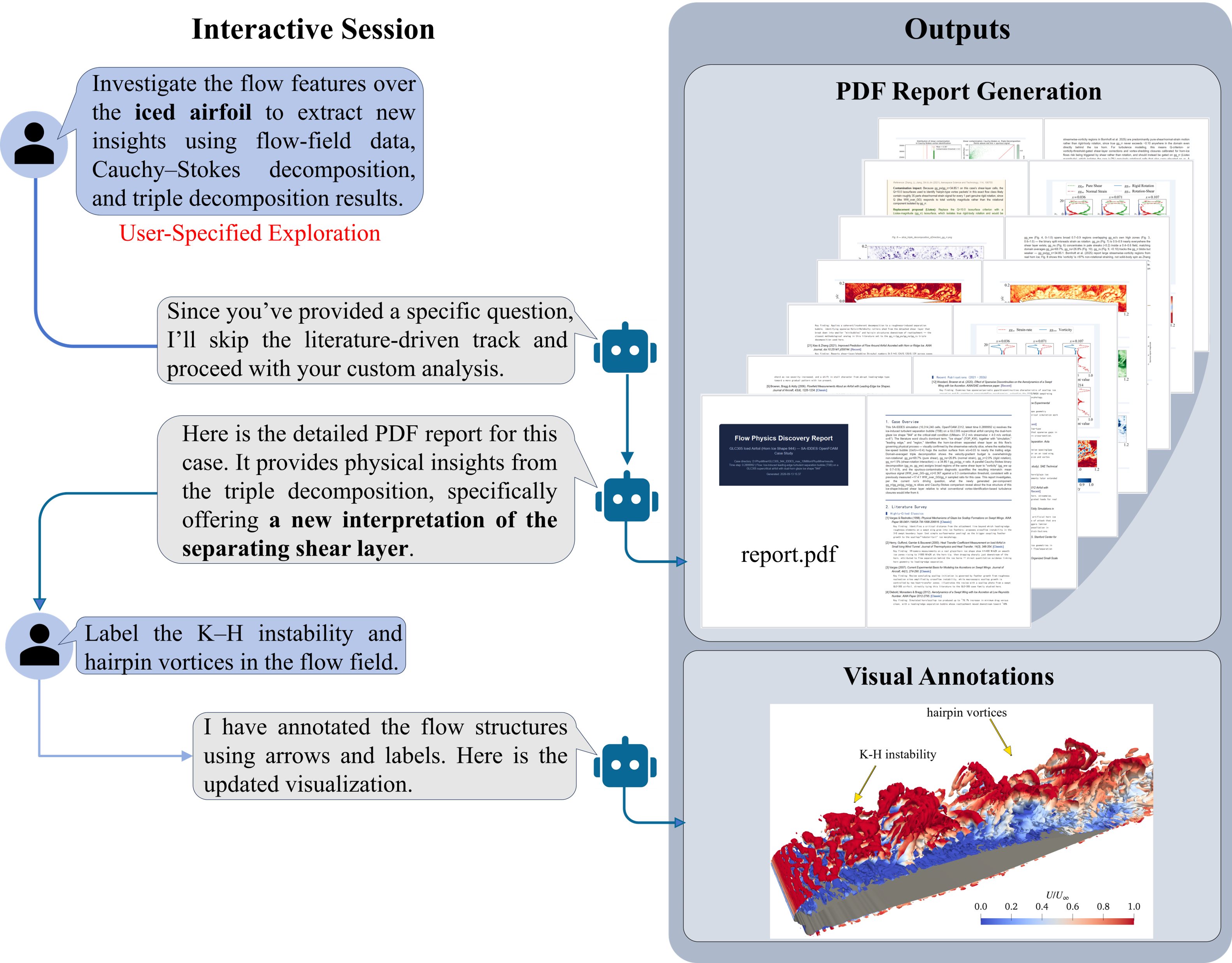}
\caption{User-specified prompt and the final PDF report generated by PhysMiner for GLC 305 with 944 ice shape case.}
\label{fig:user_specified_prompt}
\end{figure}

Figure~\ref{fig:user_specified_prompt} presents the user-specified prompt and the final PDF report generated by PhysMiner for this case. Given the complex iced-airfoil geometry, flow-structure characterization represents an important research topic. The selected prompt asks PhysMiner to investigate the flow structures in the resulting flow field. The prompt is not intended to represent a particularly challenging research question for this case. Instead, it is selected as a representative example to demonstrate that PhysMiner can accommodate research questions specified directly by users. Figure~\ref{fig:user_specified_prompt} also demonstrates the PhysMiner's ability to interactively refine an automatically generated plot by labeling specific flow structures, including K-H instability and hairpin vortices. The specified structures are automatically and accurately identified by PhysMiner. \action{Some representative figures and key findings from the automatically generated final PDF report by PhysMiner are presented below.}

\subsubsection{New Findings from PhysMiner based on the Specified Prompt}

\begin{figure}[hbt!]
\centering
\includegraphics[width=0.8\textwidth]{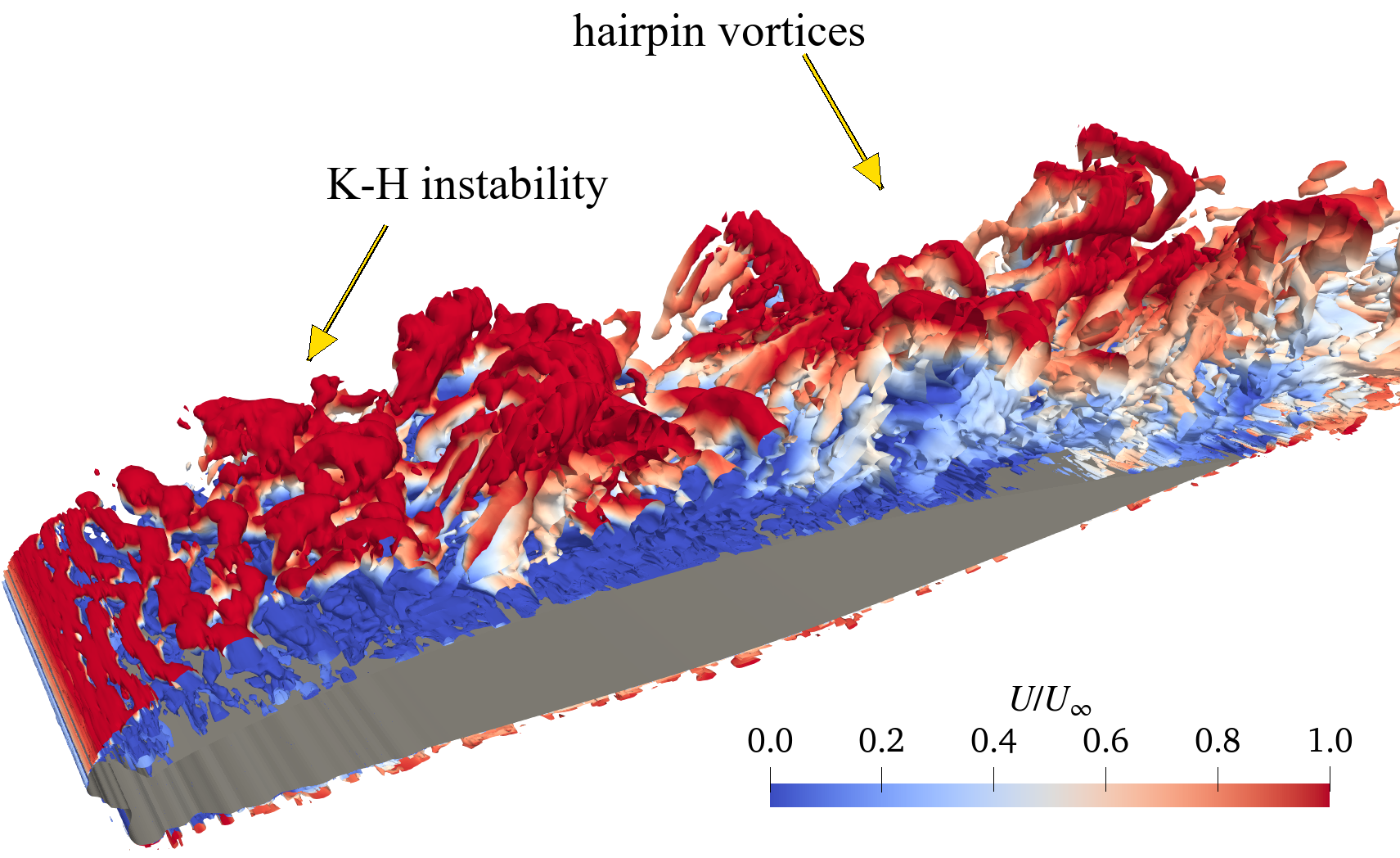}
\caption{Q isosurface colored by the dimensionless velocity $U/U_{\infty}$ for GLC 305 with 944 ice shape.}
\label{fig:iced_airfoil_Q}
\end{figure}

Ice accretion often induces large-scale flow separation accompanied by complex flow structures.
Figure~\ref{fig:iced_airfoil_Q} presents the Q isosurface colored by the dimensionless velocity $U/U_{\infty}$. A distinct separation bubble forms downstream of the leading-edge ice. The separation point remains fixed at the ice tip. Several characteristic vortex structures are observed, including Kelvin--Helmholtz instability, and hairpin vortices. This flow represents a practical engineering phenomenon with a complex geometry. A detailed understanding of these flow structures is essential for elucidating ice-induced flow mechanisms and developing configurations to mitigate ice-accretion. \action{Figure~\ref{fig:iced_airfoil_Q} was generated automatically. The flow structures were identified automatically, and the arrows and annotations were generated by the agent.}

\begin{figure}[hbt!]
\centering
\begin{minipage}[b]{0.49\textwidth} 
\centering
\includegraphics[width=\textwidth]{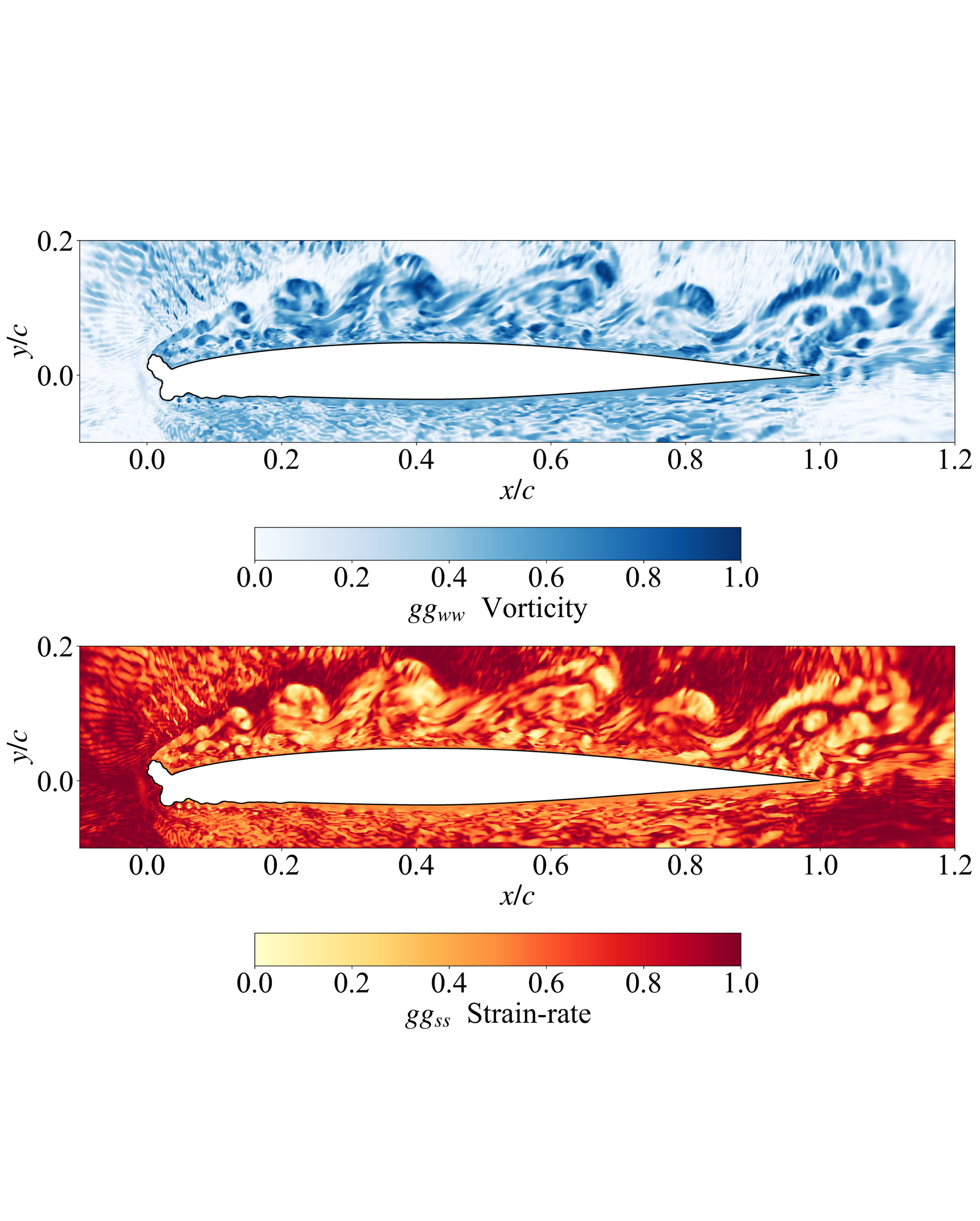}\\[2pt]
a) Cauchy--Stokes decomposition
\end{minipage}
\begin{minipage}[b]{0.49\textwidth}
\centering
\includegraphics[width=\textwidth]{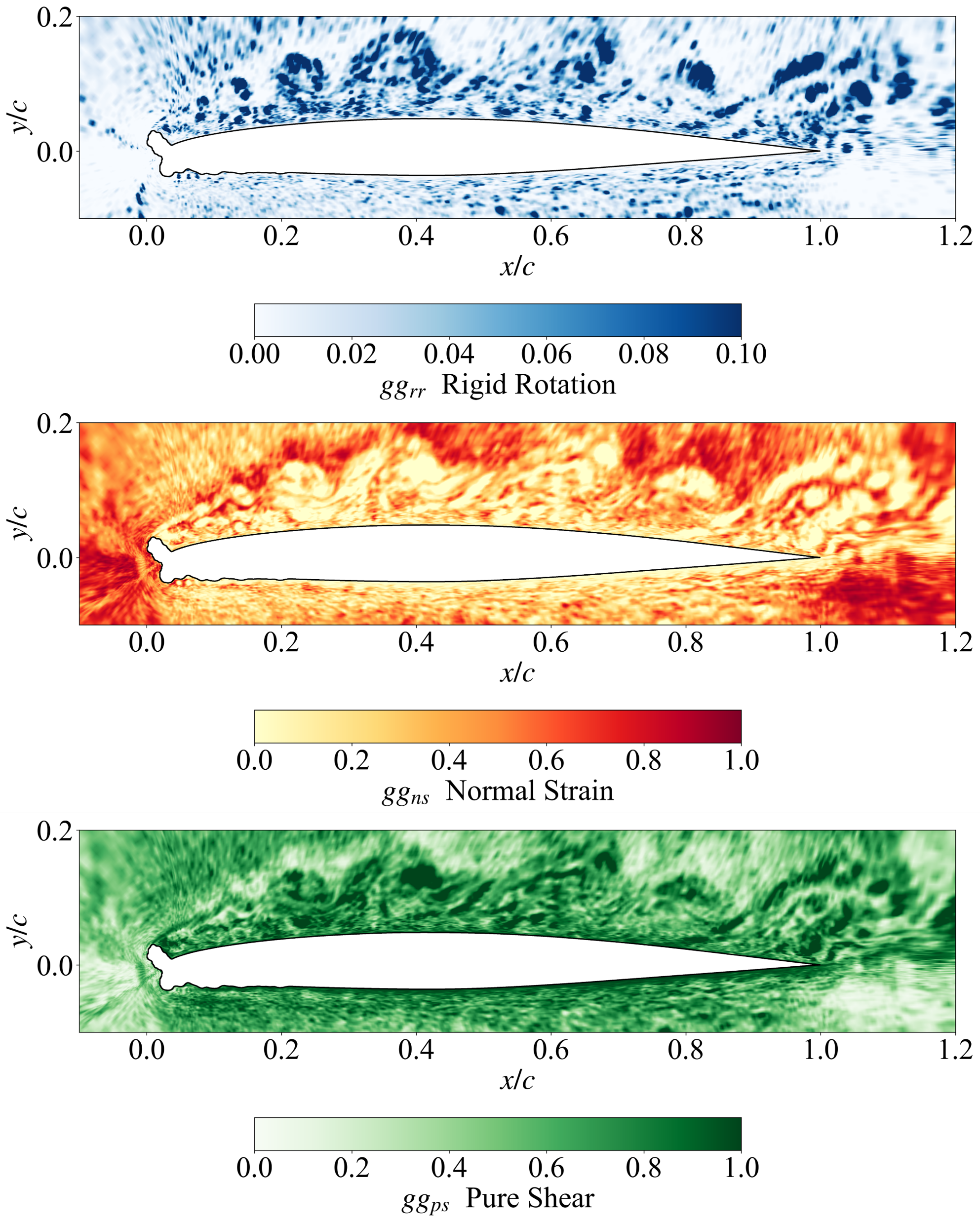}\\[2pt]
b) Triple decomposition
\end{minipage}
\caption{Differences between the Cauchy--Stokes decomposition and triple decomposition for GLC 305 with 944 ice shape.}
\label{fig:iced_airfoil_CS_vs_TD}
\end{figure}

Figure~\ref{fig:iced_airfoil_CS_vs_TD} compares the Cauchy--Stokes decomposition with the triple decomposition for the GLC305 airfoil with the 944 ice shape at a selected section. Figure~\ref{fig:iced_airfoil_CS_vs_TD}(a) shows the dimensionless magnitude of the vorticity tensor, $gg_{ww}$, which is commonly used to characterize vortical structures. The separating shear layer and the subsequent vortex-shedding process are captured. Vorticity-based plots~\cite{alam2015hybrid, zhang2025improvement} and $Q$ criteria-based plots~\cite{zhang2021analysis, xiao2021improved}  are widely used to visualize such flow structures. However, these criteria do not distinguish rigid rotation from other velocity-gradient components. Figure~\ref{fig:iced_airfoil_CS_vs_TD}(b) presents the triple decomposition at the same section. The three components, $gg_{rr}$, $gg_{ns}$, and $gg_{ps}$, are shown separately for clarity. \action{The $gg_{rr}$ plot uses a different legend range, with an upper limit of 0.1 rather than 1.0.} The $gg_{rr}$ field identifies substantially fewer rigid-rotation structures than the $gg_{ww}$ field in Figure~\ref{fig:iced_airfoil_CS_vs_TD}(a). This difference indicates that part of the structures identified by conventional vorticity-based visualization is associated with shear rather than rigid-body rotation.

\begin{figure}[hbt!]
  \centering
  \begin{minipage}[b]{0.33\textwidth}
    \centering
    \includegraphics[width=\textwidth]{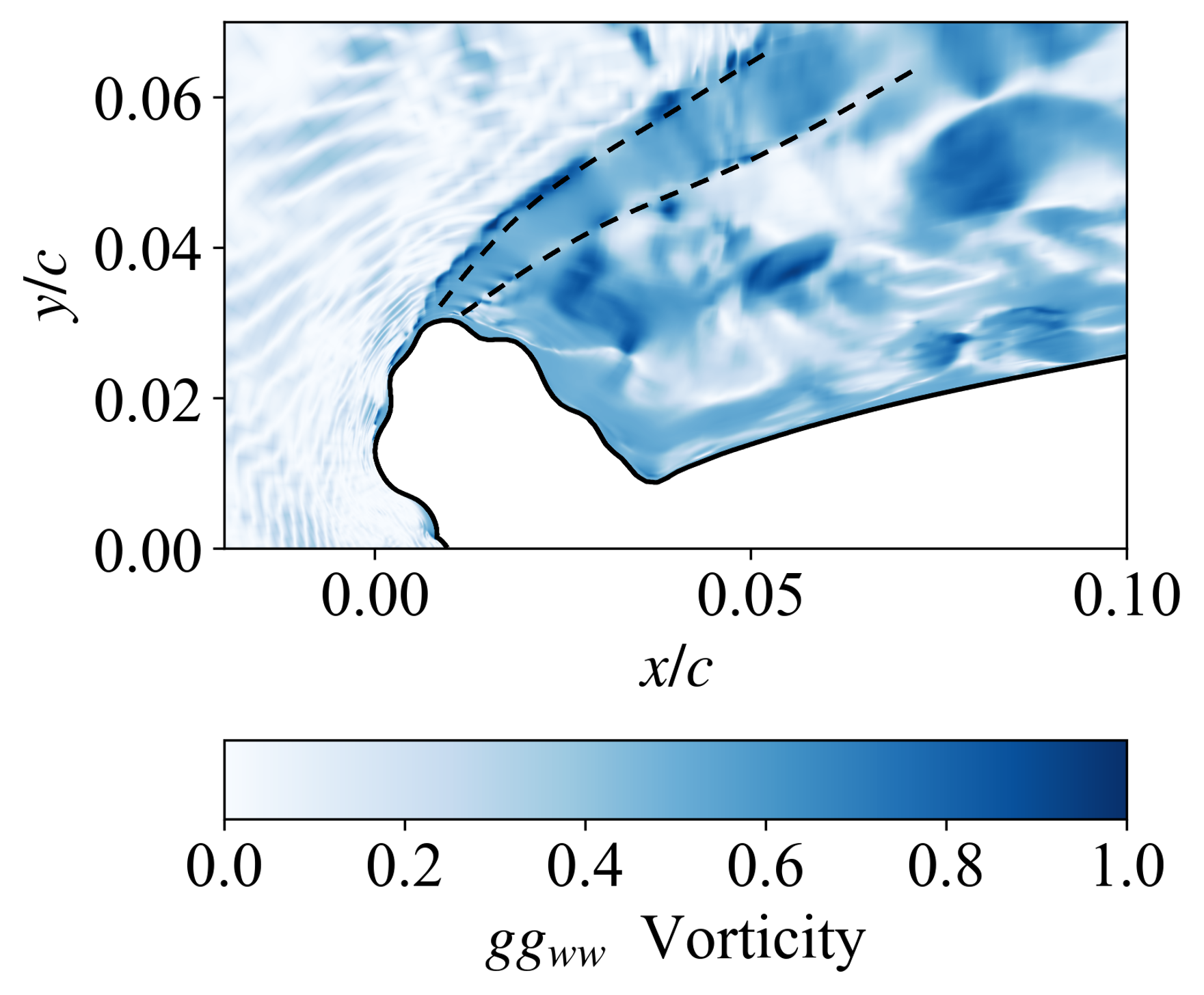}\\[2pt]
    a) Cauchy–Stokes, $gg_{ww}$
  \end{minipage}%
  \hspace{2em}
  \begin{minipage}[b]{0.33\textwidth}
    \centering
    \includegraphics[width=\textwidth]{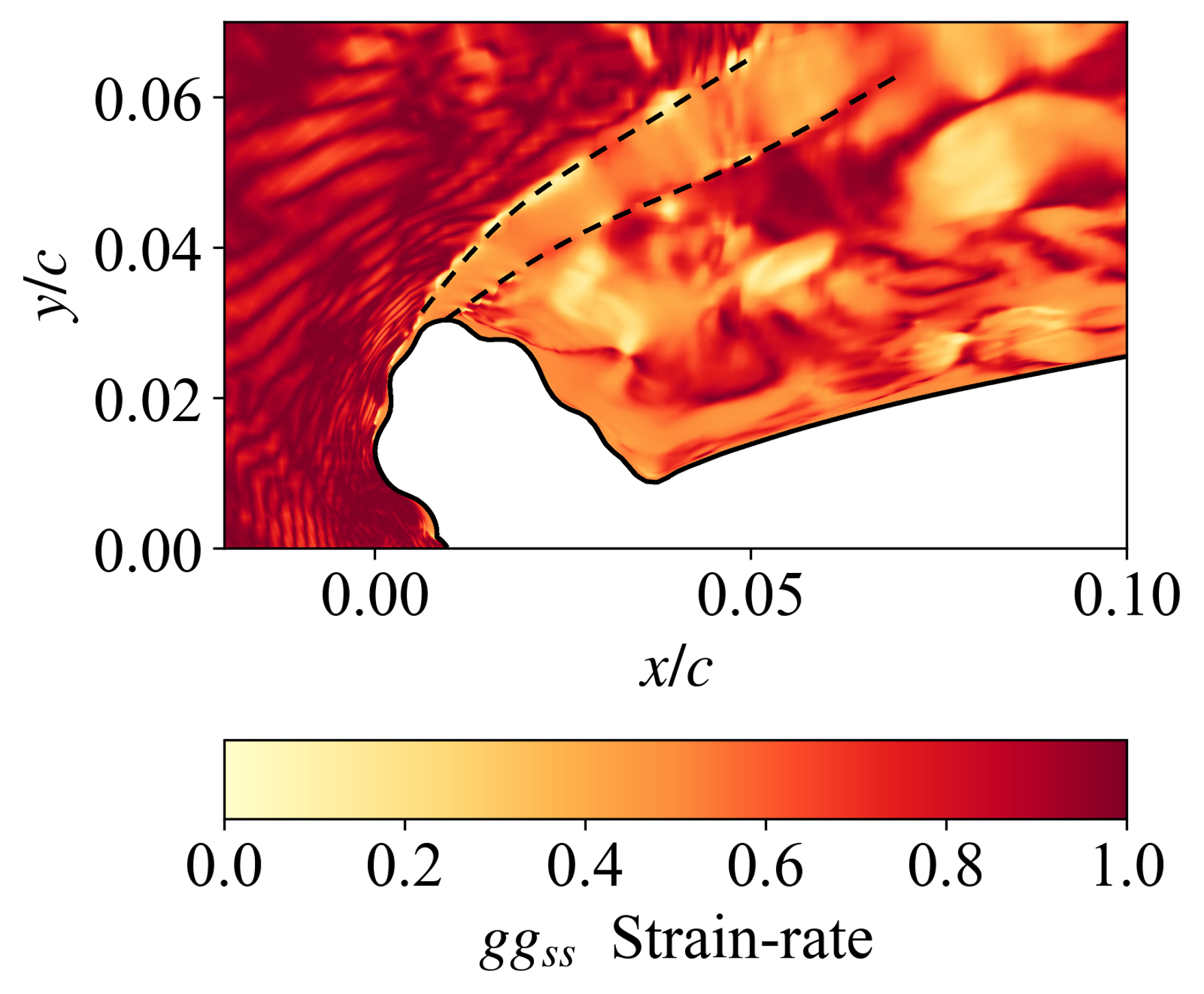}\\[2pt]
    b) Cauchy–Stokes, $gg_{ss}$
  \end{minipage}

  \vspace{1em} 

  \begin{minipage}[b]{0.32\textwidth}
    \centering
    \includegraphics[width=\textwidth]{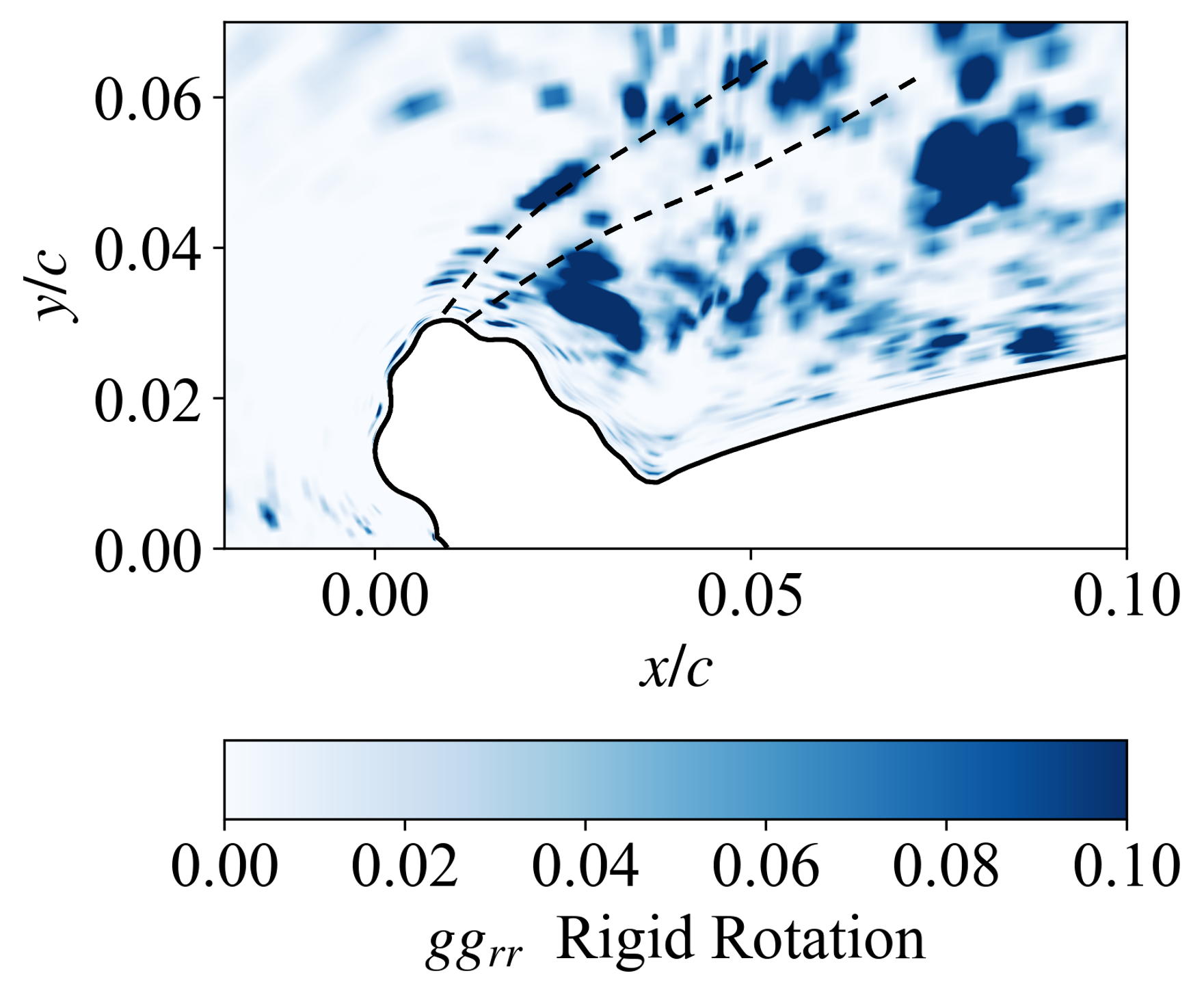}\\[2pt]
    c) Triple decomposition, $gg_{rr}$
  \end{minipage}\hfill
  \begin{minipage}[b]{0.32\textwidth}
    \centering
    \includegraphics[width=\textwidth]{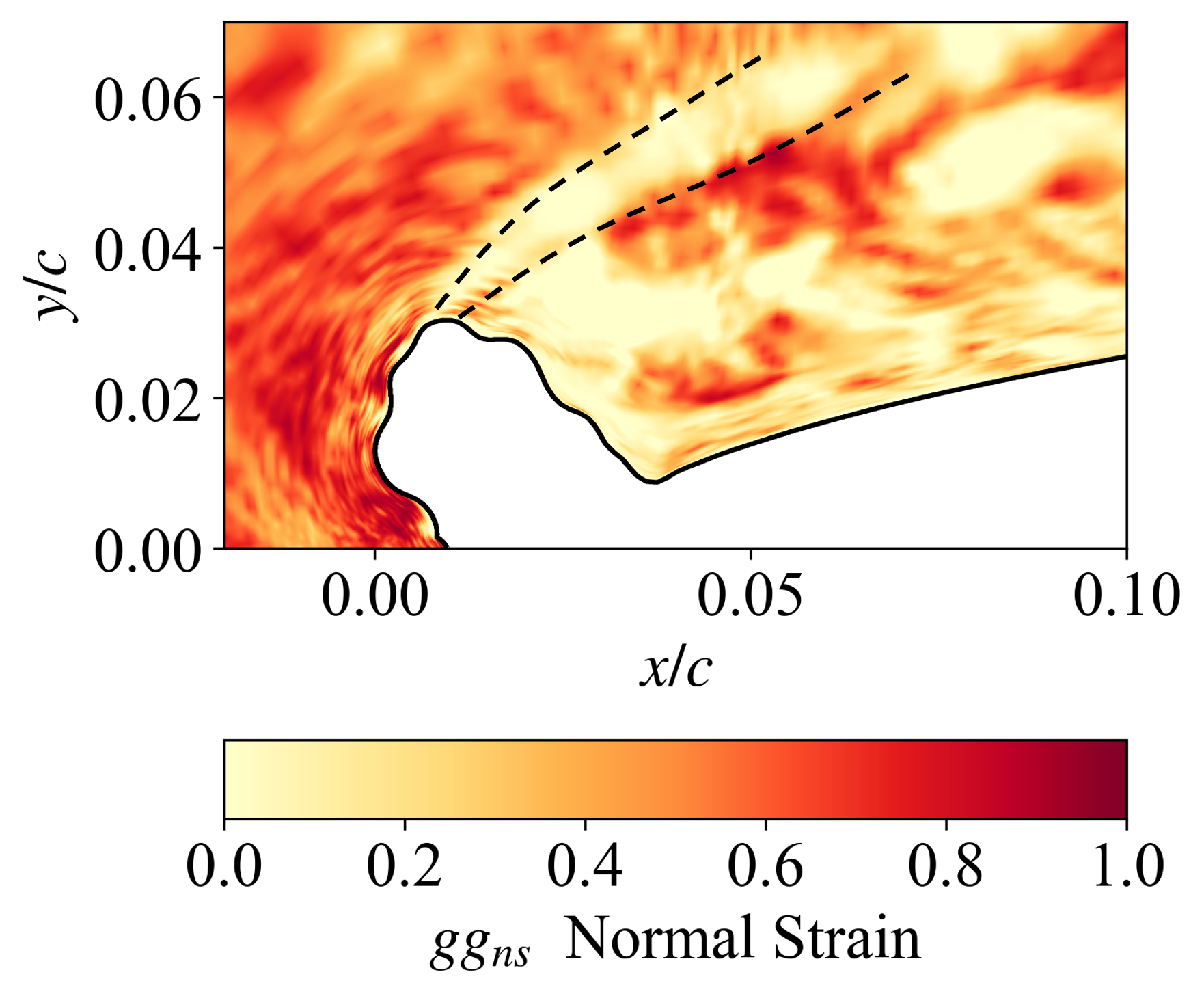}\\[2pt]
    d) Triple decomposition, $gg_{ns}$
  \end{minipage}\hfill
  \begin{minipage}[b]{0.32\textwidth}
    \centering
    \includegraphics[width=\textwidth]{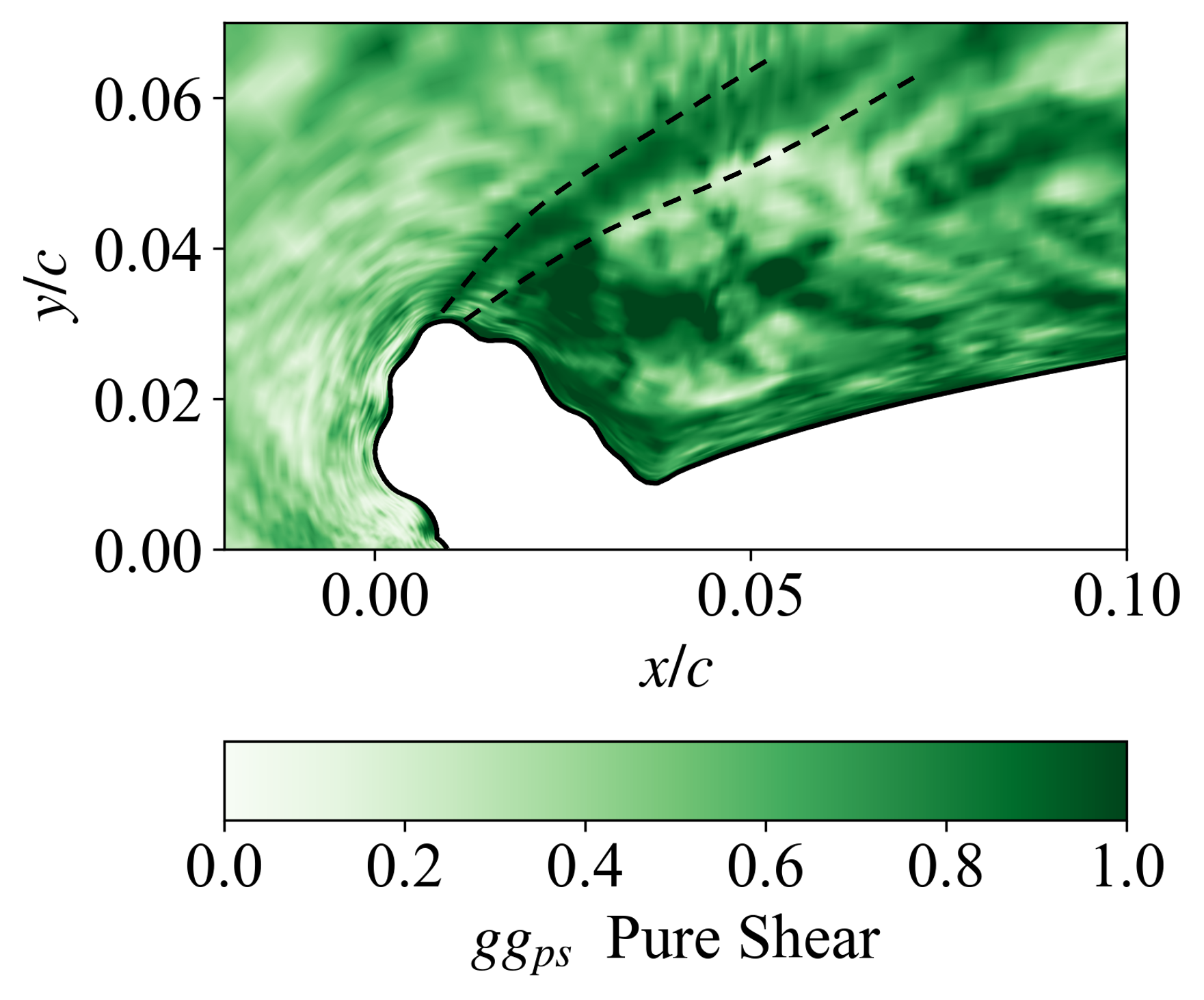}\\[2pt]
    e) Triple decomposition, $gg_{ps}$
  \end{minipage}

  \caption{Differences between the Cauchy--Stokes decomposition and triple decomposition in the leading-edge region for GLC 305 with 944 ice shape.}
  \label{fig:iced_airfoil_CS_vs_TD_leading}
\end{figure}

Figure~\ref{fig:iced_airfoil_CS_vs_TD_leading} further highlights the differences between the Cauchy--Stokes decomposition and triple decomposition in the leading-edge region. \action{The $gg_{rr}$ in Figure~\ref{fig:iced_airfoil_CS_vs_TD_leading}(c) uses a different legend range, with an upper limit of 0.1 rather than 1.0.} The Cauchy--Stokes decomposition identifies the separating shear layer through the concentrated $gg_{ww}$ region, suggesting a vortex-like flow structure. In contrast, triple decomposition provides a different physical interpretation of the separating shear layer. The dominant motion in this region is identified as pure shear rather than rigid-body rotation. This distinction indicates that the apparent vortex-like structure identified by the Cauchy--Stokes decomposition is primarily associated with strong shear, rather than coherent rotational motion. PhysMiner can accurately characterize a specific flow structure in engineering flows.

\subsection{Application of PhysMiner to case 3: propeller}

This section applies PhysMiner to the automated analysis of flow components in a rotating geometry, namely a propeller. The propeller operates at a Reynolds number of $Re_D=1{,}000{,}000$, based on the propeller diameter $D$. The flow-field data are obtained from URANS simulations using OpenFOAM~\cite{jasak2007openfoam}. The propeller flow involves complex and interacting flow structures. A user-specified analysis method is therefore employed to provide explicit prompts to PhysMiner. The prompts guide the analysis of the flow structures and facilitate the exploration of clear physical insights into the propeller flow.

\subsubsection{Effectiveness of Simplified User-Specified Prompts}

\begin{figure}[hbt!]
\centering
\includegraphics[width=1.0\textwidth]{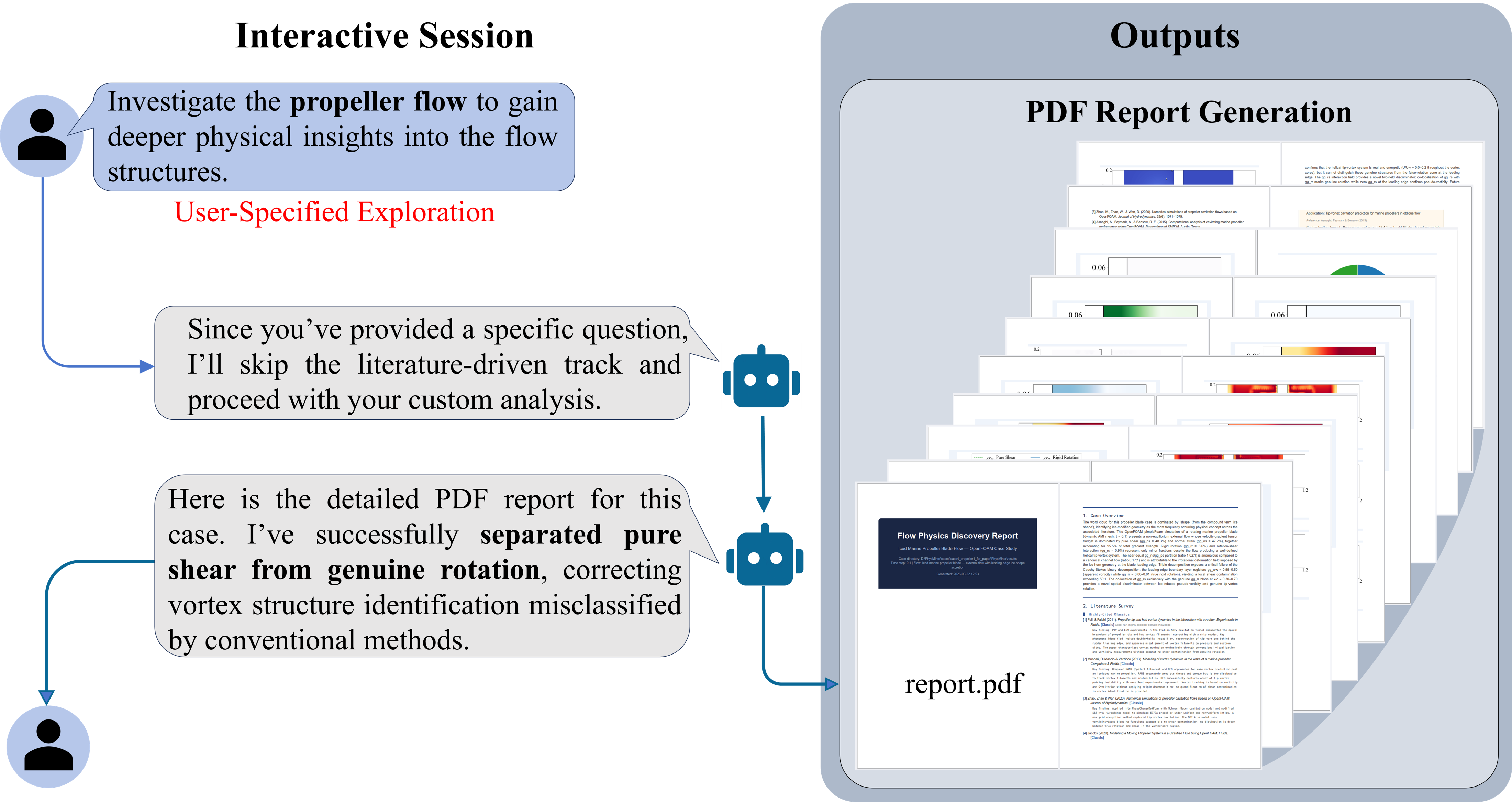}
\caption{User-specified prompt and the final PDF report generated by PhysMiner for propeller case.}
\label{fig:user_specified_prompt_propeller}
\end{figure}

The rotating propeller and its internal flow exhibit complex flow structures. Understanding these structures is therefore an important aspect of analyzing propeller flows. For this case, PhysMiner is explicitly prompted to provide a clearer physical interpretation of the flow structures. Figure~\ref{fig:user_specified_prompt_propeller} shows the user-specified prompt and the final PDF report generated by PhysMiner for the propeller flow. Compared with the prompt used for the iced-airfoil case, the present prompt is considerably simpler. It only specifies ``flow structures'' as the analysis target, without additional guidance based on the flow field, Cauchy-Stokes decomposition, or triple decomposition. Despite this minimal input, PhysMiner can still effectively identify and interpret the underlying flow structures. This result indicates that the user-specified prompt does not need to provide detailed physical guidance for PhysMiner to perform meaningful flow-physics analysis. \action{Some representative figures and key findings from the final PDF report automatically generated by PhysMiner are presented below.}

\subsubsection{New Findings from PhysMiner based on the Specified Prompt}

\begin{figure}[hbt!]
  \centering
  \begin{minipage}[b]{0.48\textwidth}
    \centering
    \includegraphics[width=\textwidth]{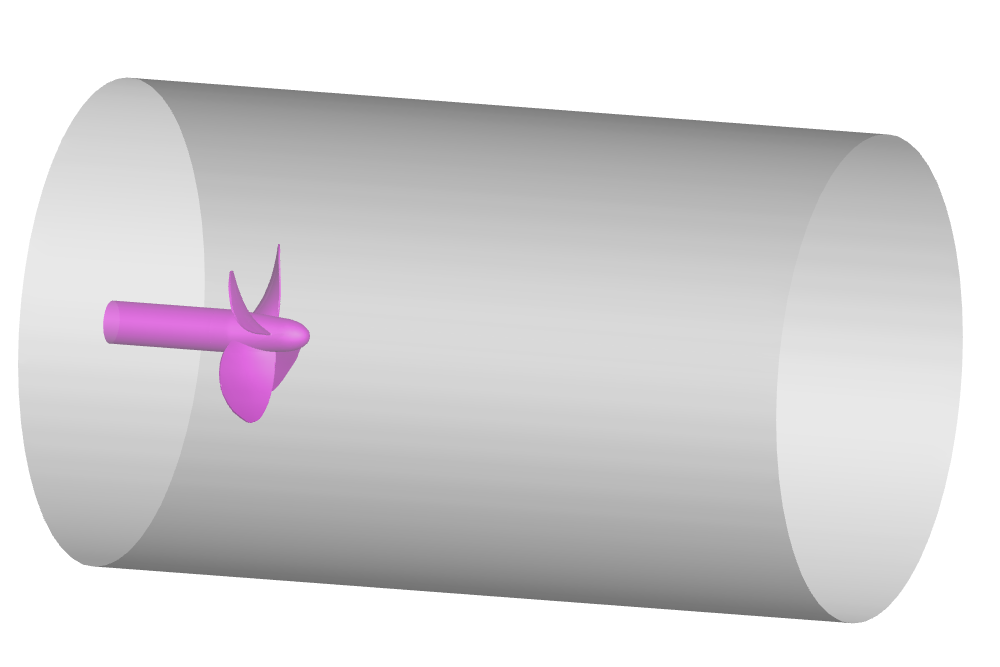}\\[2pt]
    a) Propeller geometry and computational domain
  \end{minipage}
  \\[8pt] 

  \begin{minipage}[b]{0.48\textwidth}
    \centering
    \includegraphics[width=\textwidth]{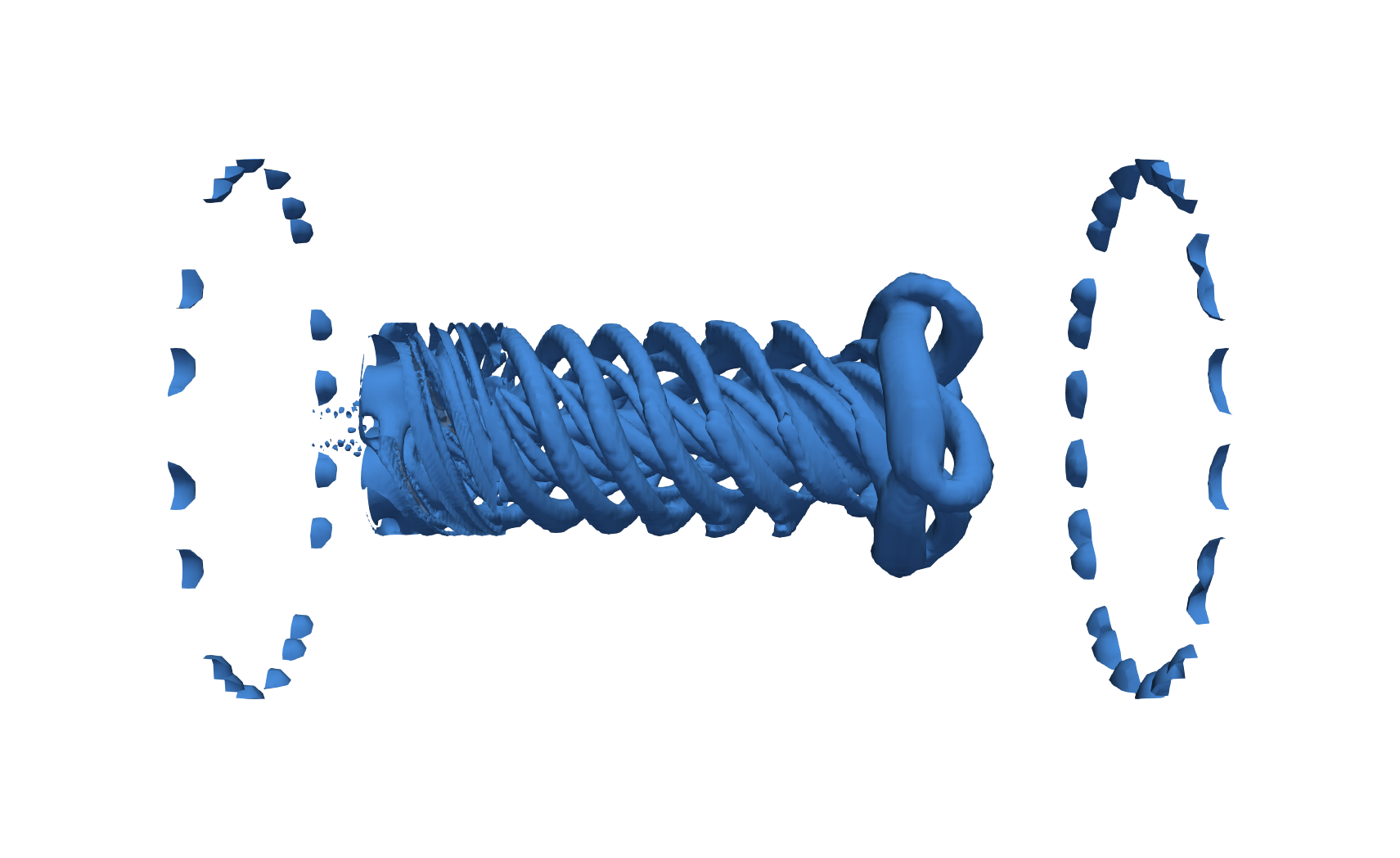}\\[2pt]
    b) Q=10
  \end{minipage}
  \hfill
  \begin{minipage}[b]{0.48\textwidth}
    \centering
    \includegraphics[width=\textwidth]{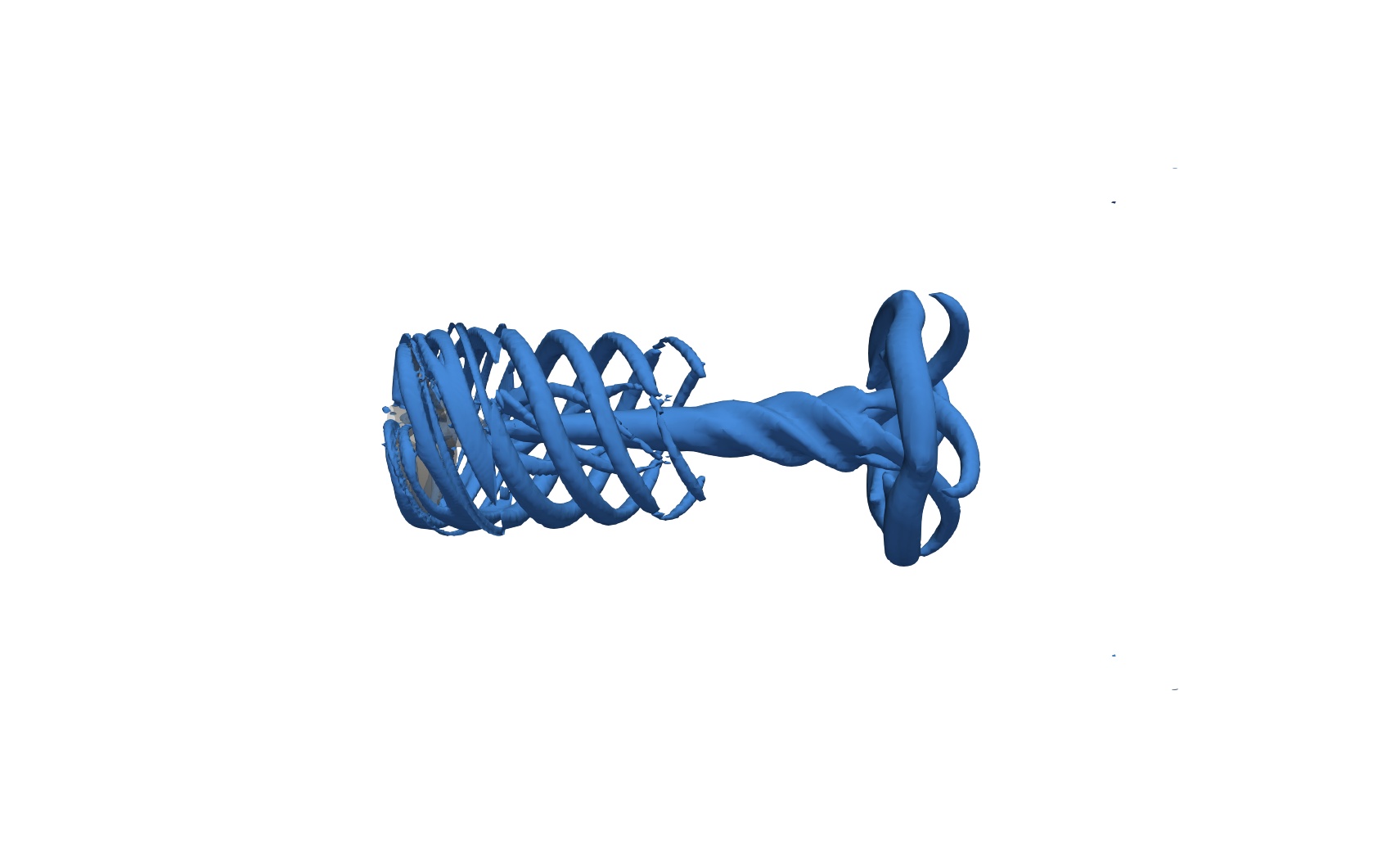}\\[2pt]
    c) Q=1000
  \end{minipage}

  \caption{Propeller configuration and the identified vortex structures using the $Q$-criterion with threshold values of 10 and 1,000.}
  \label{fig:propeller_geometry_Q}
\end{figure}

Figure~\ref{fig:propeller_geometry_Q} shows the propeller configuration and the identified vortex structures using the $Q$-criterion with threshold values of 10 and 1,000. As the threshold increases, the visualization progressively filters out weaker structures and emphasizes only the strongest vortices. At low thresholds, both strong and weak vortex structures are captured, whereas high thresholds isolate only the strong vortices. At the same downstream location, the hub vortex is consistently stronger than the tip vortex, as evidenced in Figure~\ref{fig:propeller_geometry_Q}(c). Along the streamwise direction, the tip vortex exhibits a non-monotonic evolution: it is relatively strong in the near wake, weakens in the intermediate region, and strengthens again further downstream. This behavior reflects the complex dynamics of vortex development and interaction. These results indicate that multiple $Q$-criterion thresholds are required to fully characterize vortex structures across different intensity levels. In comparison, the following triple-decomposition component $\mathrm{gg}_{rr}$ provides complementary insight by directly quantifying the strength characteristics of the vortical structures without threshold dependence.

\begin{figure}[hbt!]
\centering
\begin{minipage}[b]{0.9\textwidth}
  \centering
  \includegraphics[width=\textwidth]{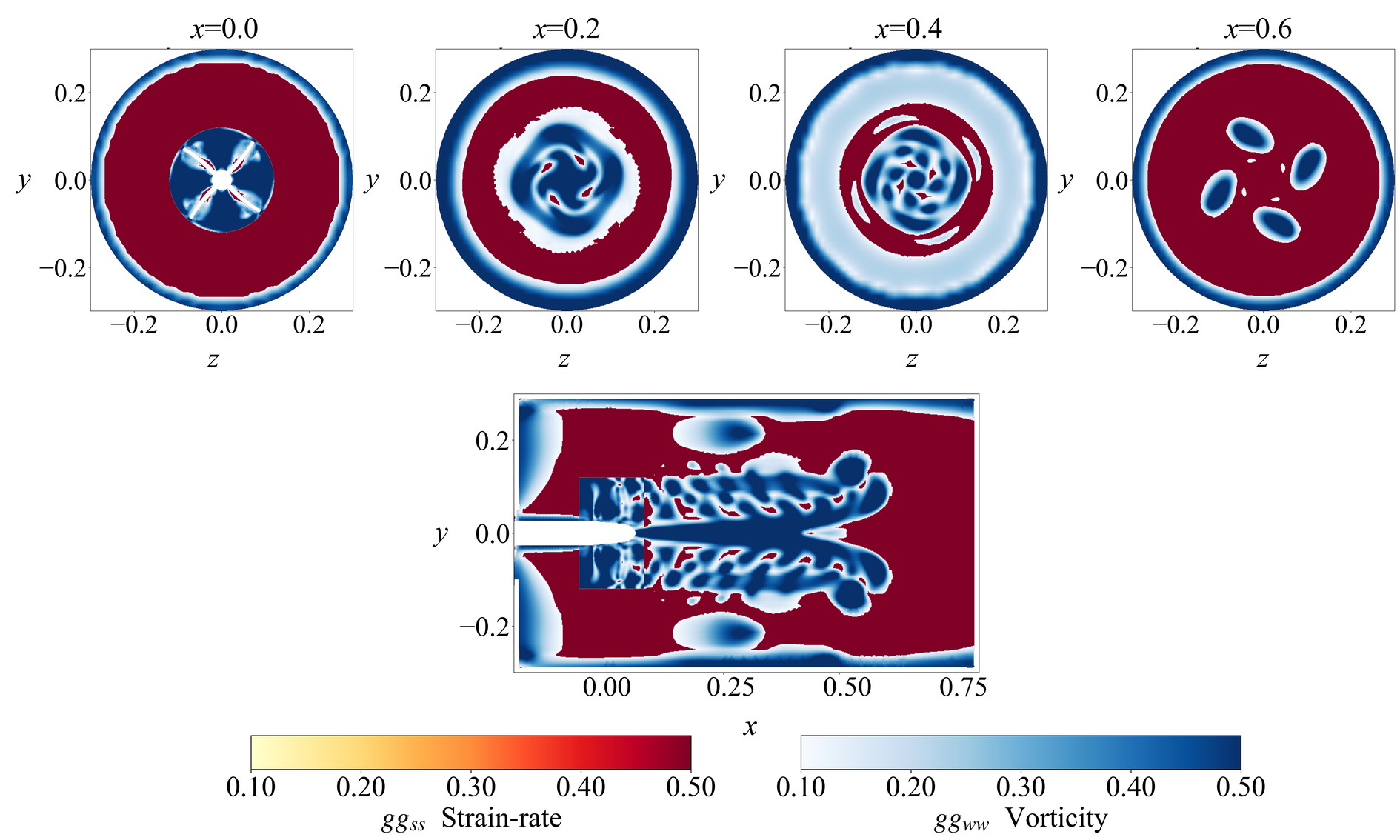}\\[2pt]
  a) Cauchy--Stokes decomposition
\end{minipage}
\hfill
\begin{minipage}[b]{0.9\textwidth}
  \centering
  \includegraphics[width=\textwidth]{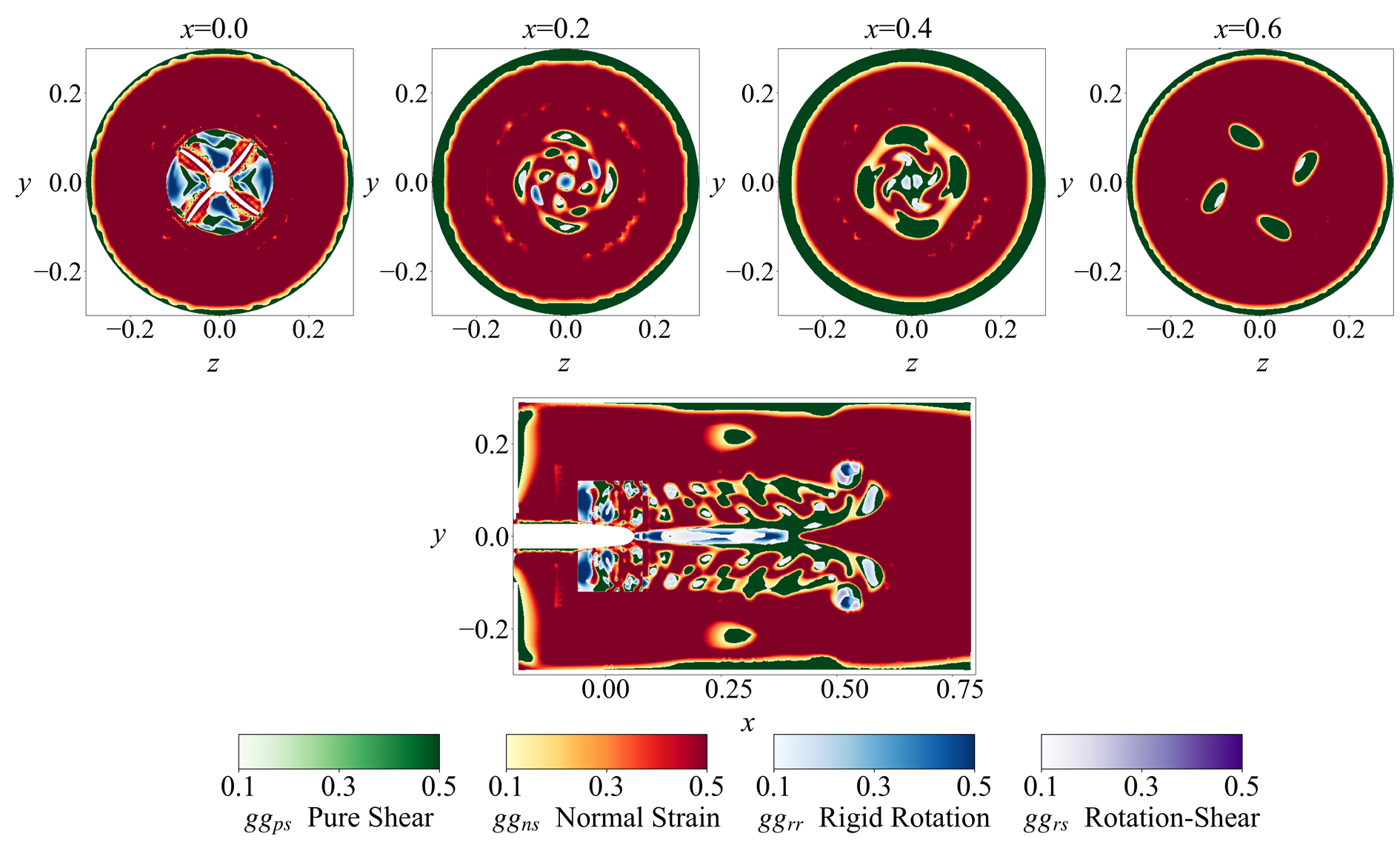}\\[2pt]
  b) Triple decomposition
\end{minipage}
\caption{Comparative vortex system analysis in the propeller flow using both Cauchy--Stokes decomposition and the triple decomposition.}
\label{fig:propeller_CS_TD}
\end{figure}

Figure~\ref{fig:propeller_CS_TD} presents a comparison between Cauchy--Stokes decomposition and triple decomposition in the propeller flow. As shown in Figure~\ref{fig:propeller_CS_TD}(a), the vorticity component is significantly contaminated by pure shear. This effect is particularly pronounced near the outer boundary and in the regions between the wake and the outer boundary, where spurious vortex structures are erroneously identified. As a result, the classical vorticity method fails to reliably distinguish genuine vortical structures from shear-dominated regions. In contrast, Figure~\ref{fig:propeller_CS_TD}(b) demonstrates that the triple decomposition effectively separates rigid rotation from pure shear. The extracted rigid rotation component accurately captures the physically meaningful vortex structures while eliminating false identifications induced by shear contamination. Furthermore, the spatial distribution of the rigid rotation component $\mathrm{gg}_{rr}$ reveals a clear evolution of the vortex system. The hub vortex exhibits a higher strength than the tip vortex at the same downstream locations.

\section{Conclusion}

This paper proposes the PhysMiner framework for automated flow-component analysis. An advanced flow-analysis approach based on triple decomposition is employed to convert raw flow fields into physically interpretable components. These components provide a physics-informed basis for LLM reasoning and support both literature-driven and user-specified physics discovery. Several key findings can be drawn from this study:

(1) The PhysMiner framework does not rely heavily on user-designed prompts. It supports both literature-driven exploration and user-specified problem exploration. In the literature-driven mode, no user input is required. PhysMiner can review the literature for a specific flow and identify challenging or unresolved problems for further investigation. This process reduces the dependence on human-defined problem formulations and minimizes the potential bias introduced by preconceived research questions. User-specified exploration is also supported when a particular research problem is of interest. Importantly, the prompt does not need to provide a detailed description of the underlying physics or the specific analysis method.  For example, the user does not need to distinguish between Cauchy--Stokes decomposition and triple decomposition or specify either one in the prompt.

(2) The triple decomposition library enables systematic cross-case comparison by benchmarking each new flow case against previously stored cases through a categorical Jaccard tree distance. For the periodic hill case, the library identifies the backward-facing step as the structurally closest historical case, with an identical ten-dimensional fingerprint. Building on this match, the flow-comparison agent compares the two cases in detail. Despite sharing separated internal-flow geometry, the periodic hill exhibits a higher pure-shear fraction and more spatially uniform streamwise profiles. The backward-facing step, in contrast, carries a persistent normal-strain background from its developing inflow boundary layer. This demonstrates that even within the same broad flow category, subtle physical differences can be effectively identified and interpreted.

(3) PhysMiner is demonstrated on three flow cases: periodic hill flow, iced-airfoil flow, and propeller wake. For periodic hill flow, an existing subgrid-scale model is modified by introducing a rotation-suppression correction term. The modified model improves Reynolds-stress predictions compared with the standard baseline model. For the iced-airfoil flow, PhysMiner provides a new interpretation of the separating shear layer. The layer is misidentified by existing methods, leading to misleading physical interpretations. For the propeller wake, PhysMiner distinguishes shear from genuine rotation. This separation corrects the identification of vortex structures that are misclassified by conventional methods.

Although this study demonstrates the potential of PhysMiner for automated flow-component analysis and turbulence-model development, several limitations remain. First, the current triple-decomposition library contains a limited number of flow cases. As the library is publicly available, its database can be progressively expanded through broader community use. Second, the current framework does not yet provide a fully automated tool for complete posterior CFD validation. Future development will extend PhysMiner to automate the entire validation process.

\section*{Funding Sources}

This material is based upon work supported by the National Science Foundation under Grant
Numbers 2415347 and 2541867. 

\section*{Acknowledgments}

Part of the computations was conducted on the Stampede~3 supercomputer at Texas Advanced Computing Center
through allocation ATM-140019 from the Advanced Cyberinfrastructure Coordination Ecosystem:
Services \& Support (ACCESS) program, which is supported by U.S. National Science Foundation
grants \#2138259, \#2138286, \#2138307, \#2137603, and \#2138296. 
Portions of this manuscript benefited from the use of artificial intelligence tools, which were employed primarily to improve readability, grammar, and clarity of language. All intellectual content, analysis, and conclusions remain the sole work of the authors.

\bibliography{physminer}

\end{document}